\documentclass[12pt]{article}
\pdfoutput=1

\usepackage{putex}
\usepackage{amsmath,amssymb,amsfonts}
\usepackage{psfrag}
\usepackage{url}
\usepackage{color}

\definecolor{darkblue}{rgb}{0.1,0.1,.7}
\definecolor{purple}{rgb}{0.6,0,0.6}
\definecolor{orange}{rgb}{0.9,0.6,0}
\usepackage[colorlinks, linkcolor=darkblue, citecolor=darkblue, urlcolor=darkblue, linktocpage]{hyperref} 
\usepackage[square, comma, compress,numbers]{natbib}
\usepackage[]{amsmath}
\usepackage[]{graphicx}
\usepackage[]{latexsym}
\usepackage[utf8]{inputenc}
\usepackage{geometry}
\usepackage{amscd}
\usepackage[all,cmtip]{xy}
\usepackage{mathrsfs}
\usepackage[margin=10pt,font=small,labelfont=bf]{caption}
\usepackage{dsdshorthand}
\usepackage{changepage}
\usepackage{setspace}
\usepackage{tikz}
\usepackage{subcaption}

\usepackage[titles]{tocloft}
\usetikzlibrary{arrows,positioning} 

\tikzset{
    >=stealth',
    punkt/.style={
           rectangle,
           rounded corners,
           draw=black, very thick,
           text width=15em,
           minimum height=2em,
           text centered},
    pil/.style={
           ->,
           thick,
           shorten <=2pt,
           shorten >=2pt,}
}

\def\tl{\tilde{\ell}}

\def\SL2{\widetilde{SL}(2,\mathbb R)}
\def\mP{\mathcal{P}}

\def\mC{\mathcal C}

\def\la{\langle}
\def\ra{\rangle}
\newcommand{\es}[2] {\begin{equation} \label{#1} \begin{split} #2 \end{split} \end{equation}}

\newcommand\mR{\mathbb{R}}
\newcommand\mZ{\mathbb{Z}}

\numberwithin{equation}{section}

\newcommand{\grp}[1]{\mathrm{#1}}

\newcommand{\abs}[1]{\left\lvert #1 \right\rvert}
\newcommand {\bes} {\begin {equation*}}
\newcommand {\ees} {\end {equation*}}

\renewcommand{\es}[2] {\begin{equation} \label{#1} \begin{split} #2 \end{split} \end{equation}}

\numberwithin{equation}{section}

\def\<{\langle}
\def\>{\rangle}

 \def\ie{\begin{equation}\begin{aligned}}
\def\fe{\end{aligned}\end{equation}}

\begin{document}

\begin{flushright}
\hfill{\tt PUPT-2584}
\end{flushright}

\institution{PU}{Joseph Henry Laboratories, Princeton University, Princeton, NJ 08544, USA}

\title{
An exact quantization of Jackiw-Teitelboim gravity
}

\authors{Luca V.~Iliesiu, Silviu S.~Pufu, Herman Verlinde, and Yifan Wang}

\abstract{
We propose an exact quantization of two-dimensional Jackiw-Teitelboim (JT) gravity by formulating the JT gravity theory as a 2D gauge theory placed in the presence of a loop defect.  The gauge group is a certain central extension of $PSL(2, \mR)$ by $\R$.  We find that the exact partition function of our theory when placed on a Euclidean disk matches precisely the finite temperature partition function of the Schwarzian theory.  We show that observables on both sides are also precisely matched: correlation functions of boundary-anchored Wilson lines in the bulk are given by those of bi-local operators in the Schwarzian theory.  In the gravitational context,  the Wilson lines are shown to be equivalent to the world-lines of massive particles in the metric formulation of JT gravity.  
}
\date{}

\maketitle
\tableofcontents

\section{Introduction and summary of results}
\label{sec:intro}

While AdS/CFT \cite{Maldacena:1997re, Witten:1998qj, Gubser:1998bc} has provided a broad framework to understand quantum gravity, most discussions are limited to perturbation theory around a fixed gravitational background. The difficulty of going beyond perturbation theory stems from our limited understanding of both sides of the duality: on the boundary side, it is difficult to compute correlators in strongly coupled CFTs, while in the bulk there are no efficient ways of performing computations beyond tree level in perturbation theory.  2D/1D holography  \cite{Maldacena:2016hyu, Polchinski:2016xgd, Mezei:2017kmw, Gross:2017vhb, Gross:2017aos, Stanford:2017thb, Witten:2016iux, Klebanov:2016xxf, kitaevTalks, Maldacena:2016upp, Jensen:2016pah, Engelsoy:2016xyb, Maldacena:2017axo, Maldacena:2018lmt, Harlow:2018tqv, Kitaev:2018wpr, yang2018quantum, Saad:2019lba} provides one of the best frameworks to understand quantum gravity beyond perturbation theory, partly because gravitons or gauge bosons in two dimensions have no dynamical degrees of freedom. Nevertheless, many of the open questions from higher dimensional holography, such as questions related to bulk reconstruction or the physics of black holes and wormholes, persist in 2D/1D holography.  

One of the simplest starting points to discuss 2D/1D holography is the two-dimensional Jackiw-Teitelboim (JT) theory \cite{teitelboim1983gravitation, Jackiw:1984je}, which, in the second-order formalism, involves a dilaton field $\Phi$ and the metric tensor $g_{\mu \nu}$. The Euclidean action is given by
 \be
 \label{eq:JT-gravity-action}
 S_{JT}[\Phi, g] = -{1\over 16\pi G}\int_\Sigma d^2 x \sqrt{g}\,\Phi(R + \Lambda) -{1\over 8\pi G} \int_{\partial \Sigma} du\,\sqrt{\gamma} \left(\Phi |_{\partial \Sigma}  \right) K \,,
\ee
where we have placed the theory on a manifold $\Sigma$ with metric $g_{\mu \nu}$ and where the  boundary of this manifold, $\partial \Sigma$, is endowed with the induced metric $\gamma$ and the extrinsic curvature $K$. The bulk equations of motion set 
 \es{BulkEoms}{
  R = -\Lambda \,, \qquad
   \nabla_\mu \nabla_\nu \Phi =  \frac{\Lambda}{2}  g_{\mu\nu} \Phi  \,,
 }
and thus, on-shell, the bulk term in \eqref{eq:JT-gravity-action} vanishes.  The remaining degrees of freedom are thus all on the boundary of some connected patch of Euclidean AdS$_2$ (or, equivalently, of the Poincar\'e disk), where one can fix $\Phi |_{\partial \Sigma} = C/\epsilon$ and the boundary metric $g_{uu} = 1/\epsilon^2$ to be constant, such that the total boundary length is given by $\beta/\epsilon$. Taking the limit $\epsilon \rightarrow 0$ is then equivalent to taking the patch to extend to the entire Poincar\'e disk. Consider  $\rho$ and $f$  as Poincar\'e disk coordinates (with $ds^2 =d\rho^2 + \sinh^2 (\rho) \,df^2$).  Thus one can rewrite the action \eqref{eq:JT-gravity-action} in terms of $\rho(u)$ and $f(u)$ on the boundary $\partial \Sigma$, which we can parametrize by the proper length $u$. In the limit $\epsilon \rightarrow 0$, one finds that after imposing all the previously specified boundary conditions, the boundary term in \eqref{eq:JT-gravity-action} is given (up to a divergent piece removed by holographic renormalization) by the Schwarzian action \cite{kitaevTalks, Maldacena:2016upp, Engelsoy:2016xyb}, 
\be 
\label{eq:schw-action}
S_{\text{Schw}}[f] &= - C \int_0^\b d u \left\{\tan \frac{\pi f}{\beta}, u\right\} \,, \qquad
  \{F, u\} \equiv \frac{F'''}{F'} -  \frac{3}2 \left(\frac{F''}{F'}\right)^2\,.
\ee 
While the equivalence between the JT-gravity action and the Schwarzian action is clear on-shell \cite{Maldacena:2016upp}, there are subtleties in quantizing and uncovering the global structure of the gravitational theory.\footnote{For example, it is unclear what measure and integration contour one should use in the gravitational path integral.
 }  Due to these subtleties, it is difficult to formally prove the equivalence between JT-gravity and the Schwarzian theory in a path integral approach.\footnote{See, however, \cite{Harlow:2018tqv, Blommaert:2018oro, Kitaev:2018wpr, yang2018quantum, Saad:2019lba} for progress in this direction.}

 An important tool that we use for quantizing the bulk gravitational theory is the equivalence between its first order formulation and a 2D gauge theory. Specifically, the frame $e^a$ and spin connection $\omega$ associated to the manifolds which are summed over in the gravitational path integral can be packaged together as a gauge field with an $\mathfrak{sl}(2, \mR)$ gauge algebra \cite{Isler:1989hq, Chamseddine:1989yz, Cangemi:1992bj, Mertens:2018fds,Grumiller:2017qao,Gonzalez:2018enk}. The bulk term in \eqref{eq:JT-gravity-action} is then captured by a topological BF theory with this gauge algebra. This equivalence is analogous to the formulation of 3D Chern Simons theory as a theory of 3D quantum gravity, where the gauge algebra is given by various real forms of $\mathfrak{so}(2, 2)$ \cite{Witten:1988hc, Witten:1989ip}.  The quantization of JT gravity in the gauge theory description was also explored recently by dimensionally reducing the Chern Simons theory that describes 3D gravity \cite{Blommaert:2018oro, Cotler:2018zff, Lin:2018xkj,Blommaert:2018iqz,  Blommaert:2019hjr, Mertens:2019bvy}. However,  obtaining the  possible boundary terms and the exact gauge group that are needed in order to recover the dual of the Schwarzian theory is, to our knowledge, still an open question that we hope to answer in this work.\footnote{A priori it is unclear whether there even exists a gauge group for which the gauge theory  would reproduce observables in the Schwarzian, which in turn are expected to capture results in JT gravity. This is due to the fact that there exist gauge field configurations where the frame is non-invertible and, consequently, such configurations do not have a clear geometric meaning in JT gravity. Note that, in the Chern-Simons description of 3D gravity, due to the non-invertibility of the frame, one does not expect to be able to capture all the desired features of 3D pure quantum gravity \cite{Witten:2007kt}. For example, given that the Chern-Simons theory is topological, and consequently has few degrees of freedom, one cannot expect to reproduce the great degeneracy of BTZ black hole states from the gravitational theory \cite{Witten:2007kt}. In contrast, as we will discuss in this paper, the 2D gauge theory formulation of JT gravity is able to reproduce all known Schwarzian observables exactly. }

When placing the gauge theory on a disk, the natural Dirichlet boundary conditions are set by fixing the gauge field or, equivalently, the frame $e^a$ and spin connection $\omega$ at the boundary of the disk. In such a case, a boundary term like that in \eqref{eq:JT-gravity-action} does not need to be added to the action in order for the theory to have a well-defined variational principle. The resulting system can be shown to be a trivial topological theory which does not capture the boundary dynamics of \eqref{eq:JT-gravity-action}. Consequently, we introduce a boundary condition changing defect whose role in the BF-theory is to switch the natural Dirichlet boundary conditions to those needed in order to reproduce the Schwarzian dynamics. With this boundary changing defect the first and second formulations of JT gravity give rise to the same boundary theory:\footnote{Possible boundary conditions for the gauge theory reformulation of JT-gravity were also discussed in \cite{Grumiller:2017qao}. A concrete proposal for the rewriting of the boundary term in \eqref{eq:JT-gravity-action} was also discussed in \cite{Gonzalez:2018enk}, however the quantization of the theory was not considered. }  
  \begin{center}
  \vspace{0.1cm}
  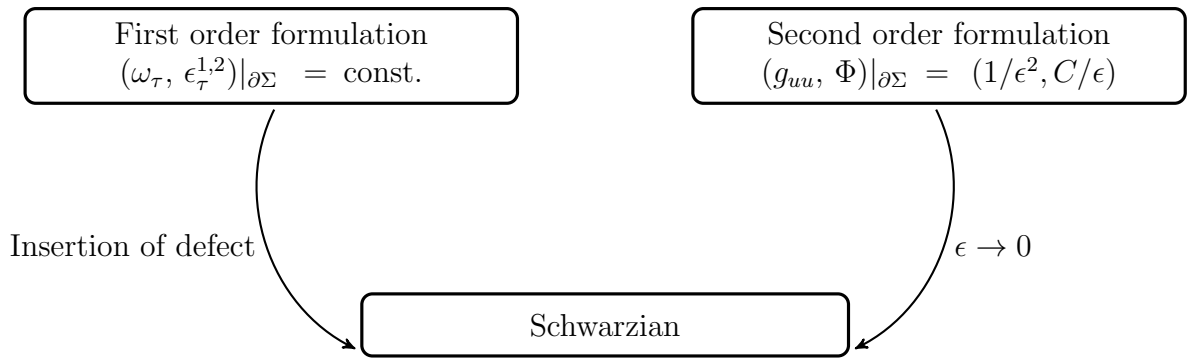
\begin{figure}[h!]
\begin{tikzpicture}[node distance=1cm, auto]
\node[punkt] (schw) {Schwarzian};
\node[above=3cm of schw](dummy) {};
 \node[punkt, inner sep=5pt,left=1cm of dummy] (first) {First order formulation \\ $(\omega_\tau, \,\e^{1,2}_{\tau})|_{\partial \Sigma}\, =\, $ const.}
 edge[pil,->, bend right=45] node[left] {Insertion of defect} (schw);
 \node[punkt, inner sep=5pt,right=1cm of dummy]
 (second) {Second order formulation \\ $(g_{uu}, \, \Phi )|_{\partial \Sigma}=\, (1/\epsilon^2, C/\epsilon)$ }
 edge[pil,->, bend left=45] node[right]{$\epsilon \to 0$} (schw);

\end{tikzpicture}\\
\vspace{0em}
\caption{Schematic representation showing that the dynamics on the defect in the gauge theory is the same as that in the Schwarzian theory, which in turn describes the boundary degrees of freedom of \eqref{eq:JT-gravity-action}.  }
\end{figure}
\vspace{-1.2cm}
\end{center} 

  In order for the equivalence between the Schwarzian and the gauge theory to continue to hold at the quantum level, we find that the gauge group needed to properly capture the global properties of the gravitational theory is given by an extension of $PSL(2, \mR)$ by $\mR$. This extension is related to the universal cover of the group $PSL(2, \mR)$, denoted by $\SL2$.\footnote{A similar observation was made in \cite{Schaller:1993np}. There it was shown that in order for gravitational diffeomorphisms to be mapped to gauge transformations in the BF-theory when placed on a cylinder, one needs to consider a gauge group given by $\SL2$, instead of the typically assumed $PSL(2, \mR)$. } With this choice of gauge group, when placing the bulk theory in Euclidean signature on a disk we find a match between its exact partition function and that computed in the Schwarzian theory \cite{Bagrets:2016cdf, Witten:2016iux, Belokurov:2018aol}. This match is obtained by demanding that the gauge field component along the boundary should vanish.\footnote{In a gauge-independent language, here we demand a trivial holonomy around the boundary of the disk. 
For general boundary holonomy, the dual is given by a non-relativistic particle moving on $H_2^+$ in a magnetic field, in the presence of an $\SL2$ background gauge field. As we point out in Appendix~\ref{sec:schw} this is slightly different than considering the Schwarzian with $SL(2, \mR)$ twisted boundary conditions, which was considered in \cite{Witten:2016iux, Mertens:2019tcm}. }

 The first natural observable to consider beyond the partition function is given by introducing probe matter in the gauge theory.  On the gauge theory side, introducing probe matter is equivalent to adding a Wilson line anchored at two points on the boundary. In the Schwarzian theory we expect that this coupling is captured by bilocal operators $\cO_\l(u_1, u_2)$. We indeed confirm that all the correlation functions of bi-local operators in the Schwarzian theory \cite{Mertens:2017mtv}  match the correlation functions of  Wilson lines that intersect the defect. More specifically, the time ordered correlators of bi-local operators in the boundary theory are given by correlators of non-intersecting defect-cutting Wilson lines, while out-of-time-ordered correlators are given by intersecting Wilson line configurations. Furthermore, by computing the expectation value of bulk Wilson lines in the gauge theory, we provide a clear representation theoretic meaning to their correlators and provide the combinatorial toolkit needed to compute  any such correlator.  As we will show these Wilson lines also have a gravitational interpretation: inserting such Wilson lines in the path integral is equivalent to summing over all possible world-line paths for a particle moving between two fixed points on the boundary of the AdS$_2$ patch.  Furthermore, we discuss the existence of further non-local gauge invariant operators which can potentially be used to computed the amplitudes associated to a multitude of scattering problems in the bulk.

The remainder of this paper is organized as follows. In Section~\ref{sec:SL(2,R)-Yang-Mills} we show the on-shell equivalence between the equations of motion of the Schwarzian theory and those in the gauge theory description of JT gravity, when boundary conditions are set appropriately. In Section~\ref{sec:quantization-and-choice-of-gg} we discuss the quantization of the gauge theory. In this process, in order to match results in the Schwarzian theory, or, alternatively in the second order formulation of JT gravity, we determine a consistent global structure for the gauge group and determine potential boundary conditions such that the partition function of the gauge theory agrees with that of the Schwarzian. In Section~\ref{sec:wilson-loop-and-bi-local-op}, we show the equivalence between Wilson lines in the gauge theory and bi-local operators in the boundary theory. Furthermore, we discuss the role of a new class of gauge invariant non-local operators and compute their expectation value. Finally we discuss future directions of investigation in Section~\ref{sec:future-directions}. In Appendix~\ref{sec:schw}, we review various properties of the Schwarzian theory and derive at the level of the path integral, its equivalence to a non-relativistic particle moving in hyperbolic space in the presence of a magnetic field. For the readers interested in details, we suggest reading Appendix~\ref{app:harmonic-analysis-SL(2,R)} and \ref{app:fusion-coeff} where we provide a detailed description of harmonic analysis on the $\SL2$ group manifold and derive the fusion coefficients for various representations of $\SL2$. Finally, we revisit the gravitational interpretation of the gauge theory observables in Appendix~\ref{sec:wilson-loop-and-external-matter} and we show that Wilson lines that intersect the defect are equivalent to probe particles in JT-gravity propagating between different points on the boundary.

\section{Classical analysis of $\mathfrak{sl}(2, \mathbb{R})$ gauge theory}
\label{sec:SL(2,R)-Yang-Mills}

\subsection{A rewriting of JT-gravity in the first-order formulation}
\label{sec:rewriting-JT-gravity}

As shown in \cite{Isler:1989hq, Chamseddine:1989yz}, JT gravity \eqref{eq:JT-gravity-action} can be equivalently written in the first-order formulation, which involves the frame and spin-connection of the manifold, as a 2D BF theory with gauge algebra $\mathfrak{sl}(2, \R)$.\footnote{Similarly, there is an equivalence between a different 2D gravitational model, the Callan–Giddings–Harvey–Strominger model and a 2D BF-theory with the gauge algebra given by a central extension of $\mathfrak{iso}(1, 1)$ \cite{ Cangemi:1992bj, Callan:1992rs}. Similar to our work here, it would be interesting to explore exact quantizations of this gauge theory.}  Let us review this correspondence starting from the BF theory.\footnote{Unlike \cite{Isler:1989hq, Chamseddine:1989yz}, we will work in Euclidean signature.} To realize this equivalence on shell, we only need to rely on the gauge algebra of the BF theory and not on the global structure of the gauge group.  Thus, the gauge group could be $PSL(2, \R)$ or any of its central extensions. For this reason, we will for now consider  the gauge group to be $\cG$ and will specify the exact nature of $\cG$ in Section~\ref{sec:quantization-and-choice-of-gg}. 

To set conventions, let us write the $\mathfrak{sl}(2, \R)$ algebra in terms of three generators $P_0$, $P_1$, and $P_2$, obeying the commutation relations
 \es{sl2R}{
  [P_0, P_1] = P_2 \,, \qquad
   [P_0, P_2] = -P_1 \,, \qquad
   [P_1, P_2] = -P_0 \,.
 }
For instance, in the two-dimensional representation the generators $P_0$, $P_1$, and $P_2$ can be represented as the real matrices
 \es{MatrixRepresentation}{
  P_0 = \frac{i \sigma_2}{2} \,, \qquad P_1 = \frac{\sigma_1}{2} \,, \qquad P_2 = \frac{\sigma_3}{2} \,.
 }
An arbitrary $\mathfrak{sl}(2, \R)$ algebra element consists of a linear combination of the generators with real coefficients. The field content of the BF theory consists of the gauge field $A_\mu$ and a scalar field $\phi$, both transforming in the adjoint representation of the gauge algebra.  Under infinitesimal gauge transformations with parameter $\epsilon(x) \in \mathfrak{sl}(2, \R)$, we have
 \es{GaugeTransf}{
  \delta \phi = [\epsilon, \phi] \,, \qquad
   \delta A_\mu = \partial_\mu \epsilon + [\epsilon, A_\mu] \,.
 }
Consequently, the covariant derivative is $D_\mu = \partial_\mu - A_\mu$ (because then we have, for instance,  $\delta ( D_\mu \phi) = [\epsilon, D_\mu \phi]$), and then the gauge field strength is $F_{\mu\nu} \equiv - [D_\mu, D_\nu] = \partial_\mu A_\nu - \partial_\nu A_\mu - [A_\mu, A_\nu]$.  In differential form notation, $F = dA  - A \wedge A$.

Ignoring any potential boundary terms, the BF theory Euclidean action is 
 \es{eq:BF-action}{
  S_\text{BF} = -i \int \tr (\phi F) \,, 
 }
where the trace is taken in the two-dimensional representation \eqref{MatrixRepresentation}, such that $\tr \phi F = \eta^{ij} \phi_i F_j/2$, where $\eta^{ij} = \diag(-1, 1, 1)$, with $i, j = 1, \,2,\,3$. 
To show that the action \eqref{eq:BF-action} in fact describes JT gravity, let us denote the components of $A$ and $\phi$ as 
 \es{AToe}{
  A(x) = \sqrt{\frac{\Lambda}{2}} e^a(x) P_a + \omega(x) P_0 \,, \qquad \phi(x) = \phi^a(x) P_a + \phi^0(x) P_0 \,,
 }
where the index $a = 1, 2$ is being summed over, $\Lambda > 0$ is a constant, and $e^a$ and $\omega$ are one-forms while $\phi^a$ and $\phi^0$ are scalar functions.  An explicit computation using $F = dA - A \wedge A$ and the commutation relations \eqref{sl2R} gives
 \es{GotF}{
  F = \sqrt{\frac{\Lambda}{2}} \left[ de^1 + \omega \wedge e^2 \right] P_1 
    + \sqrt{\frac{\Lambda}{2}}\left[ de^2 - \omega \wedge e^1  \right] P_2 + \left[ d \omega + \frac{\Lambda}{2} e^1 \wedge e^2  \right] P_0  \,.
 }
The action \eqref{eq:BF-action} becomes
 \es{ActionAgain}{
  S_\text{BF} = -\frac{i}{2} \int  \sqrt{\frac{\Lambda}{2}} 
   \left[ \phi^1  (de^1 + \omega \wedge e^2) + \phi^2 (de^2 - \omega \wedge e^1 )  \right] 
    -  \phi^0 \left( d \omega + \frac{\Lambda}{2} e^1 \wedge e^2 \right) \,.
 }
The equations of motion obtained from varying $\phi$ yields $F=0$. Specifically, the variation of $\phi^1$ and $\phi^2$ imply $\tau^a = de^a + \omega^{a}{}_{b} \wedge e^b = 0$, with $\omega^{1}{}_{2} = - \omega^{2}{}_{1} = \omega$, which are precisely the zero torsion conditions for the frame $e^a$ with spin connection $\omega^{a}{}_{b}$.  Plugging these equations back into \eqref{ActionAgain} and using the fact that for a 2d manifold $d \omega = \frac{R}{2} e^1 \wedge e^2$, with $R$ being the Ricci scalar, we obtain
 \es{ActionAgain2}{
  S_\text{BF} = \frac{i}{4} \int  d^2 x\, \sqrt{g}\, \phi^0 \left( R+ \Lambda \right)   \,,
 }
which is precisely the bulk part of the JT action with dilaton $\Phi= - i \phi^0/4$.\footnote{One might be puzzled by the fact that when $\phi^0$ is real, $\Phi$ is imaginary. However, when viewing  $\Phi$ or $\phi_0$ as Lagrange multipliers, this is the natural choice for the reality of both fields. However, note that in the second-order formulation of JT-gravity \eqref{eq:JT-gravity-action} one fixes the value of the dilaton ($\Phi$) along the boundary to be real. As we describe in Section~\ref{sec:var-principle}, we do not encounter such an issue in the first-order formulation, since we will not fix the value of $\phi$ along the boundary.    }  Here, the 2d metric is $g_{\mu\nu} = e^1_\mu e^1_\nu + e^2_\mu e^2_\nu$, and $d^2 x \sqrt{g} = e^1 \wedge e^2$.  The equation of motion obtained from varying $\phi_0$ implies $R = - \Lambda$, and since $\Lambda > 0$, we find that the curvature is negative.  Thus, the on-shell gauge configurations of the BF theory parameterize a patch of hyperbolic space (Euclidean AdS).

Note that the equations of motion obtained from varying the gauge field, namely \be 
\label{eq:eq-for-phi}
D_\mu \phi = \partial_\mu \phi - [A_\mu, \phi] =  0\,,
\ee
 can be written as 
 \es{eomsGauge}{
  d\phi_0 &=  \sqrt{\frac{\Lambda}{2}} \left( -e^1 \phi^2 + e^2 \phi^1  \right) \,, \\
  d \phi^1 &= - \omega \phi^2 + \sqrt{\frac{\Lambda}{2}} e^2 \phi_0 \,, \\
  d \phi^2 &= \omega \phi^1 - \sqrt{\frac{\Lambda}{2}} e^1 \phi_0 \,.
 }
It is straightforward to check that taking another derivative of the first equation and using the other two gives the equation for $\Phi$ in \eqref{BulkEoms}.  

The spin connection $\omega^a{}_b$ is a connection on the orthonormal frame bundle associated to a principal $SO(2)$ bundle.  For a pair of functions $\epsilon^a$ transforming as an $SO(2)$ doublet, the covariant differential acts by $D \epsilon^a = d \epsilon^a + \omega^a{}_b \epsilon^b$.  With this notation, we see that the infinitesimal gauge transformations \eqref{GaugeTransf} in the BF theory with gauge parameter $\epsilon = \sqrt{\Lambda/2} \epsilon^a P_a + \epsilon^0 P_0 $ take the form
 \es{GaugeTransfe}{
  \delta e^1 &= D \epsilon^1 - \epsilon^0 e^2 \,, \\
  \delta e^2 &= D \epsilon^2 + \epsilon^0 e^1 \,, \\
  \delta \omega &= d \epsilon^0 + \frac{\Lambda}{2} (\epsilon^2 e^1 - \epsilon^1 e^2) \,.
 }
The interpretation of these formulas is as follows.  The parameters $\epsilon^i$ act as local gauge parameters for the $SO(2)$ symmetry.  When the gauge connection is flat with $F=0$, infinitesimal gauge transformation are related to infinitesimal diffeomorphisms generated by a vector fields $\xi^\mu$  (via $\delta g_{\mu\nu} = \nabla_\mu \xi_\nu + \nabla_\nu \xi_\mu$)  
 \es{epsxiRelation}{
  \epsilon^a = e^a_\mu \xi^\mu \,, \qquad 
\epsilon^0(x) = \omega_\mu(x) \xi^\mu(x)\,.
 }
The parameter $\epsilon^0$ generates an infinitesimal frame rotation, and thus it leaves the 2d metric invariant.  Note that the gauge transformations in the BF theory preserve the zero-torsion condition and the 2d curvature because these quantities appear in the expression for $F$ in \eqref{GotF} and the equation $F = 0$ is gauge-invariant.

So far, we have solely focused on the on-shell equations of motion in the bulk. We have not yet specified the crucial ingredients that are needed to provide an exact dual for the Schwarzian theory: specifying the boundary condition  along $\partial \Sigma$ in \eqref{eq:BF-action} or determining the global structure of the gauge group.  Thus, in the next subsection we discuss possible boundary conditions and boundary terms such that the resulting theory has a well defined variational principle, while later, in Section~\ref{sec:quantization-and-choice-of-gg}, we discuss the global structure of the gauge group.

\subsection{Variational principle, boundary conditions, and string defects}
\label{sec:var-principle}

Infinitesimal variations of the action \eqref{eq:BF-action} yield
\be 
\label{eq:variational-principle}
\delta S_{BF} = \text{(bulk equations of motions) } - i  \int_{\partial \Sigma}  \tr\left(\phi \delta A_\tau\right)\,,
\ee
where $\tau$ is used to parametrize the boundary $\partial \Sigma$.  As is well-known \cite{wald1984general} and can be easily seen from the variation \eqref{eq:variational-principle}, the BF theory  has a well-defined variational principle when fixing the gauge field $A_\tau$ along the boundary $\partial \Sigma$. In the first-order formulation of JT gravity, this amounts to fixing the spin connection and the frame and no other boundary term is necessary in order for the variational principle to be well defined.\footnote{This is in contrast with the second-order formulation of JT gravity \eqref{eq:JT-gravity-action}, when fixing the metric and the dilaton along the boundary. In such a case the boundary term in \eqref{eq:JT-gravity-action} needs to be added to the action  in order to have a well defined variational principle.}   In fact, due to gauge invariance, observables in the theory will depend on $A_\tau$ only through the holonomy around the boundary, 
\be
\label{eq:holonomy-definition}
\tilde g =  \cP \exp \left(\int_{\partial \Sigma} A \right)  \in \cG\,,
\ee
instead of depending on the local value of $A_\tau$. However, solely fixing the gauge field around the boundary yields a trivial topological theory (see more in Section~\ref{sec:quantization-and-choice-of-gg}). Of course, such a theory cannot be dual to the Schwarzian. In order to effectively  modify the dynamics of the theory we consider  a defect along a  loop $I$ on $\Sigma$.  A generic way of inserting such a defect is by adding a term $S_I$, to the BF action,  
\be
\label{eq:total-action}
S_E = S_{BF} + S_{I}\,, \qquad S_{I} = e \int_0^\beta d u\,  V(\phi(u))\,.
\ee
where $u$ is the proper length parametrization of the loop $I$, whose coordinates are given by $x_I(u)$ and whose total length is $\beta$ measured with the induced background metric from the disk.\footnote{Consequently, the defect is not topological.}  

 Since, the overall action needs to be gauge invariant we should restrict $V(\phi)$ to be of trace-class; as we will prove shortly in order to recover the Schwarzian on-shell we simply set 
$V(\phi) = -\tr\phi^2/4$, with the trace in the fundamental representation of $\mathfrak{sl}(2, \mR)$. 

Note that as a result of the Schwinger-Dyson equation
\ie
	\left\la d\tr \phi^2(x) \dots \right\ra_{BF}=-2i \left \la \tr \left (\phi(x) {\delta \over \delta A(x)} \right) \dots \right\ra_{BF}=0
	\label{topop}
\fe
$\tr\phi^2$ is a topological operator in the BF theory independent of its location on the spacetime manifold, as long as the other insertions represented by $\dots$ above  do not involve $A$.\footnote{In the $\mathfrak{sl}(2, \mR)$ gravitational theory, $ -\tr \phi^2$ is usually interpreted as a black hole mass and its conservation law can be interpreted as an energy conservation law \cite{Gonzalez:2018enk}. }  

\begin{figure}[t!]
    \centering
          \includegraphics[width=0.9\textwidth]{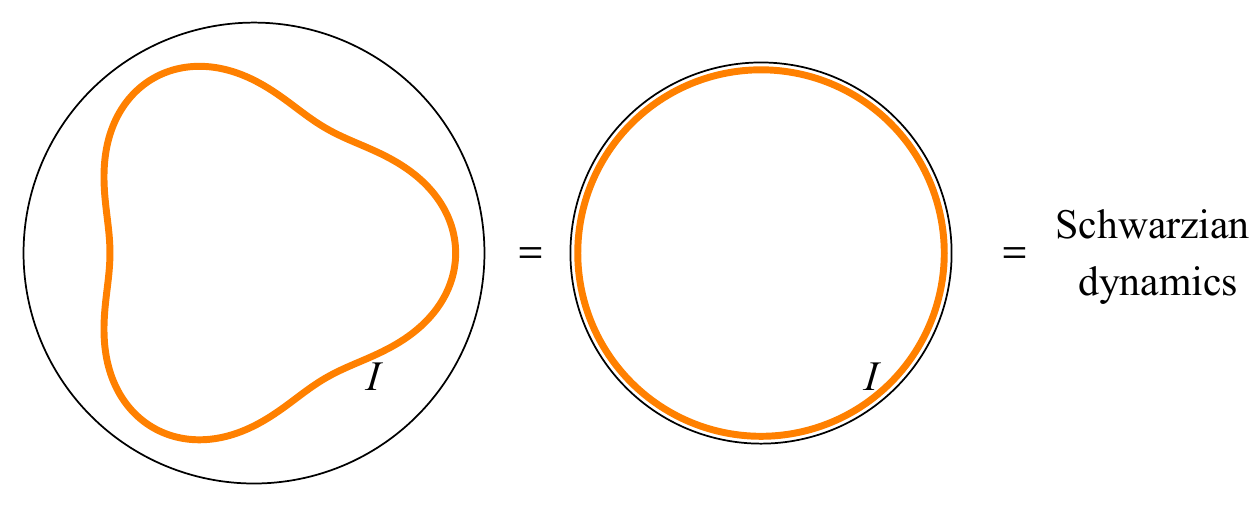} 

    \caption{Cartoon emphasizing the properties of the string defect. The resulting theory is invariant under perimeter preserving defect diffeomorphisms and thus the defect can be brought arbitrarily close to the boundary of the manifold. Furthermore, the degrees of freedom of the gauge theory defect can be captured by those in the Schwarzian theory.}
    \label{fig:string-defect}
\end{figure}

As emphasized in Figure~\ref{fig:string-defect}, due to the fact that theory is topological away from $I$ and due to the appearance of the length form in \eqref{eq:total-action} the action is invariant under diffeomorphisms that preserve the local length element on $I$.\footnote{This is similar to 2d Yang-Mills theory which is invariant under area preserving diffeomorphisms \cite{Blau:1991mp, Witten:1991we, Cordes:1994fc}. } Thus, one can modify the metric on $\Sigma$, away from $I$, in order to bring it arbitrarily close to the boundary $\partial \Sigma$.  This proves convenient for our discussion below since we fix the  component $A_\tau$ of the gauge field along the boundary and can thus easily use the equations of motion to solve for the value of $\phi$ along $I$. 

 Specifically, we choose 
\be 
A_\tau\bigg|_{\text{bdy}} \equiv \omega  \,\ell_0 + \sqrt{\frac{\L}2} e_+\, \ell_{+} + \sqrt{\frac{\L}2} e_-\, \ell_{-}\,,
\ee
where 
\be 
\label{eq:sl(2,R)-anti-hermitian-generators}
\ell_0 &\equiv i P_0\,,\qquad && \ell_+ \equiv -P_2 - i P_1\,, \qquad   &&\ell_- \equiv P_2- iP_1\,,\nn\\ \qquad \omega  &\equiv -i \omega_\tau \bigg|_\text{bdy.} \,, \qquad && e_+ \equiv \frac{ie^1_\tau - e^2_\tau}2\bigg|_\text{bdy.} \,, \qquad&& e_- \equiv \frac{ie^1_\tau + e^2_\tau}2\bigg|_\text{bdy.}\,.
\ee
The generators $\ell_0$ and $\ell_\pm$ satisfy the commutation relations
\be 
\label{eq:sl(2,R)-algebra}
[\ell_\pm, \ell_0] = \pm \ell_\pm\,, \text{\hspace{1.0cm}} [\ell_+, \ell_-] = 2\ell_0 \,.
\ee
As previously discussed, all observables can only depend on the value of the holonomy, thus without loss of generality we can set $\omega  $ and $e_\pm$ to be constants whose value we discuss in the next subsection. Fixing the value of the gauge field, in turn, sets the metric in the JT-gravity interpretation along the boundary to be $g_{\tau \tau}  = - 4 e_+ e_- $.

The equation of motion obtained by varying $A_\tau$ close to the boundary, $D_\tau \phi = \partial_\tau \phi - [A_\tau|_{\text{bdy}}, \phi] =  0$, can be used to solve for the value of $\phi$ along $I$.  It is convenient to relate the two parametrizations of the defect $I$ through the function $u(\tau)$, choosing $\tau$ in such a way that $e \phi_-(\tau)  \equiv \sqrt{\L} e_-/ \partial_\tau u(\tau) $, where $\phi  = \phi_0 \ell_0+ \phi_+ \ell_+ + \phi_- \ell_-$.  Instead of solving the equation of motion for $A_\tau$ in terms of $u(\tau)$ it is more convenient to perform a reparametrization and rewrite the equation in terms of $\tau(u)$ using $A_u = A_\tau \tau'(u)$, where $\tau'(u) \equiv \partial_u \tau(u)$.   The solution to the equation of motion for the  $\ell_-$ and $\ell_0$ components of $D_u \phi=0$ yields
\be
\label{eq:rewriting-in-terms-of-inverse}
 e \phi(u)&= \sqrt{2{\L}} e_- \ell_- \tau'  + 2\ell_0 \left(\omega  \tau'  - \frac{\tau'' }{\tau'} \right) + \sqrt{2{\L}}  \ell_+   \left(e_+ \tau' + \frac{\tau''' }{\L e_- (\tau')^2} - \frac{\omega  \tau'' }{\L e_- \tau'} - \frac{(\tau'')^2}{\L e_- (\tau')^3} \right)\,,\ee
where  $\tau(u)$ is further constrained from the component of the $D_u \phi=0$ along $\ell_+$,
 \be 
 0 &=  4 \det A_\tau (\tau')^4 \tau''  + 3 (\tau'')^3 - 4 \tau'  \tau''  \tau'''  + (\tau')^2 \tau'''' \,, \label{eq:extra-constraint}
\ee
 with  $\det  A_\tau = \left(-\omega^2 + 2\L e_- e_+ \right)/4 = (2\omega^2_\tau - \L g_{\tau \tau})/8|_{\text{bdy}}$.  When considering configurations with $\tau'(u) = 0$ (and $\tau''\neq 0$ or $\tau''' \neq 0$), $\phi(u)$ becomes divergent and consequently the action also diverges. Thus, we restrict to the space of configurations where $\tau(u)$ is monotonic, and we can set  $\tau(\beta) - \tau(0) = L$, where $L$ is an arbitrary length whose meaning we discuss shortly. Using this solution for $\phi(u)$ we can now proceed to show that the dynamics on the defect is described by the Schwarzian. 

\subsection{Recovering the Schwarzian action}
\label{sec:recovering-schw}

We can now proceed to show that the Schwarzian action is a consistent truncation of the theory \eqref{eq:total-action}. We start by integrating out $\phi$ inside the defect which sets $F=0$ and thus the nonvanishing part of the action \eqref{eq:total-action} comes purely from the region between (and including) the defect and the boundary. Next we \textit{partially} integrate out $A_\tau$ in this region using the equations of motion of $D_u\phi=0$ along the $\ell_-$ and $\ell_0$ directions, whose solution is given by \eqref{eq:rewriting-in-terms-of-inverse}.
Plugging \eqref{eq:rewriting-in-terms-of-inverse} back into the action \eqref{eq:total-action}, we find that the total action can be rewritten as\footnote{This reproduces the result in \cite{Grumiller:2017qao, Gonzalez:2018enk} where the Schwarzian action was obtained by adding a boundary term similar to that in \eqref{eq:total-action}, by imposing a relation between the boundary value of the gauge field $A_\tau$ and the zero-form field $\phi$ and by fixing the overall holonomy around the boundary. In our discussion, by using the insertion of the defect, we greatly simplify the quantization of the theory. Our method is similar in spirit to the derivation of the 2D Wess-Zumino-Witten action from 3D Chern-Simons action with the appropriate choice of gauge group \cite{moore1989taming}. 
} 
\be 
\label{eq:interm-larg-in-f}
S_E[\tau] = -\frac{1}{e} \int_0^\beta du \left( \{\tau(u), u \}+	2\tau'(u)^2  \det A_\tau 																	\right)\,,\qquad \tau(\beta)-\tau(0) = L\,,
\ee
where the determinant is computed in the fundamental representation of $\mathfrak{sl}(2, \R)$.   The equation of motion obtained by infinitesimal variations $\delta \tau(u)$ in \eqref{eq:interm-larg-in-f} yields~\cite{Maldacena:2016upp} 
  \be 
\partial_u\left[ \{\tau(u), u \}+	2\tau'(u)^2  \det A_\tau \right]= 0
\ee
which is equivalent to \eqref{eq:extra-constraint} that was obtained directly from varying all components of $A_\tau$ in the original action \eqref{eq:total-action}. This provides a check that the dynamics on the boundary condition changing defect in the gauge theory is consistent with that of the action \eqref{eq:interm-larg-in-f}. 

Finally, performing a change of variables, 
\be
\label{eq:def-F(u)}
F(u) = \tan \left(\sqrt{ \det A_\tau}\, \tau(u)\right)\,,
\ee
we recover the Schwarzian action as written in \eqref{eq:schw-action}, 
\be 
\label{eq:schw-in-terms-of-F}
S_E[F] = -\frac{1}{e} \int_0^\beta du \{F(u),\, u\}\,.  
\ee
While we have found that the dynamics on the defect precisely matches that of the Schwarzian we have not yet matched the boundary conditions for \eqref{eq:schw-in-terms-of-F} with those typically obtained from the second-order formulation of JT gravity: $\beta =L$ and $F(0) = F(\beta)$.\footnote{Instead the relation between $F(0)$ and $F(\beta)$ in  \eqref{eq:schw-in-terms-of-F}, with the boundary conditions set by those in  \eqref{eq:interm-larg-in-f}, is given by, 
\be 
\label{eq:F-twist}
F(\beta) = \frac{\cos(\sqrt{\det A_\tau} L ) F(0)+ \sin(\sqrt{\det A_\tau}L ) }{-\sin(\sqrt{\det A_\tau} L ) F(0)+ \cos(\sqrt{\det A_\tau} L) } \,.  
\ee } The relation between $L$ and $\beta$ is obtained by requiring that the field configuration is regular inside of the defect $I$: this can be achieved by requiring that the holonomy around a loop inside of $I$ be trivial. In order to discuss regularity we thus need to address the exact structure of the gauge group instead of only specifying the gauge algebra. To gain intuition about the correct choice of gauge group it will prove useful to first discuss the quantization of the gauge theory and that of the Schwarzian theory.

\section{Quantization and choice of gauge group}
\label{sec:quantization-and-choice-of-gg}

So far we have focused on the classical equivalence between the $\mathfrak{sl}(2, \R)$ gauge theory formulation of JT gravity and the Schwarzian theory.   This discussion relied only on the gauge algebra being $\mathfrak{sl}(2, \R)$, with the global structure of the gauge group not being important.  We will now extend this discussion to the quantum level, where, with a precise choice of gauge group in the 2d gauge theory, we will reproduce exactly the partition function and the expectation values of various operators in the Schwarzian theory.

 \subsection{Quantization with non-compact gauge group $\cG$}
 \label{sec:quantization}

We would like to consider the theory with action \eqref{eq:total-action} and (non-compact) gauge group $\cG$ (to be specified below), defined on a disk $D$ with the defect inserted along the loop $I$ of total length $\beta$.  The quantization of gauge theories with non-compact gauge groups has not been discussed much in the literature,\footnote{See however,  \cite{constantinidis2009quantization} and comments about non-unitarity in Yang-Mills with non-compact gauge group in \cite{Tseytlin:1995yw}.} although there is extensive literature on the quantum 2d Yang-Mills theory with compact gauge group \cite{Blau:1991mp, Witten:1991we, Cordes:1994fc, migdal1975phase, Migdal:1984gj, Fine:1991ux, Witten:1992xu, ganor1995string}.  Let us start with a brief review of relevant results on the compact gauge group case, and then explain how these results can be extended to the situation of interest to us.

What is commonly studied is the 2d Yang-Mills theory defined on a manifold ${\cal M}$ with a compact gauge group $G$, with Euclidean action 
 \es{2dYM}{
  S^\text{2d YM}\,[\phi, A] =- i \int_{\cal M}  \,  \tr (\phi F) -  g_\text{YM}^2  \int_{\cal M} d^2 x\, \sqrt{g}\, V(\phi)  \,, \qquad
   V(\phi) = \frac{1}{4} \tr \phi^2\,.
 }
After integrating out $\phi$, this action reduces to the standard form $ - \frac{1}{2 g_\text{YM}^2} \int_{\cal M} d^2 x\, \sqrt{g} \tr F_{\mu\nu} F^{\mu\nu}$.  When quantizing this theory on a spatial circle, it can be argued that due to the Gauss law constraint, the wave functions are simply functions $\Psi[g]$ of the holonomy $g = P \exp[ \oint A^a T_a]$ around the circle that depend only on the conjugacy class of $g$. Here $T^a$ are anti-Hermitian generators of the group  $G$.  The generator $T^a$ are normalized such that $\tr(T^i T^j) = N \eta^{ij}$, where for compact groups we set $\eta^{ij} = \text{diag}(-1, \,\dots,\,-1)$.  Thus, the wavefunctions $\Psi[g]$ are class functions on $G$, and a natural basis for them is the ``representation basis'' given by the characters $\chi_R(g) = \tr_R g$ of all unitary irreducible representations $R$ of $G$.

The partition function of the theory \eqref{2dYM} when placed on a Euclidean manifold $\cM$ with a single boundary is given by the path integral, 
\be 
\label{eq:partition-function-2D-YM-def}
Z^\text{2d YM}_\cM (g, g_\text{YM}^2 \cA) =  \int D\phi\, DA\,  e^{-S^\text{2d YM}\,[\phi, \,A]} 
\ee
where we impose that overall $G$ holonomy around the boundary of $\cM$  be given by $g$. Note that this partition function depends on the choice of metric for $\cM$ only through the total area $\cA$ (as the notation in \eqref{eq:partition-function-2D-YM-def} indicates, it depends only on the dimensionless combination $g_{\rm YM}^2\cA$).  The partition function can be computed using the cutting and gluing axioms of quantum field theory from two building blocks:  the partition function on a small disk and the partition function on a cylinder.  For the disk partition function $Z_{\text{disk}}^\text{2d YM}(g, g_{\rm YM}^2  \cA)$, which in general depends on the boundary holonomy $g$ and  $g_{\rm YM}^2 \cA$, the small $\cA$ limit is identical to the small $g_{\rm YM}^2$ limit in which \eqref{2dYM} becomes topological.  In this limit, the integral over $\phi$ imposes the condition that $A$ is a flat connection, which gives $g=\mathbf 1$, so \cite{Witten:1991we}
 \es{ZSmallLimit}{
  \lim_{\cA \to 0} Z_\text{disk}^\text{2d YM}(g, g_\text{YM}^2 \cA) = \delta(g)
   = \sum_R \dim R \, \chi_R(g) \,.
 }
Here, $\delta(g)$ is the delta-function on the group $G$ defined with respect to the Haar measure on $G$, which enforces that $\int dg\, \delta(g) x(g) = x(\mathbf 1)$.

To determine this partition functions at finite area, note that the action \eqref{2dYM} implies that the canonical momentum conjugate to the space component of the gauge field $A_1^i(x)$ is $\phi_i(x)$, and thus the Hamiltonian density that follows from \eqref{2dYM} is just $\frac{g_\text{YM}^2}{4} \tr (\phi_i T^i)^2$.  In canonical quantization, one find that $\pi_j = -i  N \phi_j$ and the Hamiltonian density becomes $H =- \frac{g_\text{YM}^2}{4N}\eta^{ij} \pi_i \pi_j$. Using $\pi_j = \frac{\delta}{\delta A_1^j}$, each momentum acts on the wavefunctions $\chi_R(g)$ as $\pi_i \chi_R(g) = \chi_R (T_i g)$.  It follows that the Hamiltonian density derived from the action \eqref{2dYM} acts on each basis element of the Hilbert space $\chi_R(g)$ diagonally with eigenvalue $g_\text{YM}^2 C_2(R)/(4N)$ \cite{Cordes:1994fc}, where $C_2(R)$ is the quadratic Casimir, with $C_2(R) \geq 0$ for compact groups.  One then immediately finds
 \es{ZCyl}{
  Z_\text{disk}^\text{2d YM}(g, g_\text{YM}^2 \cA) 
   &= \sum_R \dim R \, \chi_R(g) e^{- \frac{g_\text{YM}^2}{4N} \cA C_2(R)} \,.
 }

From these expressions, sticking with compact gauge groups for now, one can determine the disk partition function of a modified theory 
 \es{OurTheory}{
  S = - i \int_{\cM} \tr (\phi F) - e \int_I du\,  V(\phi) \,, \qquad 
   V(\phi) = \frac 14 \tr \phi^2 \,,
 }
where $I$ is a loop of length $\beta$ as in Figure~\ref{fig:string-defect}.  Such an action can be obtained by modifying the Hamiltonian of the theory to a time-dependent one and by choosing time-slices to be concentric to the loop $I$. \footnote{Alternatively, one can consider the gluing of a topological theory with $g_\text{YM}^2 = 0$ in the regions inside and outside $I$, and a theory of type \eqref{2dYM} in a fattened region around $I$ of a small width (so that the region does not intersect with other operator insertions such as Wilson lines).}   Applying such a quantization to the theory with a loop defect we obtain
 \es{ZdiskAgain}{
  Z(g, e \beta) = \sum_R \dim R\,  \chi_R(g) e^{-\frac{e \beta\,  C_2(R)}{4N}} \,.
 } 
One modification that one can perform in the above discussion is to consider, either in \eqref{2dYM} or in \eqref{OurTheory} a more general $V(\phi)$ than $\frac 14 \tr \phi^2$. For example, if $V(\phi) = \frac 14 \tr \phi^2 +\frac 14 \alpha (\tr \phi^2)^2$, then one should replace $C_2(R) $ by $C_2(R) + {\alpha\over N} C_2(R)^2$ in all the formulas above.

The discussion above assumed that $G$ is compact, and thus the spectrum of unitary irreps is discrete.  The only modification required in the case of a non-compact gauge group $\cG$ is that the irreducible irreps are in general part of a continuous spectrum.\footnote{For the case with non-compact gauge group we will continue to maintain the same sign convention in Euclidean signature as that shown in \eqref{eq:partition-function-2D-YM-def}. }  To generalize the proof above, we have to use the Plancherel formula associated with non-compact groups in \eqref{ZSmallLimit}
 \es{Replacement}{
   \delta(g)
   = \sum_R \dim R \, \chi_R(g) \qquad \to \qquad
     \delta(g) 
   = \int dR\,  \rho(R) \, \chi_R(g) \,,
 }
where $\rho(R)$ is the Plancherel measure.\footnote{In the case in which the spectrum of irreps has both continuous and discrete components, $\rho(R)$ will be a distribution with delta-function support on the discrete components.}  Then, following the same logic that led to the disk partition function in \eqref{ZCyl}, by determining the Hamiltonian density and applying it to the characters in \eqref{Replacement}, we find that the disk partition function of the theory \eqref{OurTheory} reduces to
 \es{ZdiskOur}{
  Z(g, e \beta) = \int dR\, \rho(R)  \chi_R(g) e^{-\frac{e \beta \,  C_2(R)}{4N}} \,.
 }
where we normalize the generators $P^i$ of the non-compact group by $\tr(P^i P^j) = N\eta^{ij}$, where $\eta^{ij}$ is diagonal with $\pm 1$ entries. In these conventions we set the Casimir of the group to be given by $\hat C_2  = - \eta^{ij} P_i P_j$.  One may worry that if the gauge group is non-compact, then it is possible for the quadratic Casimir $C_2(R)$ to be unbounded from below, and then the integral \eqref{ZdiskOur} would not converge.  If this is the case, we should think of $V(\phi)$ in \eqref{OurTheory} as a limit of a more complicated potential such that the integral \eqref{ZdiskOur} still converges.  For instance, we can add ${1\over 4}\alpha (\tr \phi^2)^2$ to \eqref{OurTheory} and consequently $\alpha C_2(R)^2$ to the exponent of \eqref{ZdiskOur} as described above.

\begin{figure}[t!]
    \centering
          \includegraphics[width=0.5\textwidth]{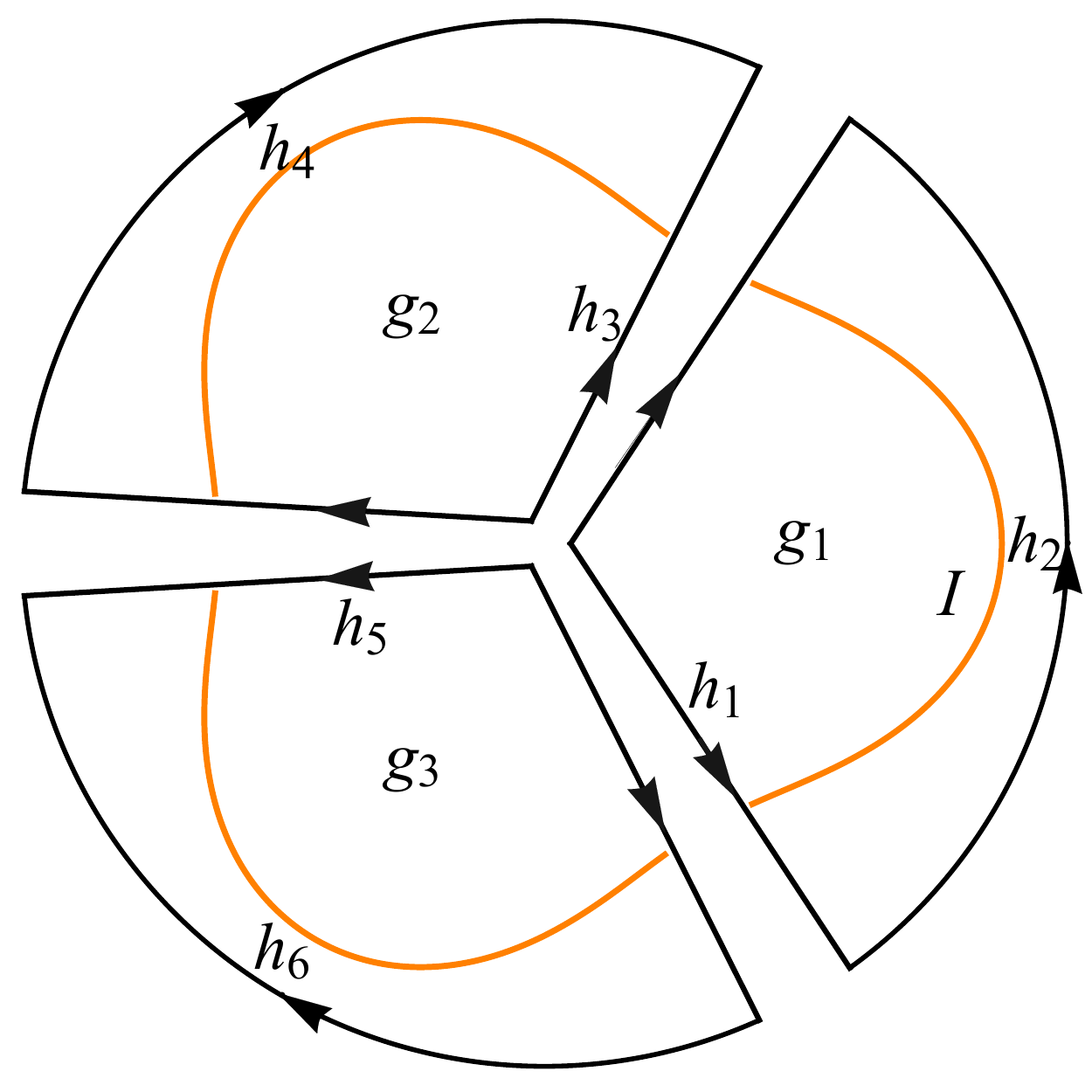} 

    \caption{Cartoon showing an example of gluing of three disk patches whose overall partition function is given by the gluing rules in \eqref{eq:gluing-rules}. Each segment has an associated group element $h_a$ and each patch has an associated holonomy $g_i$. In the case pictured above: $g_1 = h_1h_2 h_3^{-1}$, $g_2= h_3 h_4 h_5^{-1}$ and $g_3 = h_5 h_6 h_1^{-1}$. We take all edges to be oriented in the counter-clockwise direction.}
    \label{fig:example-gluing}
\end{figure}

In order to consider more complicated observables, we can glue together different segments of the boundary of the disk. In general, the gluing of $n$ disks, each containing a defect $I_j$ of length $\beta_j$, onto a different manifold $\Sigma$ with a single boundary with holonomy $g$, will formally be given by integrating over all  group elements  $h_1, \, h_2, \dots,  h_m$ associated to the $\cC_1$, \ldots , $\cC_m$ segments which need to be glued. Here, $h_i = \cP \exp \int_{\cC_i} A$. The resulting partition function is given by\footnote{Various formulae useful for gluing  in gauge theory are shown in Appendix \ref{app:comparison-compact-vs-non-compact}, where results for compact gauge groups and non-compact gauge groups are compared.}  
\be
\label{eq:gluing-rules}
Z(g,\, e\beta, \,\Sigma) = \frac{1}{\Vol(\cG)^m} \int \left(\prod_{i =1}^m dh_{i}\right) \left(\prod_{j=1}^n Z(g_j(h_a), e \beta_j)\right) \delta\left(g^{-1}\prod_{j=1}^n g_{j}(h_a)\right)\,,
\ee
where the product $i$ runs over all $m$ edges which need to be glued,  while the product $j$ runs over the labels of the $n$ disks. Each disk $j$   comes with a total  holonomy $g_j(h_a)$ depending on the group elements $h_a$ associated to each segment $\cC_a$ along the boundary of disk $j$. Thus, for instance if the edge of the disk $j$ consists of the segments $\cC_1$, \ldots , $\cC_{m_j}$ (in counter-clockwise order), then $g_j(h_a) = h_1 \cdots h_{m_j}$.  Furthermore, $ dh_{i}$ denotes the Haar measure on the group $\cG$, which is normalized by   the group volume.  The group $\delta$-function imposes that the total holonomy around the boundary of $\Sigma$ is fixed to be $g$.  An example of the gluing of three patches is given in Figure~\ref{fig:example-gluing}.

While for compact gauge groups \eqref{eq:gluing-rules} yields a convergent answer when considering manifolds $\Sigma$ with higher genus or no boundary, when studying non-compact gauge theories on such manifolds divergences can appear. This is due to the fact that the unitary representations of a non-compact group $\cG$ are infinitely dimensional.\footnote{ When setting $\cG$ to be $PSL(2, \mR)$ or one of its extensions, these divergences are in tension with the expected answers in the gravitational theory \eqref{eq:JT-gravity-action}.  This is a reflection of the fact that while the moduli space of
Riemann surfaces has finite volume, the moduli space of flat $PSL(2, \mR)$ (or other group extensions of $PSL(2, \mR)$) connections does not. See \cite{Saad:2019lba, Dijkgraaf:2018vnm} for a detailed discussion.  }

\subsection{The Schwarzian theory and $\SL2$ representations}
\label{sec:repr-theory}

In order to identify the gauge group $\cG$ for which the theory \eqref{eq:total-action} becomes equivalent to the Schwarzian theory at the quantum level, let us first understand what group representations are relevant in the quantization of the Schwarzian theory. Specifically, the partition function of the Schwarzian theory at temperature $\beta$ is given, up to a regularization dependent proportionality constant, by
\be 
\label{eq:Schw-partition-function}
Z_{\text{Schwarzian}}(\beta) \propto \int_0^\infty ds s \sinh(2\pi s) e^{-\frac{\beta}{2C} s^2}\,,
\ee
(computed using fermionic localization in \cite{Witten:2016iux}), can be written as an integral of the form
\be
\label{eq:test-partition-function}
Z_\text{Schwarzian} (\beta)  \propto \int dR\, \rho(R) e^{-\frac{\beta}{2C} \left[ C_2(R) -\frac{1}4 \right]}\,.
\ee
over certain irreps of the universal cover $\SL2$.\footnote{As already discussed in \cite{Mertens:2017mtv, Kitaev:2018wpr, yang2018quantum} and as we explain in Appendix~\ref{sec:schw}, we can interpret $H = \left(\hat C_2 -1/4\right)/C$ as the Hamiltonian of a quantum system and $\rho(R)$ as the density of states.  Such an interpretation can be made precise after noticing that the Schwarzian theory is equivalent to the theory of a non-relativistic particle in 2D hyperbolic space placed in a pure imaginary magnetic field.}

To identify the representations $R$ needed to equate \eqref{eq:Schw-partition-function} to \eqref{eq:test-partition-function}, let us first review some basic aspects of $\SL2$ representation theory, following \cite{Kitaev:2017hnr}.  The irreducible representations of $\widetilde{SL}(2, \mR)$ are labeled by two quantum numbers $\lambda$ and $\mu$. These can be determined from the eigenvalue $\lambda( 1 - \lambda)$ of the quadratic Casimir $\hat C_2 =  -\eta^{ij}P_i P_j =P_0^2- P_1^2-P_2^2 = -\ell_0^2 + (\ell_- \ell_+ + \ell_+ \ell_-)/2$, as well as the eigenvalue $e^{2\pi i \mu}$ under the generator $e^{-2 \pi i \ell_0}$ of the $\mZ$ center of the $\SL2$.  Furthermore, states within each irreducible representation are labeled by an additional quantum number $m$ which represents the eigenvalue under $\ell_0$.  Thus, 
 \es{eq:irreps-labeling-SL(2,R)}{
   \hat C_2 | \lambda, \mu, m \rangle &= \lambda(1-\lambda)| \lambda, \mu, m \rangle\,, \\ 
 \ell_0 |\lambda, \mu, m\rangle  &= - m |\lambda, \mu, m\rangle\, \qquad \text{ with } \qquad m \in \mu + \mZ\,. \\
 }
One can go between states with different values of $m$ using the raising and lowering operators: 
\be
\label{eq:irreps-states}
\ell_-  |\l,\mu, m\rangle  &=  -\sqrt{(m-\l)(m-1+\l)}  |\l,\mu, m-1\rangle \,,\\
\ell_+  |\l,\mu, m\rangle  &=  -\sqrt{(m+\l)(m+1-\l)}  |\l,\mu, m+1\rangle \,.\nn
\ee 
where the generators satisfy the $\mathfrak{sl}(2, \mR)$ algebra \eqref{eq:sl(2,R)-algebra}. 
 Using these labels and requiring the positivity of the matrix elements of the operators $L_+ L_-$ and $L_- L_+$   one finds that there are four types of irreducible unitary representations:\footnote{The two-dimensional representation (corresponding to $\l=-1/2$ and $\mu = \pm 1/2$) used in Section~\ref{sec:SL(2,R)-Yang-Mills} in order to write down the Lagrangian is not a unitary representation and therefore does not appear in the list below. }
\begin{itemize}
\item \textit{Trivial representation $I$:} $\mu = 0$ and $m = 0$;
\item \textit{Principal unitary series $\mathcal C^\mu_{\l = \frac{1}2+is}$}: $\l = \frac{1}2+i s$, $m = \mu + n$, $n \in \mathbb Z$, $-1/2 \leq \mu \leq 1/2$;
\item \textit{Positive/negative discrete series $\mathcal D_\l^{\pm}$}: $\l>0$, $\lambda = \pm \mu$, $m = \pm \l \pm n$, $n \in \mathbb Z^+$, $\mu \in \mathbb R$;
\item \textit{Complementary series $\mathcal C^\mu_\l$:} $|\mu|< \l < 1/2$,  $m = \mu + n$, $n \in \mathbb Z$,\footnote{Since in the Plancherel inversion formula the complementary series does not appear, we will not include it in any further discussion.}
\end{itemize}
Only the principal unitary series and the positive/negative discrete series admit a well defined Hermitian inner-product, so for them one can define a density of states given by the Plancherel measure (up to a proportionality constant given by the regularization of the group's volume).

As reviewed in Appendix~\ref{app:harmonic-analysis-SL(2,R)}, the principal unitary series has the Plancherel measure given by
\be
\label{eq:SL2-universal-planch-mesure}
\rho(\mu, s) \,d\mu\, ds= \frac{(2\pi)^{-2} s \sinh(2\pi s)}{\cosh(2\pi s) + \cos(2\pi \mu)}\,ds \,d\mu\,, \qquad \text{ with } -\frac{1}2 \leq \mu \leq \frac{1}2\,,
\ee 
and for the positive and negative discrete series  
\be
\label{eq:Plancherel-measure-discrete-series}
\rho(\lambda) d\lambda = (2\pi)^{-2} \left(  \lambda - \frac{1}2 \right) d \lambda\,,\qquad\qquad \text{ with } \lambda = \pm \mu,\,\, \qquad \lambda \geq \frac{1}2\,,
\ee
where $\lambda = \mu$ for the positive discrete series and $\lambda = -\mu$ for the negative discrete series.

Matching \eqref{eq:test-partition-function} to \eqref{eq:Schw-partition-function} can be done in two steps:
 \begin{enumerate}
  \item We first restrict the set of $R$ that appear in \eqref{eq:test-partition-function} to representations with fixed $e^{2 \pi i \mu}$.  As mentioned above, this quantity represents the eigenvalue under the generator of the $\Z$ center of $\SL2$.  After this step, \eqref{eq:test-partition-function} becomes
   \es{PartFnIntermediate}{
    \int_{0}^\infty d s\, \frac{(2\pi)^{-2} s \sinh(2\pi s)}{\cosh(2\pi s) + \cos(2\pi \mu)} e^{-\frac{\beta}{2C} s^2}
     +  \sum_{n  = 1}^{n_\text{max}} \frac{1}{2 \pi^2}  \left(\mu + n - \frac{1}2\right) e^{- \frac{\beta}{2C}  \left[ (\mu + n) (1 - \mu - n) - \frac 14 \right]}  \,,
   }
provided that we took $\mu \in [-\frac 12, \frac 12 )$.   In writing \eqref{PartFnIntermediate} we imposed a cutoff $n_\text{max}$ on the discrete series representations.  A different regularization could be achieved by adding the square of the quadratic Casimir in the exponent, with a small coefficient.  As a function of $\mu$, Eq.~\eqref{PartFnIntermediate} can be extended to a periodic function of $\mu$ with unit period.

  \item We analytically continue the answer we obtained in the previous step to $\mu \to i \infty$.  When doing so, the sum in \eqref{PartFnIntermediate}  coming from the discrete series goes as $e^{- \frac{\beta}{C} (\Im \mu)^2}$, and the integral coming from the continuous series goes as $e^{- 2 \pi \abs{\Im \mu}}$.  Thus, when $\Im \mu \to \infty$ the continuous series dominates, and \eqref{PartFnIntermediate} becomes proportional to the partition function of the Schwarzian.

 \end{enumerate}

As was already discussed in \cite{kitaevTalks, Kitaev:2018wpr, yang2018quantum, Mertens:2017mtv} and we review in Appendix~\ref{sec:schw}, fixing $\mu \rightarrow i \infty$ can also be understood in deriving the equivalence between the Schwarzian and a non-relativistic particle in 2D hyperbolic space, as fixing the magnetic field $\tilde B$ to be pure imaginary, $\tilde B=-{i B\over 2\pi} = { \mu }$,  with $B \rightarrow \infty$. As we shall see below, on the gauge theory side, fixing the parameter $\mu \rightarrow i \infty$ can be done with an appropriate choice of the gauge group $\cG$ and boundary conditions. 

\subsection{$PSL(2, \mR)$ extensions,  one-form symmetries, and revisiting the boundary condition}
\label{sec:search-for-the-gg}

In Section~\ref{sec:repr-theory} we have gained some insight about the $\SL2$ representations that are needed in order to write the Schwarzian partition function as in \eqref{eq:test-partition-function}. We thus seek to choose a gauge group and boundary conditions that automatically isolate precisely the same representations as in Step 1 above.  We then choose the defect potential for the 2D gauge theory to achieve the desired analytically continued gauge theory partition function presented in Step 2.

\subsubsection*{Choice of gauge group}

In a pure gauge theory the center of the gauge group gives rise to a one-form symmetry under which Wilson loops are charged \cite{Gaiotto:2014kfa}. Thus, since an $\SL2$ gauge group gives rise to a $\mathbb Z$ one-form symmetry, fixing the charge under the center of the gauge group is equivalent to projecting down to states of a given one-form symmetry charge. A well known way to restrict the one-form symmetry charges in the case of a compact gauge group $G$ is by introducing an extra generator in the gauge algebra and embedding the group $G$ into its central extension \cite{Gaiotto:2014kfa, Gaiotto:2017yup}. 

In the case of non-compact groups we proceed in a similar fashion, and consider a new gauge group which is given by the central extension of $PSL(2, \mR)$ by $\mR$,\footnote{Such extensions are classified by the \v{C}ech cohomology group $\check{\mathrm{H}}^1 (SL(2, \mathbb R), \mathbb R) \simeq \text{Hom}(\pi_1(SL(2, \mathbb R)\rightarrow \mathbb R)\simeq \text{Hom}(\mathbb Z \rightarrow \mathbb R) \simeq \mathbb R$ where $\text{Hom}(\mathbb Z \rightarrow \mathbb R)$ classifies the set of homomorphisms from $\mathbb Z$ to $\mathbb R$. In other words, all extensions by $\mathbb R$ will be given by a push-forward from the elements of $\mathbb Z$ center of $\SL2$ to elements of $\mathbb R$. A basis of homomorphisms from $\mathbb Z$ to $(\mathbb R, +)$ are given by $f_{B}(n) = Bn$ for $ B \in \mathbb R$. Such a homomorphism imposes the identification \eqref{eq:description-of-the-group} for different elements of the group \cite{tuynman1987central}. } 
\be
\label{eq:description-of-the-group}
\mathcal G_B  \equiv \frac{\SL2 \times \mR}{\mZ}\,,
\ee
where the quotient, and, consequently, the definition of the group extension, is given by the identification
 \es{Identification}{
   (\tilde g, \theta) \sim (h_n \tilde g, \theta+B n)\,.
 }
Above, $\tilde g \in \SL2$ and $\theta \in \mR$, $h_n$ is the $n$-th element of $\mZ$ and $B \in \mR$ is the parameter which defines the extension.  The resulting irreducible representations of $\cG_B$ can be obtained from irreducible representation of $\SL2 \times \mR$ which are  restricted by the quotient \eqref{eq:description-of-the-group}.  The unitary representations of $\mR$ are one-dimensional and are labeled by their eigenvalue under the $\mR$ generator, $I|k\rangle  = k |k \rangle$.  In other words, the action of a general $\R$ group element $U^\mathbb{R}(\theta) = e^{i I \theta}$ on the state $|k \rangle$ is given by multiplication by $U_k^\mathbb{R}(\theta) = e^{i k \theta}$.

  Considering the representation $U_k$ of $\mathbb R$ and a representation $U_{\lambda, \mu}$ of  the $\SL2$, evaluated on the group element $(h_n, \theta)$ we have $ U^{\SL2 \times \R}_{\lambda, \mu, k} (h_n, \theta) = U_{\lambda, \mu} (h_n) U_k^\mathbb{R}(\theta)  = e^{2\pi i \mu n + i k \theta}$. We now impose the quotient identification \eqref{Identification} on the representations, $e^{i k \theta} = e^{ 2\pi i \mu n + i k (\theta + {B}n)}$, which implies $k = -2\pi \left(\mu - p\right)/ B $, with $p \in \Z$. Thus, $\mR$ irreps labeled by $k$ restrict the label $\mu$ of representations in \eqref{eq:irreps-labeling-SL(2,R)} to be\footnote{For $ B = 0$ one simply finds the trivial extension of $PSL(2, \mathbb R)$  by $\mathbb  R$ which does not contain $\SL2$ as a subgroup. }  
\be 
\label{eq:representation-restriction}
\mu = -\frac{Bk}{2\pi}+p\,,\qquad \text{ with } p \in \mZ\,.
\ee 
Thus, by projecting down to a representation $k$ of $\mR$ in the 2D gauge theory partition function, we can restrict to representations with a fixed eigenvalue $e^{2\pi i \mu}$ for the center of the gauge group $\mZ$.   

In order to understand how to perform the projection to a fixed $k$ (or $e^{2 \pi i \mu}$) in the BF theory, it is useful to explicitly write down the $\cG_B$ gauge theory action.

To start, we write the gauge algebra $\mathfrak{sl}(2, \mR) \oplus \mR$, 
\be 
\label{eq:sl(2,R)-algebra-central-ext}
[ \tl_\pm, \tl_0] = \pm \tl_\pm\,, \text{\hspace{1.0cm}} [\tl_+, \tl_-] = 2\tl_0 -\frac{  B I}{\pi} \,, \text{\hspace{1.0cm}} e^{2\pi i \tl_0}  = 1\,,\, \text{\hspace{1.0cm}} [\tl_{0, \pm}\,, I] = 0\,,
\ee
where the condition $e^{2\pi i \tl_0}  = 1$, imposed on the group, enforces the representation restriction \eqref{eq:representation-restriction}. Of course at the level of the algebra, we can perform the redefinition $\ell_0 = \tl_0 -B I/(2\pi)$ and $\ell_\pm = \tl_\pm$ to still find that $\ell_{0,\,\pm}$ satisfy an $\mathfrak{sl}(2, \mathbb R)$ algebra \eqref{eq:sl(2,R)-algebra} from which we can once again define the set of generators $P_i$ using \eqref{eq:sl(2,R)-anti-hermitian-generators}.   Considering  a theory with gauge group $\cG_B$ in \eqref{eq:description-of-the-group}, we can write the gauge field and zero-form field $\phi$ as\footnote{Note that the normalization for the $\mR$ component of $A$ is such that the BF-action in \eqref{eq:final-action-gauge-theory} is in a standard form.}
\be
\label{eq:gauge-field-SL(2,R)-ext}
 A =  
e^a P_a + \omega P_0 + \frac{B^2}{\pi^2} A^\mR I \,, \qquad \qquad \phi = \phi^a P_a + \phi^0 P_0 + \phi^\mR I \,,
\ee
where $a = 1,\, 2$ and where $\a$ is the $\mR$ gauge field. Thus, the gauge invariant action \eqref{eq:total-action} can be written as
\be 
\label{eq:final-action-gauge-theory}
S_E = -{i}\int_\Sigma  \left(\frac{\phi^a F_a + \phi^0  F_0}2 +  \phi^\mR  F^\mR \right) -  e \int_{\partial\Sigma} du \,  \,\, V(\phi^0,\, \phi^\pm)\,.
\ee
Since the $\mathfrak{sl}(2, \mR)$ generators form a closed algebra, it is clear that under a general gauge transformation the $e^a$ and $\omega$ transform under the actions of $\mathfrak{sl}(2, \mR)$, while $\a$ transforms independently under the action of $\mR$. Thus one can fix the holonomy of the $\mathfrak{sl}(2,\mR)$ gauge components independently from that of $\mR$.\footnote{We now briefly revisit the equivalence between the gauge theory and JT-gravity, as discussed in Section \eqref{sec:rewriting-JT-gravity}. One important motivation for this is that Section~\ref{sec:rewriting-JT-gravity} solely focused on an $\mathfrak{sl}(2, \mR)$ gauge algebra while $\cG_B$ has an $\mathfrak{sl}(2, \mR) \oplus \mR$ algebra.   The equations of motion for the $\mathfrak{sl}(2, \mR)$ components are independent from those for the $\mR$ components, namely $F^\mR = 0$ and $\phi^\mR =\, \text{constant}$. Thus, the  $\mathfrak{sl}(2, \mR)$ and  $\mR$ sectors are fully decoupled and, since $F_\mR = 0$, the $\mR$ sector does not contribute to the bulk term in the action. Finally, note that $\cG_B$ indeed has a two-dimensional representation with $(\l,  \mu, k) = (-1/2, \pm 1/2, \mp \pi/B)$, as discussed in Section~\ref{sec:recovering-schw} when recovering the Schwarzian action. Since we will be considering the limit $B \rightarrow \infty$ throughout this paper, the contribution from  the $\mR$ component to $\tr \phi^2$ in this two dimensional representation is suppressed.    Thus, the classical analysis in Section~\ref{sec:SL(2,R)-Yang-Mills} is unaffected by the extension of the group.}

\subsubsection*{Revisiting the boundary condition}

Since the two sectors are decoupled, we can independently fix the holonomy $\tilde g$ of the $\mathfrak{sl}(2, \mR)$ components of the gauge field,  as specified in Section~\ref{sec:SL(2,R)-Yang-Mills}, and fix the value of $\phi^\mR = k_0$ on the boundary. In order to implement such boundary conditions and in order for the overall action to have a well-defined variational principle,  one can add a boundary term 
\be 
\label{eq:bdy-term-R-component}
S_\text{bdy.} = i \oint_{\partial \Sigma} \phi^\mR A^\mR\,.
\ee
to the action \eqref{eq:final-action-gauge-theory}. The partition function when fixing this boundary condition can be related to that in which the $\cG_B$ holonomy $g = (\tilde g, \theta)$, is fixed, with  $\tilde g= \oint_{\partial \Sigma} A^i P_i \in \SL2$ and $\theta = \oint_{\partial \Sigma} A^\mR \in \mR$, as 
\be
\label{eq:fourier-partition-func}
Z_{k_0} (\tilde g, \,e\beta) = \int d\theta Z ((\tilde g, \theta),\, e\beta) e^{-i k_0 \theta} \,.
\ee
More generally, without relying on \eqref{eq:fourier-partition-func}, following the decomposition of the partition function into a sum of irreducible representation of $\cG_B$, fixing $\phi^\mR = k_0$, isolates the contribution of the $\mR$ representation labeled by $k_0$, in the partition function, or equivalently fixes $e^{2 \pi i \mu}$ with $\mu = - \frac{Bk_0}{2 \pi} + \text{integer}$.  This achieves the goal of Step 1 in the previous subsection \ref{sec:repr-theory}.

To achieve Step 2, namely sending $\mu \rightarrow i \infty$, or equivalently $k B \rightarrow i \infty$, we can choose
\be 
\label{eq:limit-for-kB}
\cG \equiv \cG_B\qquad \text{with}\qquad B \rightarrow \infty\,, \qquad \phi^\mR = k_0 = -i\,.
\ee  
Note that all the groups $\cG_B$ with $B \neq 0$ are isomorphic. Therefore, one can make different choices when considering the limits in \eqref{eq:limit-for-kB} as long as the invariant quantity $k B \rightarrow i \infty$.

Alternatively, instead of fixing the value of $\phi^\mR $ on the boundary, the change in boundary condition \eqref{eq:fourier-partition-func} can be viewed as the introduction of a 1D complexified Chern-Simons term for the $\mR$ gauge field component $\alpha$, $S_{CS} = i k_0 \oint_{\partial \Sigma} A^\mR$, which is equivalent to the boundary term in \eqref{eq:bdy-term-R-component}. By adding such a term to the action and by integrating over the $\mR$ holonomy we once again recover the partition function given by \eqref{eq:fourier-partition-func} .

Thus, the choice of gauge group $\cG$ (with $B \to \infty$) together with the boundary condition for the field $\phi^\mR$ or through the addition of the boundary Chern-Simons discussed above,  will isolate the contribution of representations with $k= k_0$ in the partition function.\footnote{Note that in such a case the representations of $\mR$ with $k \in \mathbb C \setminus \mR$ are not $\delta$-function normalizable.}  Finally, note that in order to perform the gluing procedure described in Section~\ref{sec:quantization}, one first computes all observables in the presence of an overall $\cG$ holonomy. By using \eqref{eq:fourier-partition-func} one can then fix $ \phi^\mR = k_0 $ along the boundary and obtain the result with $k_0 = -i$ by analytic continuation.\footnote{This analytic continuation is analogous to the one needed in Chern-Simons gravity when describing Euclidean quantum gravity \cite{Witten:2010cx}.}  

\subsubsection*{Higher order corrections to the potential $V(\phi)$}

 Finally, as shown in Section~\ref{sec:SL(2,R)-Yang-Mills} in order to reproduce the Schwarzian on-shell the potential $V(\phi^0, \phi^\pm, \phi^\mR) $ needed to be quadratic to leading order. However, as we shall explain below, one option is to introduce higher order terms, suppressed in $\cO(1/B)$, in order to regularize the contribution of discrete series representations whose energies (given by the quadratic Casimir) are arbitrarily negative. Thus, we choose
\be 
 V(\phi^0, \phi^\pm, \phi^\mR) &=\frac{1}2 + \frac{1}4 \,\tr_{({\bf 2}, -\frac{\pi}{B})} \,\phi^2  \nn \\ &+\,\text{higher order terms in $\phi$ suppressed in }1/B\,,
 \ee
 where $\tr_{({\bf 2}, -\frac{ \pi}{B})}$ is the trace taken in the two-dimensional representation with $k=-\frac{ \pi}{B}$, and the shift in the potential is needed in order to reproduce the shift for the Casimir seen in \eqref{eq:test-partition-function}.  Note that in the limit $B \to \infty$,  the trace only involves the $\mathfrak{sl}(2, \mR)$ components of $\phi$.  While observables are unaffected by the exact form of these higher order terms, their presence regularizes the contribution of such representations to the partition function.\footnote{An example for such a higher-order term is given by $e^{(2)} \left(\left(\phi^0\right)^2+ 2\phi^+ \phi^-+\frac{1}4\right)/B$ where $e^{(2)} \sim O(1) $ is a new coupling constant in the potential .}

\subsection{The partition function in the first-order formulation }
\label{sec:partition-function}
Since we have proven that the degrees of freedom in the second-order formulation of JT-gravity can be mapped to those in the first-order gauge theory formulation, we expect that with the appropriate choice of measure and boundary conditions, the two path integrals agree:
\be 
 \int D\phi\, DA\,  e^{-S_E[\phi, \,A]} \cong \int Dg_{\mu \nu} \, D \Phi\,   e^{-S_{JT}[\Phi, \, g] } \,.
 \label{gaugeJTequiv}
\ee

Using all the ingredients in Section~\ref{sec:search-for-the-gg}, we can now show  that the partition function of the gravitational  theory \eqref{gaugeJTequiv} matches that of the Schwarzian. We first compute the partition function in the presence of a fixed $\cG$ holonomy is given by
\be 
\label{eq:parition-function-fixed-GB-holonomy}
Z(g,\, e\beta) &\propto \int_{-\infty}^\infty dk \int_0^\infty ds 
\frac{ s \sinh(2\pi s)}{\cosh(2\pi s) + \cos(B k)} \chi_{(s, \mu = -\frac{Bk}{2\pi}, k)}(g) e^{-\frac{e \beta s^2}2} \nn \\ &+ \text{ discrete series contribution}\,,
\ee
where, we remind the reader that the generators $P_i$ satisfying the $\mathfrak{sl}(2, \mR)$ algebra are normalized by $\tr_{\pmb 2}(P^i P^j)=-\eta^{ij}/2$ with $\eta^{ij} = \text{diag}(-1,\, 1,\, 1)$. When using the symbol ``$\propto$'' in the computation of various observables in the gauge theory we mean that the result is given up to a regularization dependent, but $\beta$-{\it independent}, proportionality constant.

Using this result, we can now understand the partition function in the presence of the mixed boundary conditions discussed in the previous subsection. To leading order in $B$ the partition function with a fixed holonomy $\tilde g$ and a fixed value of $\phi^\mR=k_0 = -i$ is dominated by terms coming from the  principal  series representations, 
\be \label{eq:parition-func-w-holonomy}
Z_{k_0}(\tilde g, \,e\beta) &\propto e^{-B}\int_0^\infty ds \,s\, \sinh(2\pi s) \chi_{s,\mu = -\frac{Bk_0}{2\pi}}(\tilde g) e^{-\frac{e \beta  s^2}2} + O(e^{-2B}) \,,
\ee
 where $\chi_{s, \mu}(\tilde g)$ is the character of the $\SL2$ principal series representation labeled by $(\lambda=1/2+is,\mu)$ evaluated on the group element $\tilde g$, which can be parametrized by exponentiating the generators in \eqref{eq:sl(2,R)-algebra} as $\tilde g = e^{\phi P_0} e^{\xi P_1} e^{-\eta P_0}$. For $\phi - \eta\in [2\pi(n-1),\, 2\pi n)$, the character for the continuous series representation $s$ is given by
\be 
\label{eq:cont-series-character-main-text}
\chi_{s, \mu}(\tilde g) =  \begin{cases}
  e^{2\pi i \mu n}\left(\frac{|x|^{1 - 2 \lambda} + |x|^{-1 + 2 \lambda}}{|x - x^{-1}|}  \right)\,,\qquad \text{ for } \tilde g \text{ hyperbolic,}\\
  0 \,, \hspace{4.72cm}   \text{ for } \tilde g \text{ elliptic.}
  \end{cases}
\ee 
Here,  $x$ (and $x^{-1}$) are the eigenvalues of the group element $\tilde g$, when expressed in the two-dimensional representation (see Appendix~\ref{app:harmonic-analysis-SL(2,R)}).   Note that for hyperbolic elements, $x  \in \R$, with $|x|>1$, and the character is non-vanishing, while for elliptic elements, we have $\abs{x} = 1$ (with $x \notin \R$) and the character is always vanishing.\footnote{In Appendix~\ref{sec:schw} we confirm the expectation that \eqref{eq:parition-func-w-holonomy} reproduces the partition function in the Schwarzian theory when twisting the boundary condition for the field $F(u)$ by an $\SL2$ transformation $\tilde g$.  We expect such configurations with non-trivial holonomy to correspond to singular gravitational configurations. }

Note that since in the partition function only representations with a fixed value of $\mu$ contribute, when the holonomy is set to different center elements $h_n$ of $\cG$, the partition function will only differ by an overall constant $e^{2\pi i \mu n}$ as obtained from \eqref{eq:cont-series-character-main-text}. 
For simplicity we will consider $\tilde g = \mathbf 1$. The character in such a case can be found by setting $n=0$ and taking the limit $x \rightarrow 1^+$ from the hyperbolic side in \eqref{eq:cont-series-character-main-text}. In this limit, the character is divergent, yet the divergence is independent of the representation, $s$.  Thus,  as suggested in Section~\ref{sec:search-for-the-gg}, we find that after setting  $k_0=-i$ via analytic continuation in the limit $B \to \infty$,
\be
\label{eq:trivial-holonomy-partition-function}
Z_{k_0}\propto \,\Xi \ \int_0^\infty ds\, \rho(s) e^{-\frac{e \beta s^2}2 }\,, \qquad \rho(s) \equiv \,s\, \sinh(2\pi s)\,,
\ee
where $\Xi = \lim_{x \to 1^+, n=0} \chi_{s, \mu}(g)$ is the divergent factor mentioned above, which comes from summing over all states in each continuous series irrep $\lambda = 1/2 + is$. Note that we have absorbed the factor of $e^{-B}$ in \eqref{eq:parition-func-w-holonomy} by redefining our regularization scheme, thus changing the partition function by an overall proportionality constant.  In the remainder of this paper we will use this regularization scheme in order to compute all observables. 

Performing the integral in \eqref{eq:trivial-holonomy-partition-function} we find
\be 
Z_{k_0}  = \Xi \left(\frac{2\pi}{e \beta}\right)^{\frac{3}2} e^{\frac{2\pi^2}{e \beta}} \,.
\ee
Thus, up to an overall regularization dependent factor, we have constructed a bulk gauge theory whose energies and density of states (\ref{eq:trivial-holonomy-partition-function}) match that of the Schwarzian theory (\ref{eq:Schw-partition-function}) for
$
\frac{1}{C} = {e}$, reproducing the relationship suggested in the classical analysis. 

\section{Wilson lines, bi-local operators and probe particles}
\label{sec:observables-and-duality}
\label{sec:wilson-loop-and-bi-local-op}

An important class of observables in any gauge theory are Wilson lines and Wilson loops,
\be
\label{eq:Wilson-loop-definiton} 
\hat \cW_R(\mathcal C)= \chi_R \left(\mP \exp  \int_\mC A\right)\,,
\ee
where $R$ is an irreducible representation of the gauge group, $\mC$ denotes the underlying path or loop, and $ \chi_R(g)$ is the character of $\cG$. When placing the theory on a topologically trivial manifold all Wilson loops that do not intersect the defect are contractible and therefore have trivial expectation values. A more interesting class of non-trivial non-local operators in the gauge theory are the Wilson lines that intersect the defect loop and are anchored  on the boundary. 

To determine the duals of such operators, we start by focusing on Wilson lines in the positive or negative discrete series irreducible representation of $\cG$, with $R = (\l^\pm, \mp\frac{2\pi \l}{B})$ where the $\pm$ superscripts distinguish between the positive and negative discrete series. In the $B \rightarrow \infty$ limit, this representation becomes $R = (\l^\pm, 0)$. As we will discuss in detail below, in order to regularize the expectation value of these boundary-anchored Wilson lines, we will replace the character $\chi_R(g)$ in \eqref{eq:Wilson-loop-definiton}  by a truncated sum $\bar \chi_R(g)$ over the diagonal elements of the matrix associated to the infinite-dimensional representation $R$.
 
\begin{figure}[t!]
    \centering
          \includegraphics[width=0.4\textwidth]{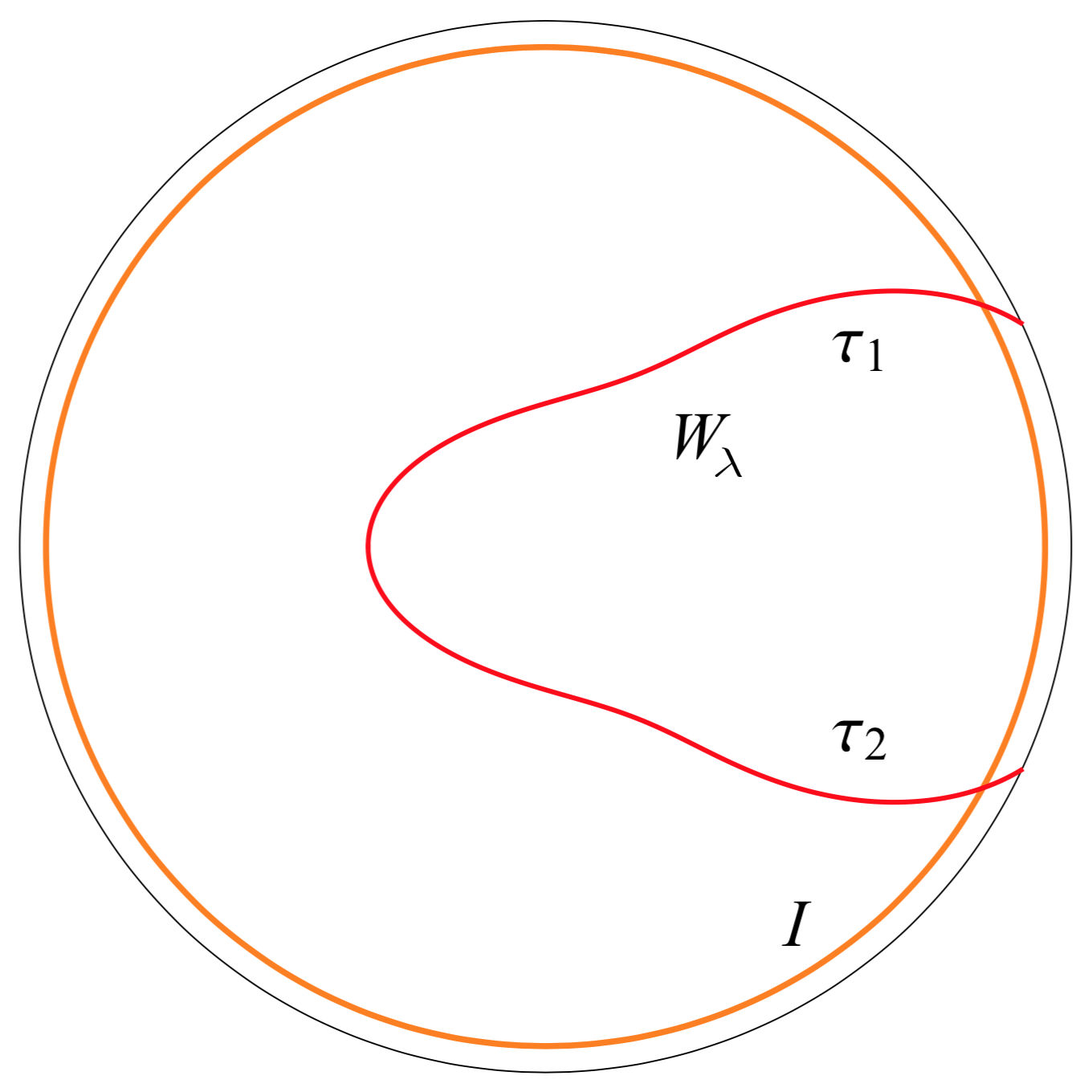}
        \includegraphics[width=0.4\textwidth]{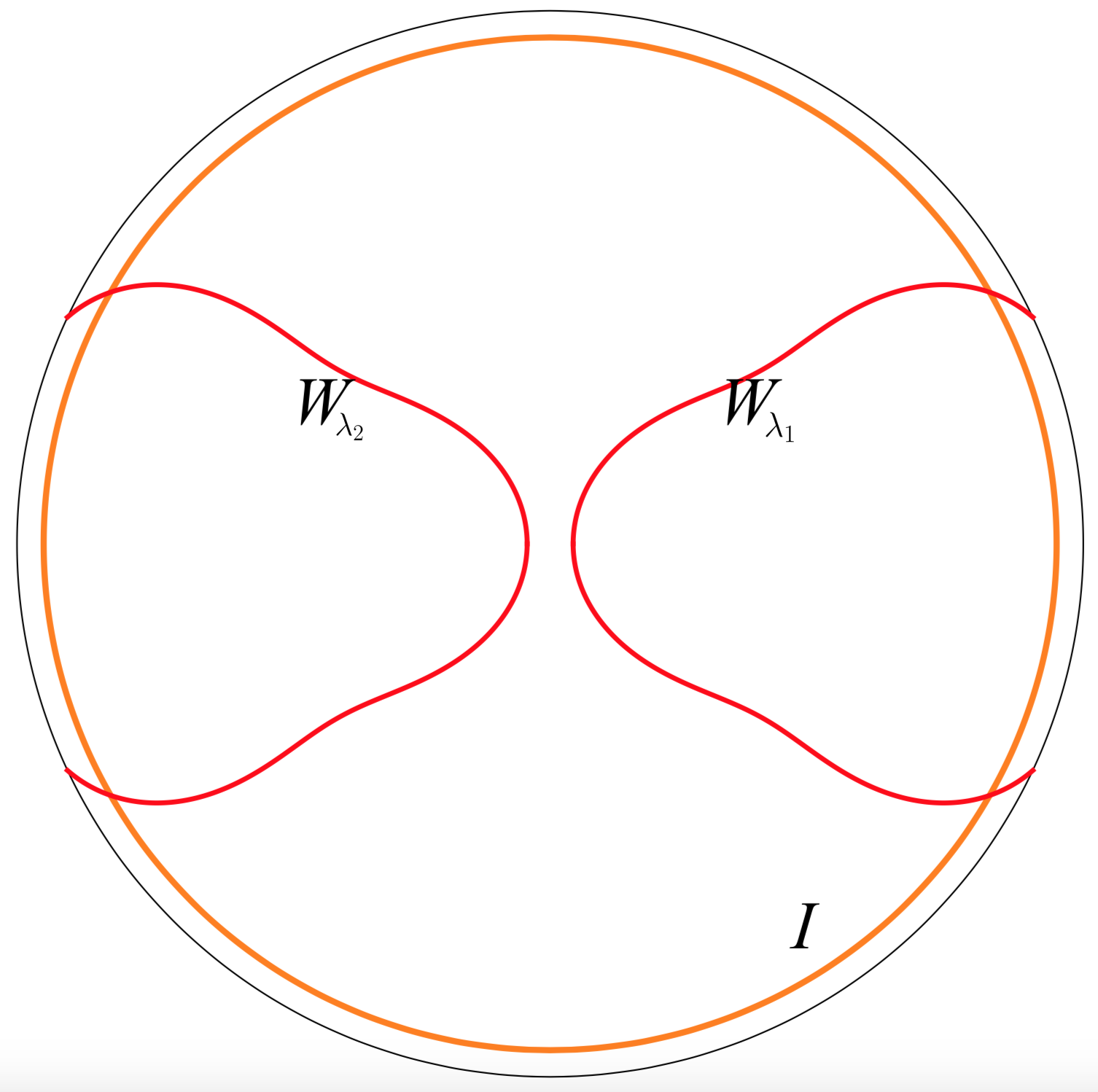}\\
         \includegraphics[width=0.4\textwidth]{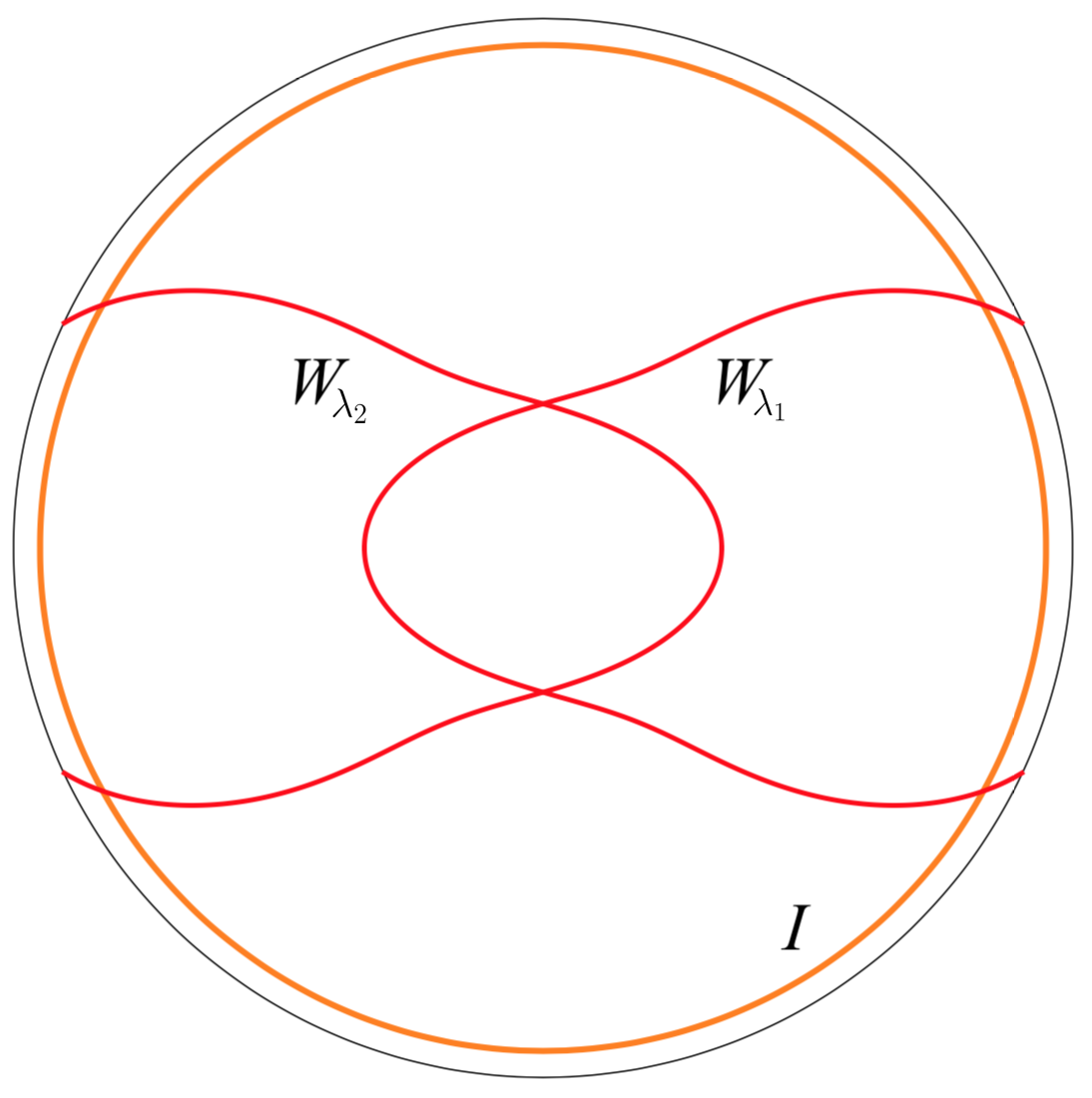}
           \includegraphics[width=0.4\textwidth]{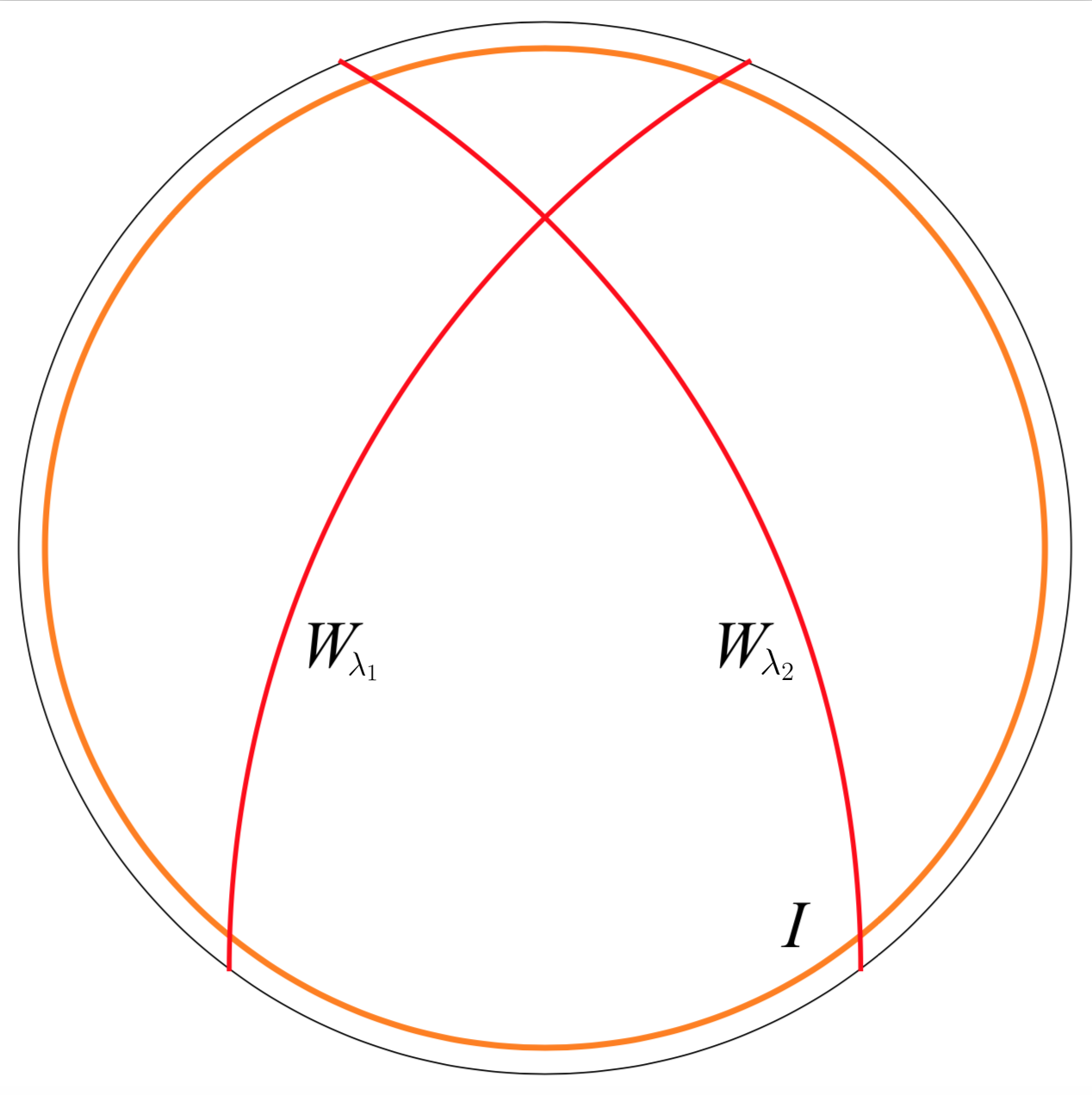}
    \caption{Several Euclidean Wilson line configurations, equivalent to different finite temperature correlation functions of the bi-local operator $\cO_\l(x_1, x_2)$: the top-left figure shows $\langle \cO_\l(\tau_1, \tau_2) \rangle_\beta = \langle \mathcal W_\l(\mC_{\tau_1, \tau_2}) \rangle$, the top-right figure yields the equality of the time-ordered correlators $\langle \cO_{\l_1}(\tau_1, \tau_2)  \cO_{\l_2}(\tau_3, \tau_4)\rangle_\beta = \langle \mathcal W_{\l_1}(\mC_{\tau_1, \tau_2}) \mathcal W_{\l_2}(\mC_{\tau_3, \tau_4}) \rangle$, the bottom-left figure shows a pair of intersecting Wilson lines that can be disentangled to the top-right configuration, while the bottom-right figure gives the out-of-time-ordered configurations. Note that the results are independent of the trajectory of the Wilson line inside of the bulk and only depend on the location where the Wilson lines intersect the defect. }
    \label{fig:wilson-loops-duals}
\end{figure}

 We propose the duality between such Wilson lines,  ``renormalized'' by an overall constant $N_R$, 
\be  
\label{eq:definitions-wilson-lines}
\cW_\l \equiv \hat \cW_R(\mC_{\tau_1, \tau_2})/N_R = \bar \chi_R \left(\mP \exp  \int_\mC A\right)/N_R \,,
\ee
and bi-local operators $\cO_\l(\tau_1, \tau_2)$ in the Schwarzian theory, defined in terms of the field $F(u)$ appearing in \eqref{eq:schw-action}
\be 
\cO_\l(\tau_1, \tau_2) \equiv \left(\frac{\sqrt{F'(\tau_1) F'(\tau_2)}}{F(\tau_1) - F(\tau_2)}\right)^{2\l} \,.
\ee
Our goal in this section will thus be to provide evidence that \footnote{As we will elaborate on shortly, when using the proper normalization, both Wilson lines in the positive or negative discrete series representation $\cD_{\l}^\pm$  will be dual to  insertions of $\cO_\l(\tau_1, \tau_2)$. For intersecting Wilson-line insertions we will consider the associated representations to be either all positive discrete series or all negative discrete series.  Note that the gauge theory has a charge-conjugation symmetry due to the $\mZ_2$ outer-automorphism of the $\mathfrak{sl}(2,\mR)$ algebra that acts as $(P_0,P_1,P_2)\to(-P_0,P_1,-P_2) $. In particular,  the principal series representations are self-conjugate, but the positive and negative series representations $\cD_{\l}^\pm$ are exchanged under this $\mZ_2$. Since the boundary condition $A_\tau=0$ preserves the charge-conjugation symmetry, the Wilson lines associated to the representations $\cD_{\l}^\pm$ have equal expectation values.}
\be
\label{eq}
\cO_\l(\tau_1, \tau_2)\quad  \Longleftrightarrow \quad \cW_{\l}(\mC_{\tau_1, \tau_2})\,,
\ee
for any boundary-anchored path  $\mC_{\tau_1, \tau_2}$ on the disk $D$ that intersects $I$ at points $\tau_1$ and $\tau_2$ (see the bottom-left diagram in Figure \ref{fig:wilson-loops-duals}).\footnote{Similar Wilson lines have been previously considered for compact gauge group \cite{Blommaert:2018oue}. They have also been considered in the context of a dimensional reduction from 3D Chern-Simons gravity \cite{Blommaert:2018oro, Blommaert:2018iqz}. } 

If imposing that gauge transformations are fixed to the identity along the boundary, the group element $g = \mP \exp  \int_\mC A$ is itself gauge invariant. While so far it was solely necessary to fix the holonomy around the boundary, to make the boundary-anchored Wilson lines \eqref{eq:definitions-wilson-lines} well-defined, we have to now specify the value of the gauge field on the boundary.\footnote{More precisely we have to specify the holonomy between any two points at which the Wilson lines intersect the boundary. } For this reason throughout this section we will set $A_\tau= 0$. With this choice of boundary conditions, we will perform the path integral with various Wilson line insertions and match with the corresponding correlation functions of the bilocal operators computed using the equivalence between the Schwarzian theory and a suitable large $c$ limit of 2D Virasoro CFT \cite{Mertens:2017mtv}.    We then generalize our result to any configuration of Wilson lines and reproduce the general diagrammatic `Feynman rules' conjectured in \cite{Mertens:2017mtv} for correlation functions of bi-local operators in the Schwarzian theory .

\subsection{Gravitational interpretation of the Wilson line operators}
\label{sec:gravitational-interp-wilson-line}

The matching between correlation functions of the bilocal operator and of boundary-anchored Wilson lines  should not come as a surprise. On the boundary side, the bilocal operator should be thought of as coupling the Schwarzian theory to matter. After rewriting  JT-gravity as  the bulk gauge theory, the Wilson lines are described by coupling a point-probe particle to gravity. A similar situation has been studied when describing  3D  Einstein gravity in terms of a 3D Chern-Simons theory with non-compact gauge group \cite{witten1989topology, carlip1989exact, vaz1994wilson, de1990spin, skagerstam1990topological, Ammon:2013hba}, and the relation is analogous in 2D, in the rewriting presented in Section~\ref{sec:rewriting-JT-gravity}. Specifically, as we present in detail in Appendix~\ref{sec:wilson-loop-and-external-matter}, the following two operator insertions are equivalent in the gauge theory/gravitational theory:\footnote{Note that the discussion in appendix  \ref{sec:wilson-loop-and-external-matter} shows the equivalence of the two insertions beyond the classical level. Typically, in 3D Chern-Simons theory the equivalence has been shown to be on-shell. See for instance \cite{Ammon:2013hba}.  }
\be 
\label{eq:correspondence-wilson-loop-particle-on-a-line}
{\cal W}_\l({\cal C}_{\tau_1\!\tau_2}) \cong \!\!\int\limits_{{\rm paths}\, \sim \, {\cal C}_{\tau_1\!\tau_2}}\!\!\!\!\!\! [dx] \, e^{-m \int_{{\cal C}_{\tau_1\!\tau_2}}\! ds\, \sqrt{g_{\alpha\beta} \dot{x}^\alpha \dot{x}^\beta}}, 
\ee
The right-hand side represents the functional integral over all paths $x(s)$ diffeomorphic to the curve ${\cal C}_{\tau_1\!\tau_2}$ weighted with the standard point particle action (with $\dot{x}^\alpha = \frac{d x^\alpha}{ds}$). In turn, this action is equal to the mass $m$ times the proper length of the path, where the mass $m$ is determined by the representation $\l $ of the Wilson line, $m^2= -C_2(\l) = \l(\l-1)$. In computing their expectation values, the mapping between the gauge theory and the gravitational theory should schematically yield 
\be 
\int D\phi DA e^{-S_E[A]} \frac{\chi_R(g)}{\cC_R} = \int Dg_{\mu \nu} D \Phi \int\limits_{{\rm paths}\, \sim \, {\cal C}_{\tau_1\!\tau_2}}\!\!\!\!\!\! [dx]  e^{-S_{JT}[g, \, \Phi] -m \int_{{\cal C}_{\tau_1\!\tau_2}}\! ds\, \sqrt{g_{\alpha\beta} \dot{x}^\alpha \dot{x}^\beta}} \,.
\ee
Thus, the expectation value of Wilson lines does not only match the expectation value of bi-local operators on the boundary, but it also offers the possibility to compute the exact coupling to probe matter in JT-gravity (see \cite{yang2018quantum} for an alternative perspective).

\subsection{Two-point function}
\label{sec:2pt}

The correlation function for a single Wilson line that ends on two points on the boundary, in a 2D gauge theory placed on a disk $D$, is given by the gluing procedure described in Section~\ref{sec:quantization}. Specifically,  for the group $\cG$, the un-normalized expectation value is given by 
\be 
\label{eq:Wilson-loop-exp-value}
\langle \hat \cW_{\l^\pm, k }(\cC_{\tau_1, \tau_2}) \rangle( g)= \int   dh Z\left(h, e\tau_{21}\right)\chi_{\l, k}^\pm(h) Z\left( g h^{-1}, e\tau_{12}\right) \,,
\ee
where $\tau_{21} = \tau_{2} - \tau_1$ is the length of $I$ enclosed by the boundary-anchored Wilson line $\cC_{\tau_1, \tau_2}$ and $\tau_{12} = \beta-  \tau_{2} +\tau_1$ is the complementary length of $I$. Here and below, $Z(h, e\tau)$ is the partition function computed in \eqref{eq:parition-function-fixed-GB-holonomy} on a patch of the disk, in the presence of a defect of length $\tau$ inside the patch, when setting the holonomy to be $h$ around the boundary of the patch.  The total $\cG$  holonomy around the boundary holonomy of the disk is set to $g$. Since we are interested in the case in which the gauge field along the boundary is trivial, we will want to consider the limit $\tilde g \to \mathbf 1$ at the end of this computation.  As was previously mentioned, the Wilson line is in the positive or negative discrete series representation $(\l^\pm, k = 0)$  of $\cG$, where $k=\mp {2\pi \lambda\over B}$ is the $\mR$ representation mentioned in Section~\ref{sec:quantization-and-choice-of-gg} that becomes $0$ due to the $B \rightarrow \infty$ limit.  Expanding \eqref{eq:Wilson-loop-exp-value} in terms of characters by using \eqref{eq:parition-function-fixed-GB-holonomy}, we find 
\be
 \label{eq:Wilson-loop-exp-value-expanded-out}
\langle \hat \cW_{\l^\pm, k }(\cC_{\tau_1, \tau_2}) \rangle( g)& = \int   dh \int_{-\infty}^\infty dk_1 dk_2 \int_0^\infty ds_1 ds_2\, \rho \left(\frac{Bk_1}{2\pi}, \, s_1\right) \rho\left(\frac{ B k_2}{2\pi}, \,s_2\right) \nn \\ & \times\, \chi_{(s_1, \mu_1 = -\frac{Bk_1}{2\pi}, k_1)}(h) \,\bar \chi_{\l, k}^\pm(h)  \,\chi_{(s_2, \mu_2 = -\frac{Bk_2}{2\pi}, k_2)}(g h^{-1}) e^{-\frac{e}2\left[  s^2_1 \tau_{21}+ s^2_2 \tau_{12} \right]}   \nn \\ 
&+ \text{ discrete series contributions }\,.
\ee
As in the previous sections, we are interested in obtaining observables in the presence of mixed boundary conditions in which we set $\phi^\mR = k_0 =  -i$. This isolates the representations with $k_1=k_2 = -i$ and, the limit $B \rightarrow \infty$ sets the $\mR$ representation of the Wilson line $ k = {\mp 2\pi \l}/{B} \rightarrow 0$.\footnote{In this limit, all contributions appearing as sums over the discrete series representations in \eqref{eq:Wilson-loop-exp-value-expanded-out} once again vanish. }  However, an order of limits issue appears: since the $\cG$ representation of the Wilson line is infinite dimensional we have to consider the $B \rightarrow \infty$ limit carefully. Thus, instead of inserting the full character in \eqref{eq:Wilson-loop-exp-value} we truncate the number of states in the positive or negative discrete series using the cut-off $\Xi$, with $\Xi \ll B$, 
\be
\label{eq:regularized-character}
\bar \chi_{\l^\pm, 0}(g) = \sum_{k = 0}^\Xi U_{\l, \pm(\l+ k)}^{\pm(\l+k)}(\tilde g)\,,
\ee
where $g = (\tilde g, \theta)$ with $\tilde g$ an element of $\SL2$ and $\theta$ an element of $\mR$,  $U_{\l, \pm(\l+ k)}^{\pm(\l+k)}(\tilde g)$ is the $\SL2$ matrix element computed explicitly in Appendix~\ref{app:SL(2,R)-rep-th}. 

Since the values of $k_i$ are fixed and the integral over the $\mR$ component of $h$ is trivial, we are thus left with performing the integral over the $\SL2$ components $\tilde h$ of $h$. In order to perform this integral,  we use the $\SL2$ fusion coefficients between two continuous series representations and a discrete series representation that we computed in Appendix~\ref{app:fusion-coeff} in the limit $\mu_1, \, \mu_2 \rightarrow i\infty$. When expanding the product of an $\cC_{s_1}^{\mu \rightarrow i \infty}$ continuous series and a $\cD_{\l^\pm}$ discrete series character into characters of the continuous series $\cC_{s_2}^{\mu \pm \l} = \cC_{s_2}^{\mu \rightarrow i \infty} $, we find the fusion coefficients between the three representations, $N^{\l}{\,}_{s_1, s_2}=N^{s_2}{\,}_{s_1, \l}$. Specifically, as we describe in great detail in Appendix~\ref{app:fusion-coeff},
\be
\label{eq:fusion-coeff-def}
\int d\tilde h \chi_{(s_1, \mu_1 \rightarrow i \infty)}(\tilde h)\bar \chi_{\l^\pm}(\tilde h) \chi_{(s_2, \mu_2 \rightarrow i \infty)}(\tilde g \tilde h^{-1})   &=N_{\l^\pm} \, N^{s_2}{\,}_{s_1, \l} \, \chi_{(s_2, \mu_2 \rightarrow i\infty)}(\tilde g) \ \,\\ &+\, \text{discrete series contributions }\,,  
\ee  
where $N^{s_2}{\,}_{s_1, \l}$ is given by 
\be
\label{eq:N-fusion-value}
N^{s_2}{\,}_{s_1, \l} = \frac{|\Gamma(\l+ is_1 - is_2)\Gamma(\l+ is_1 + is_2) |^2}{\Gamma(2\l)} = \frac{\Gamma(\l\pm is_1 \pm is_2)}{\Gamma(2\l)} \,,
\ee
where $\Gamma(x\pm y\pm z) \equiv \Gamma(x+y+z)\Gamma(x-y-z)\Gamma(x+y-z)\Gamma(x-y+z)$. The fusion coefficient has an overall normalization coefficient, $N_{\l^\pm}$, that appears in \eqref{eq:fusion-coeff-def} and is computed in Appendix~\ref{app:fusion-coeff} and is independent of $s_1$ and $s_2$. We can thus properly define the ``renormalized'' Wilson line, as previously mentioned in \eqref{eq:definitions-wilson-lines},  
\be
 \cW_{\l}(\cC_{\tau_1, \tau_2})  \equiv \frac{\hat \cW_{\l^\pm, k \rightarrow 0}(\cC_{\tau_1, \tau_2})}{N_{\l^\pm}}\,,
\ee
for which the associated fusion coefficient $N^{s_2}{\,}_{s_1, \l} $ is independent of whether the discrete series representation is given $\cD_{\l^+}$ or $\cD_{\l^-}$.  Furthermore, since all unitary discrete series representations appearing in the partition function are suppressed in the $B\to \infty$ limit, they do not contribute in the thermal correlation function of any number of Wilson lines. Consequently,   plugging (\ref{eq:fusion-coeff-def}) and (\ref{eq:parition-func-w-holonomy}) into (\ref{eq:Wilson-loop-exp-value-expanded-out}) we find
\be 
\label{eq:Wilson-loop-2pt-function-w-holonomy}
 \langle \cW_{\l}(\cC_{\tau_1, \tau_2})   \rangle_{k_0} (\tilde  g) \propto \int ds_1 \rho(s_1) ds_2 \rho(s_2)N^{s_2}{\,}_{s_1, \l}\,\chi_{s_2}(\tilde g) e^{-\frac{e}2 \left[(\tau_2 - \tau_1) s_1^2 + (\beta - \tau_2 + \tau_1) s_2^2\right]}\,.
\ee
where we have set the value of $\phi^\mR=-i$ along the boundary. When taking the limit $\tilde g \to \mathbf 1$, one can evaluate the limit of the $\SL2$ characters to find the normalized expectation value
\be
\label{eq:Wilson-loop-2pt-function-trivial}
 \frac{\langle \cW_{\l}(\cC_{\tau_1, \tau_2})  \rangle_{k_0}}{Z_{k_0}} &\propto  \left(\frac{e \beta}{2\pi}\right)^{3/2} e^{-\frac{2\pi^2}{e \beta }} \int ds_1 \rho(s_1) ds_2 \rho (s_2)\,N^{s_2}{\,}_{s_1, \l}\,e^{-\frac{e}2  \left[(\tau_2 - \tau_1) s_1^2 + (\beta - \tau_2 + \tau_1)s_2^2\right]} \nn\\ 
& \propto \left(\frac{e \beta}{2\pi}\right)^{3/2} e^{-\frac{2\pi^2}{e \beta }}  \int ds_1^2 ds_2^2 \sinh(2\pi s_1) \sinh(2\pi s_2) \frac{\Gamma(\l\pm is_1\pm i s_2)}{\Gamma(2\l)}  \nn \\ &\qquad \qquad \qquad  \times e^{-\frac {e}2 \left[(\tau_2 - \tau_1) s_1^2 + (\beta - \tau_2 + \tau_1)s_2^2\right]} \,.
\ee
where $\Gamma(\l\pm is_1\pm i s_2)$ was defined after \eqref{eq:N-fusion-value}. Using the correspondence $e = 1/C$, the result agrees precisely with the computation \cite{Mertens:2017mtv} of the  expectation value of a single bi-local operator $\langle \cO_\l(\tau_1, \tau_2) \rangle$ in the Schwarzian theory. The result there was obtained using  the  equivalence between the Schwarzian theory and a suitable large $c$ limit of 2D Virasoro CFT and had no direct interpretation in terms of $\SL2$ representation  theory.\footnote{However, the recent paper of \cite{Blommaert:2018iqz, Blommaert:2018oro} offer an interpretation in terms of representations of the semigroup $SL^+(2, \mR)$. }   Here we can generalize their result and study more complicated Wilson line configurations to reproduce the conjectured Feynman rules \cite{Mertens:2017mtv} in the Schwarzian theory.

\subsection{Time-ordered correlators}
\label{sec:time-ordered-correlators}
For instance, we can consider $n$ non-intersecting Wilson lines inserted along the contours $\cC_{\tau_1, \tau_2}$, ..., $\cC_{\tau_{2n-1}, \tau_{2n}}$ with  $\tau_1 < \tau_2 < \dots < \tau_{2n}$. As an example, the Wilson line configuration for the time-ordered correlator of two bi-local operators is represented in the center column of Figure \ref{fig:wilson-loops-duals}. The $n$-point function is given by, 
\be 
\label{eq:Wilson-loop-exp-value-time-ordered}
\left\langle \prod_{i=1}^n\hat \cW_{\l^\pm_i, k_i }(\cC_{\tau_{2i-1}, \tau_{2i}}) \right\rangle (g) & = \int   \left(\prod_{i = 1}^{n}dh_i \right)\left(\prod_{i = 1}^{n} Z\left(h_i, \,e\tau_{2i,\,2i-}\right)\bar \chi_{\l_i, k }^\pm(h_i)\right) \nn \\  &\qquad \times Z\left( g (h_1 \dots h_n)^{-1}, e\tau_{1,2n}\right) \,,
\ee
 where $\tau_{2i,\,2i-1} = \tau_{2i} - \tau_{2i-1}$ is the length of an individual segment  along $I$ enclosed by the contour $\cC_{\tau_{2i-1}, \tau_{2i}}$, while  $\tau_{2n, 1} = \beta - \tau_{12} -\ldots - \tau_{2n-1, 2n}$ is the length of the segment  along $I$ complementary to the union of $\cC_{\tau_1, \tau_2},$  $\dots$,  $\cC_{\tau_{2n-1}, \tau_{2n}}$.  Once again, all  Wilson lines are in the positive or negative discrete series representation $(\l^\pm_i, k_i) = \lim_{
B \rightarrow \infty}(\l^\pm_i, \mp 2 \pi \l_i/B) = (\l^\pm_i, 0)$.
Following the procedure presented in the previous subsection, we set the overall holonomy for the $\mathfrak{sl}(2, \mR)$ components of the gauge field to $\tilde g \to \mathbf 1$ and isolate the representations with $k_0 = \phi^\mR =  - i$.  We find
\be 
\label{eq:product-of-wilson-loops}
\frac{\langle \prod_{i=1}^n\cW_{\l_i}(\cC_{\tau_{2i-1}, \tau_{2i}})  \rangle_{k_0}}{Z_{k_0} } &= \left(\frac{e \beta}{2\pi}\right)^{3/2} e^{-\frac{2\pi^2}{e \beta }} \int ds_0 \rho(s_0) \left(\prod_{i=1}^n ds_1 \rho(s_1)\right) \left(\prod_{i=1}^n  N^{s_0}{}_{s_i, \l_i}\right) \nn \\
& \times \exp \left\{{-\frac{e}2\left[\left(\sum_{i=1}^{n} s_i^2(\tau_{2i} - \tau_{2i-1})\right) + s_0^2 \left(\beta-\sum_{i=1}^{n} (\tau_{2i} - \tau_{2i-1})\right) \right]}\right\}\,.
\ee 
 This result does not only agree with the time-ordered correlator of two bilocal operators in the Schwarzian theory, but it also reproduces the conjectured Feynman rule for any time-ordered bi-local correlator \cite{Mertens:2017mtv} and gives them an interpretation in terms of $\SL2$ representation theory. Specifically, to each segment between two anchoring points on the boundary  we can associate an $\widetilde{SL}(2, \mR)$ principal series representation labeled by $s_i$. Furthermore, at each anchoring point of the Wilson line, or at each insertion point of the bi-local operator, we associate the square-root of the fusion coefficient. Diagrammatically \cite{Mertens:2017mtv}, 
\be 
\label{eq:Schw-feynman-rules}
\begin{tikzpicture}[scale=0.57, baseline={([yshift=-0.1cm]current bounding box.center)}]
\draw[thick] (-0.2,0) arc (170:10:1.53);
\draw[fill,black] (-0.2,0.0375) circle (0.1);
\draw[fill,black] (2.8,0.0375) circle (0.1);
\draw (3.4, 0) node {\footnotesize $\tau_1$};
\draw (-0.7,0) node {\footnotesize $\tau_2$};
\draw (1.25, 1.6) node {\footnotesize $s$};
\draw (6.5, 0) node {$\raisebox{6mm}{$\ \ =\ \ e^{- s^2  (\tau_2-\tau_1)}$}$};
\end{tikzpicture}, ~~~~~~~~~~\ \ \begin{tikzpicture}[scale=0.7, baseline={([yshift=-0.1cm]current bounding box.center)}]
\draw[thick] (-.2,.9) arc (25:-25:2.2);
\draw[fill,black] (0,0) circle (0.08);
 \draw[thick](-1.5,0) -- (0,0);
\draw (.3,-0.95) node {\footnotesize $\textcolor{black}{s_2}$ };
\draw (.3,0.95) node {\footnotesize $\textcolor{black}{s_1}$};
\draw (-1,.3) node {\footnotesize $\l$};
\draw (2.5,0.1) node {$\mbox{$\hspace{2.0cm}\ =\  \, \sqrt{N^{s_1}{}_{s_2,\l}}\,.$}$}; \end{tikzpicture}
\ee
Finally, we integrate over all principal series representation labels $s_i$ associated to boundary segments using the Plancherel measure $\rho(s_0) \cdots \rho(s_n)$. Since for time-ordered correlators, both anchoring points of any Wilson line contributes the same fusion coefficient, we square the contribution of the right vertex in \eqref{eq:Schw-feynman-rules}, in agreement with our expression in \eqref{eq:product-of-wilson-loops}.

\subsection{Out-of-time-ordered correlators and intersecting Wilson lines}
\label{sec:OTO4pt}
While for time-ordered correlators we have considered disjoint Wilson lines,\footnote{We will revisit this assumption shortly. } in order to reproduce correlators of out-of-time-ordered correlators we have to discuss intersecting Wilson line configurations.  As an example, we show the Wilson line configuration associated to the correlator of two out-of-time-ordered bi-locals in Figure \ref{fig:wilson-loops-duals} in the bottom-right.  The correlator of intersecting Wilson loops in Yang-Mills theory with a compact gauge group has been determined in \cite{Witten:1991we}. Using the gluing procedure, the expectation value of the intersecting Wilson lines in the bottom-right of Figure \ref{fig:wilson-loops-duals}, when fixing the overall boundary $\cG$ holonomy, is given by\footnote{Once again the $\pm$ signs for the two  discrete series representation of the two  lines are uncorrelated. }
\be 
\label{eq:intersecting-Wilson-loop-exp-value}
\langle \hat \cW_{\l^\pm_1, 0}(\cC_{\tau_1, \tau_2}) \hat \cW_{\l^\pm_2, 0}(\cC_{\tau_3, \tau_4})\rangle (g)&= \int   dh_1 dh_2 dh_3 dh_4\, Z\left(h_1 h_2^{-1},\,e\tau_{31}\right) Z\left(h_2 h_3^{-1},\,e\tau_{32}\right) \times\nn\\&\qquad  \times Z\left(h_3 h_4^{-1}, \,e\tau_{42}\right)  Z\left(g\,h_4h_1^{-1},\,e\tau_{41}\right) \times\nn\\&\qquad  \times  \bar \chi_{\l_1^\pm,  0}(h_1 h_3^{-1}) \bar \chi_{\l_2^\pm, 0}(h_2 h_4^{-1})\,, 
\ee
where we consider the ordering $0<\tau_1 < \tau_3< \tau_2< \tau_4< \beta$, with $\tau_{41} = \beta-\tau_4+\tau_1$, and we are once again interested in the limit $\tilde g \rightarrow \mathbf 1$. Using the formula (\ref{eq:parition-func-w-holonomy}) for the partition function, one finds that performing the group integrals over $h_1, \, \dots, \, h_4$ gives eight Clebsch-Gordan coefficients associated to the representations of the four areas separated by Wilson lines and to the two representations of the Wilson lines themselves (see Appendix~\ref{app:6-j symbols} for a detailed account).   Collecting the Clebsch-Gordan coefficients associated to the bulk vertex one finds the 6-j symbol of $\widetilde{SL}(2, \mR)$, which we call $ R_{s_a s_b}\! \left[\; {}^{s_2}_{s_1} \;{}^{\l_2}_{\l_1}\right]$, which can  schematically be represented as  
\be
\begin{tikzpicture}[scale=.6, baseline={([yshift=0cm]current bounding box.center)}]
\label{eq:6j-symbol-cross-figure}
\draw[thick, red] (-2.0,2.0) -- (-.15,.15);
\draw[thick, red] (-.15,.15) -- (2.0,-2.0);
\draw[thick, red] (-2.0,-2.0) -- (2.0,2.0);
\draw (1.7,0) node {\scriptsize $s_4$};
\draw (-1.7,0) node {\scriptsize $s_3$};
\draw (-.75,.33) node {\scriptsize $\l_2$};
\draw (.78,.33) node {\scriptsize $\l_1$};
\draw (0,1.7) node {\scriptsize  $s_1$};
\draw (0,-1.7) node {\scriptsize $s_2$};
\end{tikzpicture}\raisebox{-3pt}{$\ \ \  = \ \ R_{s_3 s_4}\! 
 \left[\, {}^{s_2}_{s_1} \,{}^{\l_2}_{\l_1}\right] $}\,.
\ee
 As we discuss in detail in Appendix~\ref{app:6-j symbols}, the 6-j symbol is given by \cite{groenevelt2003wilson, groenevelt2006wilson}
\be
\label{eq:6-j symbol-def-main-text}
R_{s_a s_b}\! \left[\; {}^{s_2}_{s_1} \;{}^{\l_2}_{\l_1}\right] \, & \, = \,\,\mathbb{W}(s_a, s_b ; \l_1 + i s_2,\l_1 - i s_2, \l_2 - i s_1,\l_2 + i s_1) \\[3.5mm]
 &\times\,\sqrt{\Gamma(\l_2 \pm i s_1 \pm is_a)\Gamma(\l_1 \pm i s_2 \pm is_a)\Gamma(\l_1 \pm is_1\pm is_b)\Gamma(\l_2 \pm i s_2 \pm i s_b)}\nonumber\,,
\ee
where $\mathbb{W}(s_a, s_b ; \l_1 + i s_2,\l_1 - i s_2, \l_2 - i s_1,\l_2 + i s_1) $ denotes the Wilson function which is defined by a linear combination of ${}_4 F_3$ functions.  Thus, the expectation value of two intersecting Wilson lines when setting the holonomy for the $\mathfrak{sl}(2, \mR)$ components to $\tilde g \to \mathbf 1$ and setting $ \phi^\mR= -i$ is   given by
\be
\label{eq:OTO-four-pt-function-Wilson-loops}
\langle \cW_{\l_1}(\cC_{\tau_1, \tau_2}) \cW_{\l_2}(\cC_{\tau_3, \tau_4})\rangle_{k_0} (\tilde g) &\propto \int R_{s_3\, s_4}\! \left[\; {}^{s_2}_{s_1} \;{}^{\l_2}_{\l_1}\right]  \sqrt{N^{s_4}{}_{\l_1, s_1}  N^{s_3}{}_{\l_1, s_2}  N^{s_3}{}_{\l_2, s_1}  N^{s_4}{}_{\l_2, s_2} } \nn\\ & \times \chi_{s_b}(\tilde g)\, e^{-\frac{e}2\left[s_1^2(\tau_3- \tau_1) + s_3^2(\tau_3-\tau_2)  + s_2^2(\tau_4-\tau_2) + s_4^2(\beta -  \tau_4+\tau_1)\right] }  \prod_{i=1}^4 ds_i \rho(s_i)   
\ee 
where the exponential factors are those associated to each disk partition function $Z(h,\,e\tau_{ij})$ appearing in (\ref{eq:intersecting-Wilson-loop-exp-value}), while the factors $N^{s_i}{}_{\l_k, s_k}$ are the remainder from the fusion coefficients after collecting all factors necessary for the 6-j symbol. Evaluating the correlator with a $A_\tau=0$ on the boundary and dividing by the partition function, we find 
\be
\label{eq:wilson-loop-intersecting-wilson-loops} 
\frac{\langle \cW_{\l_1}(\cS_{\tau_1, \tau_2}) \cW_{\l_2}(\cS_{\tau_3, \tau_4})\rangle}{Z_{k_0}} &=  \left(\frac{e \beta}{2\pi}\right)^{3/2} e^{-\frac{2\pi^2}{e \beta }} \int R_{s_3 s_4}\! \left[\; {}^{s_2}_{s_1} \;{}^{\l_2}_{\l_1}\right]  \sqrt{N^{s_4}{}_{\l_1, s_1}  N^{s_3}{}_{\l_1, s_2}  N^{s_3}{}_{\l_2, s_1}  N^{s_4}{}_{\l_2, s_2}} \nn\\ & \times  e^{-\frac{e}2\left[s_1^2(\tau_3- \tau_1) + s_3^2(\tau_3-\tau_2)  + s_2^2(\tau_4-\tau_2) + s_4^2(\beta -  \tau_4+\tau_1)\right] } \prod_{i=1}^4 ds_i\, \rho(s_i) \,,  
\ee
which is in agreement with the result for the out-of-time order correlator for two bi-local operators obtained in the Schwarzian theory in \cite{Mertens:2017mtv}. 

The result \eqref{eq:wilson-loop-intersecting-wilson-loops} is easily generalizable to any intersecting Wilson line configuration as one simply needs to associated the symbol $R_{s_3 s_4}\! \left[\; {}^{s_2}_{s_1} \;{}^{\l_2}_{\l_1}\right]$ to any intersection.\footnote{Note that in the compact case discussed in \cite{Witten:1991we} the gauge group 6-j symbol appears squared. This is due to the fact that when considering two Wilson loops which are not boundary-anchored they typically intersect at two points in the bulk.  }  This reproduces the conjectured Feynman rule for the Schwarzian bi-local operators,  
\be
\begin{tikzpicture}[scale=.6, baseline={([yshift=0cm]current bounding box.center)}]
\label{eq:Schw-crossed-to-uncrossed}
\draw[thick] (-1.05,1.05) -- (-.15,.15);
\draw[thick] (.15,-.15) -- (1.05,-1.05);
\draw[thick] (-1.05,-1.05) -- (1.05,1.05);
\draw[thick] (0,0) circle (1.5);
\draw[fill,black] (-1.05,-1.05) circle (0.08);
\draw[fill,black] (1.05,-1.05) circle (0.08);
\draw[fill,black] (-1.05,1.05) circle (0.08);
\draw[fill,black] (1.05,1.05) circle (0.08);
\draw (1.85,0) node {\scriptsize $s_4$};
\draw (-1.85,0) node {\scriptsize $s_3$};
\draw (-.75,.33) node {\scriptsize $\l_2$};
\draw (.78,.33) node {\scriptsize $\l_1$};
\draw (0,1.75) node {\scriptsize  $s_1$};
\draw (0,-1.75) node {\scriptsize $s_2$};
\end{tikzpicture}\raisebox{-3pt}{$\ \ \  = \ \ R_{s_3 s_4}\! 
 \left[\, {}^{s_2}_{s_1} \,{}^{\l_2}_{\l_1}\right] $}\begin{tikzpicture}[scale=.6, baseline={([yshift=0cm]current bounding box.center)}]
\draw[thick] (0,0) circle (1.5);
\draw[thick] (1.3,.7) arc (300:240:2.6);
\draw[thick] (-1.3,-.7) arc (120:60:2.6);
\draw[fill,black] (-1.3,-.68) circle (0.08);
\draw[fill,black] (1.3,-.68) circle (0.08);
\draw[fill,black] (-1.3,0.68) circle (0.08);
\draw[fill,black] (1.3,0.68) circle (0.08);
\draw (1.85,0) node {\scriptsize $s_4$};
\draw (-1.85,0) node {\scriptsize $s_3$};
\draw (0,.75) node {\scriptsize $\l_1$};
\draw (0,-.75) node {\scriptsize $\l_2$};
\draw (0,1.75) node {\scriptsize $s_1$};
\draw (0,-1.75) node {\scriptsize $s_2$};
\end{tikzpicture}\,.
\ee
where one multiplies the diagram on the right by the 6-j symbol before performing the integrals associated to the $\SL2$ representation labels along the edges.\footnote{Note that the right diagram in  \eqref{eq:Schw-crossed-to-uncrossed} is just a useful mnemonic for performing computations that involve intersecting Wilson lines. It does not correspond to a configuration in the gauge theory since the representations $s_3$ and $s_4$ are kept distinct even though they would correspond to the same bulk patch in the gauge theory. } 

Finally, as a consistency check we verify that correlation functions are insensitive to Wilson lines  intersections that can be uncrossed in the bulk, without touching the defect loop $I$ (as that in the bottom-left figure~\ref{fig:wilson-loops-duals}). Diagrammatically, we want to prove for instance the Feynman rule
\be
\begin{tikzpicture}[scale=.6,  baseline={([yshift=0cm]current bounding box.center)}]
\draw[thick] (0,0) circle (1.5);
\draw[thick] (-1.45,.41) arc (220:320:1.95);
\draw[thick] (-1.45,-.41) arc (140:40:1.95);
\draw[fill,black] (-1.45,-.41) circle (0.08);
\draw[fill,black] (1.45,-.41) circle (0.08);
\draw[fill,black] (-1.45,.41) circle (0.08);
\draw[fill,black] (1.45,.41) circle (0.08);
\draw (1.85,0) node {\scriptsize $s_4$};
\draw (0,0) node {\scriptsize $s$};
\draw (-1.85,0) node {\scriptsize $s_3$};
\draw (0,.75) node {\scriptsize $\l_1$};
\draw (0,-.75) node {\scriptsize $\l_2$};
\draw (0,1.75) node {\scriptsize $s_1$};
\draw (0,-1.75) node {\scriptsize $s_2$};
\end{tikzpicture}\,\raisebox{-3pt}{$\ \ \  = \ \ \ {\delta(s_3-s_4) \over \rho(s_3,\mu_3) }$}\begin{tikzpicture}[scale=.6, baseline={([yshift=0cm]current bounding box.center)}]
\draw[thick] (0,0) circle (1.5);
\draw[thick] (1.3,.7) arc (300:240:2.6);
\draw[thick] (-1.3,-.7) arc (120:60:2.6);
\draw[fill,black] (-1.3,-.68) circle (0.08);
\draw[fill,black] (1.3,-.68) circle (0.08);
\draw[fill,black] (-1.3,0.68) circle (0.08);
\draw[fill,black] (1.3,0.68) circle (0.08);
\draw (1.85,0) node {\scriptsize $s_3$};
\draw (-1.85,0) node {\scriptsize $s_3$};
\draw (0,.75) node {\scriptsize $\l_1$};
\draw (0,-.75) node {\scriptsize $\l_2$};
\draw (0,1.75) node {\scriptsize $s_1$};
\draw (0,-1.75) node {\scriptsize $s_2$};
\end{tikzpicture}\, .
\label{eq:uncrossing-Wilson-loops}
\ee
We will denote the contours of two such Wilson lines as $\tilde \cC_{\tau_1, \tau_2}$ and $\tilde \cC_{\tau_3, \tau_4}$, where we assume that $0<\tau_1< \tau_2< \tau_3< \tau_4<\beta$. The expectation value in such a configuration is given by 
\be 
\label{eq:wilson-lines-disentangle}
\langle \hat \cW_{\l^\pm_1, 0}(\tilde \cC_{\tau_1, \tau_2}) \hat \cW_{\l^\pm_2, 0}(\tilde \cC_{\tau_3, \tau_4})\rangle (g)&= \int   dh_1 dh_2 dh_3 dh_4 dh_5 dh_6 \, Z\left(h_1 h_2^{-1},\, e\tau_{41}\right) Z\left(h_5^{-1} h_3^{-1} h_1^{-1},\,e\tau_{12}\right)\nn\\&\qquad  \times Z\left(h_6^{-1} h_5,\,e \tau_{23}\right)  Z\left(g\,h_2h_4h_6,\,e \tau_{43}\right) Z\left(h_3h_4^{-1},0\right) \nn\\&\qquad  \times  \bar \chi_{\l_1^\pm,  0}(h_1h_4 h_5) \bar \chi_{\l_2^\pm, 0}(h_2 h_3 h_6)\,.
\ee
Using \eqref{eq:parition-func-w-holonomy}, we will associate the representation labeled by $s_4$, $s_2$, $s_3$, $s_1$, and $s$, in this order, to the five disk partition  functions in \eqref{eq:wilson-lines-disentangle}.
Performing all the group integrals we once again obtain a contracted sum of  Clebsch-Gordan coefficients each of which is associated to a Wilson line representation and the representations labelling two neighboring regions. Performing the contractions for all of the Clebsch-Gordan coefficients we find two 6-j symbol symbols, $R_{s_3 s}\! \left[\; {}^{s_2}_{s_1} \;{}^{\l_2}_{\l_1}\right]$ and $ R_{s_4 s}\! \left[\; {}^{s_2}_{s_1} \;{}^{\l_2}_{\l_1}\right] $, each associated to the 6 representations that go around each of the two vertices. The remaining sums over Clebsch-Gordan coefficients yield the product of four fusion coefficients, $\sqrt{N^{s_4}{}_{\l_1, s_2} N^{s_2}{}_{\l_1, s_3} N^{s_3}{}_{\l_2, s_1} N^{s_1}{}_{\l_2, s_4}}$.

Using the orthogonality relation for the 6-j symbol that follows from properties of the Wilson function (see {\cite{groenevelt2003wilson, groenevelt2006wilson}}) 
\be
\label{eq:6-j symbol-ortho-main-text }
\int d s \rho(s, \mu)  R_{s_3 s}\! \left[\; {}^{s_2}_{s_1} \;{}^{\l_2}_{\l_1}\right] R_{s_4 s}\! \left[\; {}^{s_2}_{s_1} \;{}^{\l_2}_{\l_1}\right] + \text{ discrete series contribution} = \frac{\delta(s_3 - s_4)}{\rho(s_3, \mu_3)}\,,
\ee
where  $\rho(s, \mu)$ is the Plancherel measure defined in \eqref{eq:SL2-universal-planch-mesure},  we find that if there's a bulk region enclosed by intersecting Wilson that does not overlap with the defect loop, one can always perform the integral over the corresponding representation label $s$ to eliminate this region.  The integral over $s_3$ or $s_4$ then becomes trivial due to the delta-function in  \eqref{eq:6-j symbol-ortho-main-text } and thus the remaining fusion coefficients reproduce those in \eqref{eq:product-of-wilson-loops} for two non-intersecting Wilson lines.

Thus, putting together \eqref{eq:uncrossing-Wilson-loops}, \eqref{eq:Schw-crossed-to-uncrossed}, and \eqref{eq:Schw-feynman-rules}, we have re-derived  the diagrammatic rules needed to compute the expectation value of any bi-local operator configuration.   These rules are simply reproduced  combinatorially in the gauge theory  starting from the basic axioms presented in Section~\ref{sec:quantization}. 

\subsection{Wilson lines and local observables}

While one can recover the correlation functions of some local observables by considering the zero length limit for various loop or line operators, it is informative to also independently compute correlation functions of  local operators. In this section, we consider the operator $\tr \phi^2(x)$ which is topological (see \eqref{topop}). Consequently correlators of $\tr \phi^2(x)$ are independent of the location of insertion. Indeed they can be easily obtained by insertions of the Hamiltonian operator at various points  in the path integral, the un-normalized correlation function is given by 
\be
\label{eq:correlator-tr-phi-sq}
\langle \tr\phi^2(x_1) \dots \tr \phi^2(x_n) \rangle_{k_0} &= \left(e/4\right)^{-n}\langle H(x_1) \dots H(x_n) \rangle_{k_0} \nn \\ &\propto \Xi\int ds\,\rho(s) s^{2n} e^{-e \beta s^2/2}\,, 
\ee
where we first evaluated the correlator for a generic value of the boundary $\cG$ holonomy and then fixed the value of the field $\phi^\mR$ on the boundary and send $B \to \infty$ as described in Section~\ref{sec:quantization-and-choice-of-gg}.  At separated points, the correlator \eqref{eq:correlator-tr-phi-sq} agrees with that of $n$ insertions of the Schwarzian operator \cite{Stanford:2017thb, Mertens:2017mtv}, thus showing that the Schwarzian operator and $\tr \phi^2$ are equivalent, as shown classically in Section~\ref{sec:recovering-schw}.\footnote{However, the contact terms associated with these correlators are different.  We hope to determine the exact bulk operator dual to the Schwarzian in future work.}  This computation explains why the correlators of the Schwarzian operator at separated points are given by moments of the energy $E$ computed with the probability distribution $\rho(\sqrt{E}/e)$, as first observed in \cite{Stanford:2017thb}.

In the presence of Wilson line insertions, the operator $\tr \phi^2$ remains topological as long as we do not move it across a Wilson line. Consequently the correlation functions of $\tr \phi^2$ depend only on the number of $\tr \phi^2$ insertions within each patch separated by the Wilson lines. For instance, we can consider the insertion of $p=p_{0}+p_1+p_2+\dots+p_n$ $\tr \phi^2$ operators in the non-intersecting Wilson lines correlator considered in Section~\ref{sec:time-ordered-correlators}, as follows.  Let us put $p_{0}$ operators in the bulk and outside of the contour of any of the Wilson lines, together with $p_1$ $\tr \phi^2$, operators  enclosed by $\cC_{\tau_1, \, \tau_2}$, $p_2$ such operators enclosed by $\cC_{\tau_3, \, \tau_4}$, and so on.  The separated point correlator is then
\be 
\label{eq:product-of-wilson-lines-w-local-ops}
&\frac{\langle\left(\prod_{j=1}^p \tr \phi^2(x_j)\right)\left( \prod_{i=1}^n\cW_{\l_i}(\cC_{\tau_{2i-1}, \tau_{2i}})  \right)\rangle_{k_0}}{Z_{k_0} } =   \left(\frac{e \beta}{2\pi}\right)^{3/2} e^{-\frac{2\pi^2}{e \beta }} \int ds_0\rho(s_0) \left(\prod_{i=1}^n ds_i \rho(s_i)\right)  \nn \\ &\times s_1^{p_1} \dots s_n^{p_n} s_0^{p_{n+1}}\left(\prod_{i=1}^n  N^{s_0}{}_{s_i, \l_i}\right)
 e^{{-\frac{e}2\left[\left(\sum_{i=1}^{n} s_i^2(\tau_{2i} - \tau_{2i-1})\right) + s_0^2 \left(\beta-\sum_{i=1}^{n} (\tau_{2i} - \tau_{2i-1})\right) \right]}}\,.
\ee
In the Schwarzian theory, such a correlator is expected to reproduce the expectation value
 \be 
\bigg\langle\left[\prod_{j=1}^p \{F, u\}|_{u=\tilde \tau_j}\right]\left[ \prod_{i=1}^n\cO_{\l_i}(\tau_{2i-1}, \tau_{2i})  \right]\bigg\rangle \,,
\ee 
where $\tau_1 < \tilde \tau_1 < \ldots < \tilde \tau_{p_1} < \tau_2 < \ldots$. Such a computation can also be performed using the Virasoro CFT following the techniques outlined  \cite{Mertens:2017mtv}. Following similar reasoning, one can consider the correlators of the operator $\tr \phi^2$ in the presence of any other Wilson line configurations. 

\subsection{A network of non-local operators}

\begin{figure}[t!]
    \centering
          \includegraphics[width=0.4\textwidth]{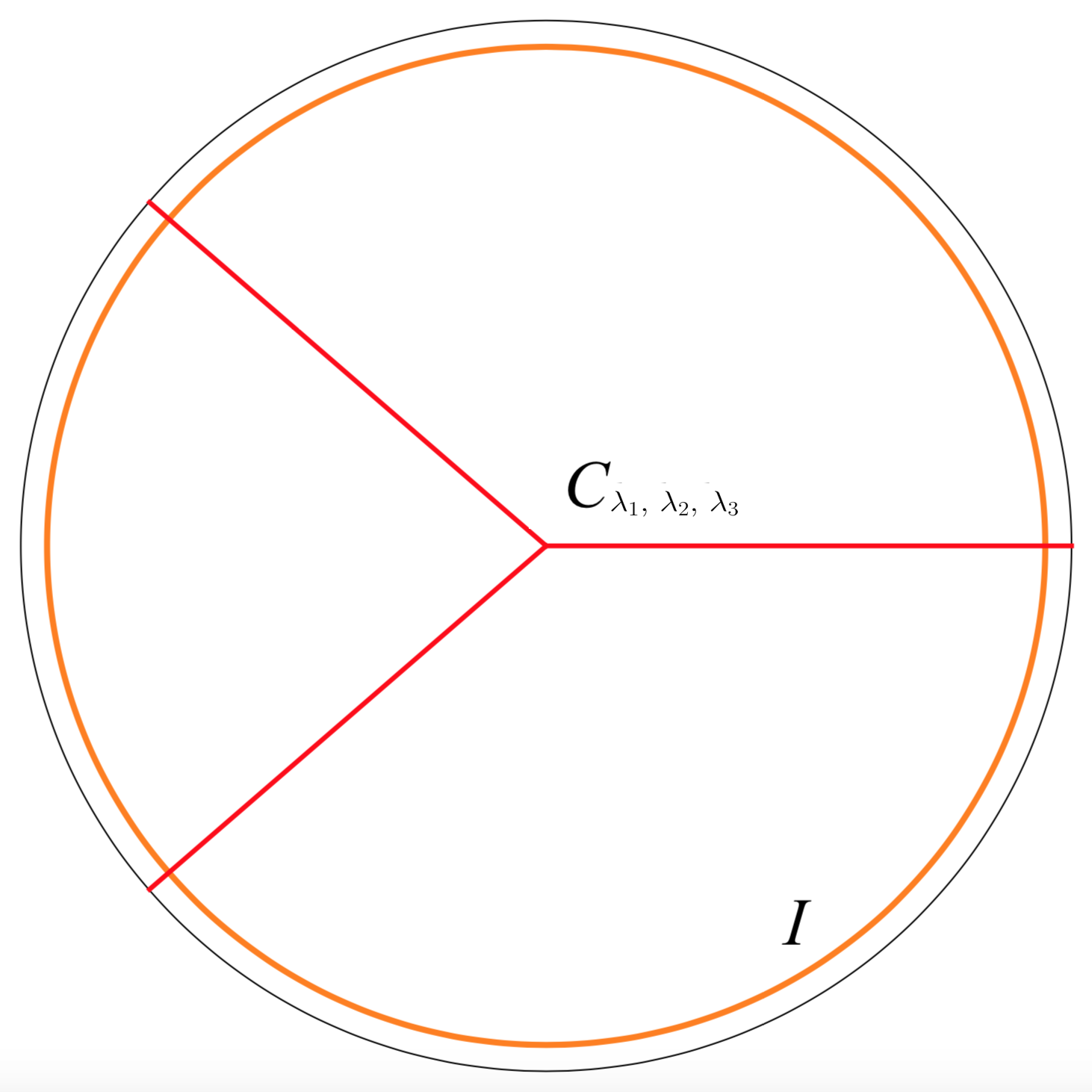}
    \caption{An example of a three-particle bulk interaction vertex corresponding to the junction of three Wilson lines defined by a Clebsch-Gordan coefficient at the vertex.}
    \label{fig:non-local-op-object}
\end{figure}

While so far we have focused on Wilson lines that end on the boundary, we now compute the expectation values of more complex non-local operators that are invariant under bulk gauge transformations that approach the identity on the boundary. Such objects, together with the previously discussed Wilson lines, serve as the basic building blocks for constructing ``networks'' of Wilson lines that capture various scattering problems in the bulk. The simplest such operator that includes a vertex in the bulk is given by the junction of three Wilson lines 
\be 
\label{eq:C-3-vertex}
C_{\l_1, \l_2, \l_3}(g_{\mC_{\tau_1, v}}, g_{\mC_{\tau_2, v}},g_{\mC_{\tau_3, v}}) &= \sum_{\substack{m_1 = \l_1 + \mZ^+_{<\Xi},\\ m_2 = \l_2+ \mZ^+_{<\Xi}}}  \,\,\sum_{\substack{n_1 = \l_1+\mZ^+, \\ n_2 = \l_2+\mZ^+}} \frac{C^{\l_1^+, \l_2^+, \l_3^+}_{m_1, m_2, m_1+m_2} (C^{\l_1^+, \l_2^+, \l_3^+}_{n_1, n_2, n_1+n_2}  )^*}{\cN_{\l_1^+, \l_2^+, \l_3^+}}\nn \\ &\times U_{(\l_1^{+}, 0),\,n_1}^{m_1}(g_{\mC_{\tau_1, v}}) U_{(\l_2^+, 0),\,n_2}^{m_2}(g_{\mC_{\tau_2, v}}) U_{(\l_3^+, 0),\,n_1+n_2}^{m_1+m_2}(g_{\mC_{\tau_3, v}}) \,,
\ee
with
\be
g_{\mC_{\tau_i, v}} = \cP \exp\left(\int_{\mC_{\tau_i, v}} A\right)\,,
\ee
where $\mC_{\tau_i, v}$ is a contour which starts on the boundary, intersects the defect at a point $\tau_i$, and ends at a bulk vertex point $v$. As indicated in \eqref{eq:C-3-vertex}, the sums over $m_1$ and $m_2$ are truncated by the cut-off $\Xi$.  Such a non-local object is schematically represented in Figure~\ref{fig:non-local-op-object}. For simplicity, we assume $0<\tau_1< \tau_2< \tau_3<\beta$ and we consider $\l_1$, $\l_2$, $\l_3$ labelling the Wilson lines to be positive discrete series representations. Once again, $U_{(\l^{+}, 0),\,n}^{m}(g) $ is the $\cG$ matrix element for the discrete representation $(\l^{+}, 0)$, $C^{\l_1^+, \l_2^+, \l_3^+}_{m_1, m_2, m_3} $ is the $\SL2$ (or, equivalently, $\cG$)  Clebsch-Gordan coefficient for the representations $\l_1, \l_2$, and $\l_3$, and $\cN_{\l_1^+, \l_2^+, \l_3^+}$ is a normalization coefficient for the Clebsch-Gordan coefficients discussed in Appendix~\ref{app:fusion-coeff}\@. Note that the operator \eqref{eq:C-3-vertex} is invariant under bulk gauge transformations. This follows from combining the fact that a gauge transformation changes $g_{\mC_{\tau_i, v}}\to g_{\mC_{\tau_i, v}} h_v$, where $h_v$ is an arbitrary $\cG$ element, with the identity
\be 
\sum_{m_1, m_2} U_{(\l_1^{+}, 0),\,n_1}^{m_1}(h_v) U_{(\l_2^+, 0),\,n_2}^{m_2}(h_v) U_{(\l_3^+, 0),\,n_1+n_2}^{m_1+m_2}(h_v) C^{\l_1^+, \l_2^+, \l_3^+}_{m_1, m_2, m_1+m_2} = C^{\l_1^+, \l_2^+, \l_3^+}_{n_1, n_2, n_1+n_2} \,.
\ee 

Using the gluing rules specified in Section~\ref{sec:quantization}, the expectation value of the operator \eqref{eq:C-3-vertex} with holonomy $g$ between the defect intersection points 3 and 1, and trivial holonomy between all other intersection points,  is given by 
\be
\label{eq:C-3-vert-exp-value-gluing}
\langle C_{\l_1, \l_2, \l_3} \rangle(g) &= \int dh_1 dh_2 dh_3 Z(h_1 h_2^{-1}, \,e\tau_{12})  Z(h_2 h_3^{-1}, \,e\tau_{12})  Z(g h_3 h_1^{-1}, \,e\tau_{12}) \nn \\ &\times C_{\l_1, \l_2, \l_3}(h_1, h_2, h_3) \,.
\ee
As before, we are interested in the case where we fix the $\SL2$ component of $\cG$ to $\tilde g \to \mathbf 1$.  Expanding \eqref{eq:C-3-vert-exp-value-gluing} into $\cG$ matrix elements we find the product of eight Clebsch-Gordan coefficients. Summing up the Clebsch-Gordan coefficients that have unbounded state indices (those that involve that $n_i$ indices instead of the $m_i$ indices in \eqref{eq:C-3-vertex}) we obtain the 6-j symbol with all representations  associated to the bulk vertex, $R_{\l_1 s_1}\! \left[\; {}^{\l_2}_{s_2} \;{}^{\l_3}_{s_3}\right]$, which is also related to the Wilson function as shown in  \cite{groenevelt2006wilson}. Setting the boundary condition $\phi^\mR = -i$ and take $\tilde g \to \mathbf 1$ we find that the 6-j symbol together with the sum over the remaining four Clebsch-Gordan coefficients yield
\be 
\frac{\langle C_{\l_1, \l_2, \l_3} \rangle}{Z_{k_0}} &=  \left(\frac{e \beta}{2\pi}\right)^{3/2} e^{-\frac{2\pi^2}{e \beta }}  N_{\l_1^+, \l_2^+, \l_3^+}  \int ds_1 \rho(s_1) ds_2 \rho(s_2) ds_3 \rho(s_3) \sqrt{N^{s_1}{}_{\l_1, s_2}  N^{s_2}{}_{\l_2, s_3}  N^{s_3}{}_{\l_3, s_1}}\nn \\ &\times R_{\l_1 s_1}\! \left[\; {}^{\l_2}_{s_2} \;{}^{\l_3}_{s_3}\right] e^{-\frac{e}2\left[s_1^2(\tau_2- \tau_1) + s_2^2(\tau_3-\tau_2)  + s_3^2(\beta - \tau_3+\tau_1)\right] }\,,
\ee
where in the limit in which all continuous representations have $\mu_1, \mu_2, \mu_3 \rightarrow i \infty$, $N_{\l_1^+ \l_2^+ \l_3^+}$ is a normalization constant independent of the representations $s_1$, $s_2$ or $s_3$ that can be absorbed in the definition of the operator $\cC_{\l_1, \l_2, \l_3}$.

We expect that the same reasoning as that applied for boundary-anchored Wilson lines should show that such a non-local operator corresponds to inserting the world-line action of three particles which intersect at a point in AdS$_2$ in the gravitational path integral (summing over all possible trajectories diffeomorphic to the initial paths shown in Figure~\ref{fig:non-local-op-object}).\footnote{It would be interesting to understand if this can be proven rigorously following an analogous approach to that presented in Appendix~\ref{sec:wilson-loop-and-external-matter}. } Thus, such insertions of non-local operators should capture the amplitude corresponding to a three-particle interaction in the bulk, at tree-level in the coupling constant between the three particles, but exact in the gravitational coupling. Similarly, by inserting a potentially more complex network of non-local gauge invariant operators in the path integral of the BF theory one might hope to capture the amplitude associated to any other type of interaction in the bulk.

\section{Discussion and future directions}
\label{sec:future-directions}
\label{sec:Lorentzian-signature}
We have thus managed to formulate a comprehensive holographic dictionary between the Schwarzian theory and the $\cG$ gauge theory: we have shown that the dynamics of the Schwarzian theory is equivalent to that of a defect loop in the $\cG$ gauge theory. Specifically, we have matched the partition function of the two theories, and have shown that bi-local operators in the boundary theory are mapped  to boundary-anchored defect-cutting Wilson lines. The gluing methods used to compute the correlators of Wilson lines provide a toolkit to compute the expectation value of any set of bi-local operators and reveal their connections to $\SL2$ representation theory.

There are numerous directions that we wish to pursue in the future. As emphasized in Section~\ref{sec:SL(2,R)-Yang-Mills}, while the choice of gauge algebra was sufficient to understand the on-shell equivalence between the gauge theory and JT-gravity, a careful analysis about the global structure of the gauge group was necessary in order to formulate the exact duality between the bulk and the boundary theories. While we have resorted to the gauge group $\cG$ with a simple boundary potential for the scalar field $\phi$, it is possible that there are other gauge group choices which reproduce observables in the Schwarzian theory or in related theories.  For instance, it would be instructive to further study the reason for the apparent equivalence between representations of the group  $\cG$ in the $B \rightarrow \infty$ limit and representations of the non-compact subsemigroup  $SL^+(2, \mR)$ which was discussed in \cite{Blommaert:2018oro, Blommaert:2018iqz, Lin:2018xkj}. Both gauge theory choices seemingly reproduce correlation functions in the Schwarzian theory. However, in the latter case the exact formulation of a two-dimensional action seems, as of yet, unclear.  Another interesting direction is to study the role of $q$-deformations for the 2d gauge theory associated to a non-compact group, which  have played an important role in the case of compact groups \cite{Aganagic:2004js}. Such a deformation is also relevant from the boundary perspective, where \cite{Berkooz:2018jqr} have shown that correlation functions in the large-$N$ double-scaled limit of the SYK model can be described in terms of representations of $q$-deformed $SU(1, 1)$. 

It is likely that one can generalize the 2D gauge theory/1D quantum mechanics duality for different choice of gauge groups and scalar potentials \cite{Gonzalez:2018enk}. A semi-classical example was given in \cite{Mezei:2017kmw}, where various 1D topological theories were shown to be semi-classically equivalent to 2D Yang-Mills theories with more complicated potentials for the field strength. It would be interesting to further understand the exact duality between such systems \cite{workInProgress2019}.

Finally, one would hope to generalize our analysis to the two other cases where the BF-theory with an $\mathfrak{sl}(2, \mR)$ gauge algebra is relevant: in understanding the quantization of JT-gravity in Lorentzian AdS$_2$ and in dS$_2$.\footnote{See \cite{Harlow:2018tqv, Maldacena:2019cbz} for a recent analysis of the quantization of the two gravitational systems. Furthermore, recently a set of gauge invariant operators was identified in the Schwarzian theory whose role is to move the bulk matter in the two-sided wormhole geometry relative to the dynamical boundaries \cite{Lin:2019qwu}. It would be interesting to identify the existence of such operators in the gauge theory context.  } By making   appropriate choices of gauge groups and boundary conditions in the two cases, one could once again hope to exactly compute observables in the gravitational theory by first understanding their  descriptions and properties in the corresponding gauge theory. We hope to address some of these above problems in the near future.

\subsection*{Acknowledgements}
We thank Po-Shen Hsin, Ho Tat Lam, Juan Maldacena, Thomas Mertens, Douglas Stanford, Gustavo J.~Turiaci, and Zhenbin Yang for several valuable discussions.  LVI and SSP were supported in part by the US NSF under Grant No. PHY-1820651 and by the Simons Foundation Grant No. 488653. The work of YW
was supported in part by the US NSF under Grant No. PHY-1620059 and by the Simons
Foundation Grant No. 488653.  The research of HV is supported by NSF grant PHY-1620059.

\appendix

\section{A review of the Schwarzian theory}
\label{sec:schw}
In this section, we review the Schwarzian theory, its equivalence to the particle on the hyperbolic plane $H_2^+$ placed in a magnetic field and the computation of observables in both theories.  The partition function for the Schwarzian theory on a Euclidean time circle of circumference $\beta$ is given by
\be 
\label{eq:schw-path}
Z_{\text{Schw.}}(\beta) &= \int_{f \in \frac{\text{Diff}(S^1)}{SL(2, \mR)}} \frac{D\mu[f]}{SL(2, \mR)} \exp\left[C \int_0^\b d u \left(\{f, u\} + \frac{2\pi^2}{\b^2}(f')^2\right)\right] \,, 
\ee 
where $C$ is a coupling constant with units of length, $\{f, u\}$ denotes the Schwarzian derivative, $f' = \partial_u f(u)$ and the path integral measure $D\mu[f]$ will be defined shortly. The field $f(u)$ obeying $f(u+ \b) =f(u) + \beta$ parameterizes the space $\text{Diff}(S^1)$ of diffeomorphisms of the circle. By performing the field redefinition $F(u) = \tan \left(\pi f(u)/\beta\right)$ with the consequent boundary condition $F(0) = F(\beta)$, as suggested in \eqref{eq:schw-action}, one can rewrite \eqref{eq:schw-path} as
 \es{SF}{
   S[F] = -C \int_0^\beta du \{F, u\} \,.
 }
 Classically, the action in (\ref{eq:schw-path}) can be seen to be invariant under $SL(2, \mathbb R)$ transformations\footnote{$SL(2, \mathbb R)$ is the naive symmetry when performing the transformation \eqref{eq:schw-SL(2,R)-action} at the level of the action. We will discuss the exact symmetry at the level of the Hilbert space shortly. }  
\be 
\label{eq:schw-SL(2,R)-action}
F \rightarrow \frac{a F + b}{c F + d}\,.
\ee
 In the path integral \eqref{eq:schw-action} one simply mods out by $SL(2, \mR)$ transformations \eqref{eq:schw-action} which are constant in time (the $SL(2, \mR)$ zero-mode). As we will further discuss in Section~\ref{sec:SL(2,R)-background}, such a quotient in the path integral is different from dynamically gauging the $SL(2, \mR)$ symmetry.  An appropriate choice for the measure on diff$(S^1)/SL(2, \mR)$ which can be derived from the symplectic form of the Schwarzian theory is given by, 
 \be
 \label{eq:Schw-measure}
 D\mu[f]  =  \prod_{u} \frac{df(u)}{f'(u)} = \prod_{u} \frac{dF(u)}{F'(u)}\,.
\ee
where the product is taken over a lattice that discretizes the Euclidean time circle.

Finally, the Hamiltonian associated to the action (\ref{SF}) is equal to the $\mathfrak{sl}(2, \mR)$ quadratic Casimir, $H = 1/C \, \left[-\ell_0^2 + (\ell_{-} \ell_{+} + \ell_{+} \ell_{-} )/2 \right]$, where $\ell_0$ and $\ell_\pm$ are the $\mathfrak{sl}(2, \mR)$ charges associated to the transformation \eqref{eq:schw-SL(2,R)-action}, which can be written in terms of $F(u)$ as  
\be 
\ell_0 &= 
\frac{i C}{\sqrt{2}} \left[ \frac{F''' F}{F'^2} - \frac{F F''^2}{F'^2}-\frac{F''}{F'}\right] \,,\nn \\  \ell_+ &=  \frac{i C}{\sqrt{2}}\left[\frac{F''' F^2}{F'^2} - \frac{F''^2 F^2}{F'^3} - \frac{2F F''}{F'} + 2F'\right] \,, \nn  \\ 
\ell_- &= \frac{i C}{\sqrt{2}} \left[\frac{F'''}{(F')^2} - \frac{(F'')^2}{(F')^3}\right]\,,
\ee 
The equality between the Hamiltonian and the Casimir suggests a useful connection between the Schwarzian theory and a non-relativistic particle on the hyperbolic upper-half plane, $H_2^+$, placed in a constant magnetic field $B$. In the latter the system, the Hamiltonian is also given by an $\mathfrak{sl}(2, \mR)$ quadratic Casimir. Below we discuss the equivalence of the two models at the path integral  level.

\subsubsection*{An equivalent description}
\label{sec:particle-in-magnetic-field}

The quantization of the non-relativistic particle on the hyperbolic plane, $H_2^+$, placed in a constant magnetic field $\tilde B$  was performed in \cite{Comtet:1984mm, Comtet:1986ki}. Writing the $H_2^+$ metric as $ds^2 =  d\phi^2 + e^{-2\phi} dF^2$ where both $\phi$ and $F$ take values in $\mR$, the non-relativistic action in Lorentzian time\footnote{For convenience, we distinguish Lorentzian time derivative $\dot f$ from Euclidean time derivatives $f'$. }
\be 
\label{eq:particle-in-magnetic-field-Lagr}
S_{\tilde B} =  \int  dt \, \left( \frac 1 4 {(\dot \phi)}^2+ \frac 14 e^{-2\phi} {{(\dot F)}^2}  + \tilde B {\dot F}e^{-\phi} + \tilde B^2 + \frac{1}4 \right) \,.
\ee
 The Hamiltonian written in terms of the canonical variables $(\phi, \pi_\phi)$ and $(F, \pi_F)$, is given by\footnote{We have shifted both the Lagrangian and the Hamiltonian by a factor of $\pm B^2$ in order to set the zero level for the energies of the particle on $H_2^+$ to be at the bottom of the continuum.} 
 \es{HamiltB}{
  H_{\tilde B} = \pi_\phi^2 + \pi_F^2 e^{2 \phi} - 2 \tilde B \pi_F e^{\phi} - \frac{1}4 \,.
 }
The thermal partition function at temperature $T= 1/\beta$ can be computed by analytically continuing \eqref{eq:particle-in-magnetic-field-Lagr} to Euclidean signature by sending $t \to -i u$ and computing the path integral on a circle of circumference $\beta$ with periodic boundary conditions $\phi(0) = \phi(\beta)$ and $F(0) =F(\beta)$.  At the level of the path integral, the partition function with such boundary conditions is given by
\be
\label{eq:part-function-particle-in-magnetic-field}
Z_{\tilde B}(\b)  =& \int_{\phi(0)=\phi(\beta),F(0) = F(\beta)} D\phi DF  \, e^{-\int_0^\beta du\left( 
{1\over 4}\phi'^2+{1\over 4}(e^{-\phi} F'-2 i \tilde B)^2
 \right) }\,.
\ee
with the $\mathfrak{sl}(2, \mR)$ invariant measure,
 \ie
 D\phi D f \equiv \prod_{u \in [0,\beta]}{d\phi}(u) dF(u) e^{-\phi(u)}
 \fe

For the purpose of understanding the equivalence between this system and the Schwarzian we will be interested in the analytic continuation to an imaginary background magnetic field $\tilde B=-{iB\over 2\pi}$ with $B\in \mathbb R$,
 \ie
 Z_{B}(\b)
 =& \int_{\phi(0)=\phi(\beta),F(0) = F(\beta)} D\phi DF  \, e^{-\int_0^\beta du\left( 
{1\over 4}\phi'^2+{1\over 4}(e^{-\phi} F'- B/\pi)^2
 \right) }\,.
  \\
 \sim & \int_{\phi(0) = \phi(\beta),\,F(0) = F(\beta)} D\phi DF  \, e^{-\int_0^\beta du\left( 
{1\over 4}\phi'^2+ {B^2\over 4\pi^2}e^{-2\phi} (F'-e^{\phi} )^2
 \right) }\,,
 \fe
 where we have shifted $\phi \to \phi-\log {B\over \pi}$ in the second line above and dropped an overall factor that only depends on $B$.
 
 The Schwarzian theory emerges as an effective description of this quantum mechanical system in the limit $B\to \infty$. Indeed, we can apply a saddle point approximation in this limit to integrate out $\phi$. This sets $F'=e^\phi$ and gives,  after taking into account the one-loop determinant for $\phi$ around the saddle,
 \ie
  Z_{B}(\b) \sim &  \int_{F(0) = F(\beta)}  \prod_{u}{dF(u)\over F'(u)}  \, e^{-\int_0^\beta du\left( 
{1\over 4} \left({F''\over F'}\right)^2
 \right) } =
   \int_{F(0) = F(\beta)}  D\mu[F] \, e^{{1\over 2}\int_0^\beta du \{F,u\}}
   \,,
   \label{eq:saddle-point-integrals}
   \fe
   where to obtain the second equality we have shifted the action by a total derivative. 
 
 Thus, as promised, we recover the Schwarzian partition function with the same measure for the field $F(u)$ in the $B\to \infty$ limit (and $\tilde B \to i \infty$), when setting the coupling $C =\frac{1}2$.\footnote{Note that the meaningful dimensionless parameter $ \beta\over C$ is unconstrained.} However, the space of integration for $F(u)$ in \eqref{eq:saddle-point-integrals} is different from that in the Schwarzian path integral \eqref{eq:schw-action}. This is most obvious after we transform to the other field variable $f(u) = \frac{\beta}\pi \tan^{-1}F(u)$ and
  \ie
  Z_{B}(\b) \sim &   \sum_{n\in  \mZ}\int_{f(0) = f(\beta)+n\beta }  D\mu[f] \, e^{{1\over 2}\int_0^\beta du \left(
  \{f,u\}
  +
  {2\pi^2\over \beta^2} (f')^2
  \right)
  }
   \,.
   \label{eq:saddle-point-integrals2}
   \fe
 While for the Schwarzian action $f(u)\in \text{Diff}(S^1)$, obeying the boundary condition $f(u+\beta) = f(u)\, +\,\beta$,  the path integral \eqref{eq:saddle-point-integrals2} consists of multiple topological sectors labeled by a winding number $n \in \mZ$ such that $f(u+\beta) = f(u) + \beta\, n$. In other words, the (Euclidean) Schwarzian theory is an effective description of the quantum mechanical particle in the $n=1$ sector.

 Reproducing the partition function of the Schwarzian theory from the particle of magnetic field thus depends on the choice of integration cycle for $F(u)$ (or $f(u)$). 
 As we explain below, the integration cycle needed in order for the partition function of the particle of magnetic field to be convergent is given by $\tilde B = i B \rightarrow i \infty$.  In order to do this it is useful to consider how the wave-functions in this theory transform as representations $\SL2$. 
 
When quantizing the particle on $H_2^+$ in the absence of a magnetic field, the eigenstates of the Hamiltonian transform as irreducible representations of $PSL(2, \mR)$ \cite{Comtet:1986ki}.  When turning on a magnetic field, the Hamiltonian eigenstates transform as projective representations of $PSL(2, \mR)$, which are the proper representations of $\widetilde{SL}(2, \mR)$ mentioned in Section~\ref{sec:search-for-the-gg}  \cite{Comtet:1986ki}.\footnote{Note that not all unitary irreducible representations of $\SL2$ need to appear in the decomposition of the Hilbert space under $\SL2$.  While there exist states transforming in any continuous series representation of $\SL2$,  there are also states transforming in the discrete series representations as long as $\lambda = -\tilde B + n$ with, $n \in \mZ$ and $0 \leq n \leq | \tilde B| -1$.}  
Specifically, the wavefunctions for the particle in magnetic field $\tilde B \in \mR$ transform in a subset of irreducible representations of $\widetilde{SL}(2, \mR)$ with fixed eigenvalues under the center of the group $e^{2\pi i \mu}  = e^{2\pi i \tilde B}$.\footnote{The fact that states transform in projective representations of the classical global symmetry can be understood as an anomaly of the global symmetry. An straightforward example of this phenomenon happens when studying a charged particle on a circle with a $\theta$-angle with $\theta = \pi$ \cite{Gaiotto:2017yup}. Note that when $B = \frac{p}q \in \mathbb Q$ states transform in absolute irreps of the $q$-cover of $PSL(2, \mR)$, which are also abolute irreps of $\widetilde{SL}(2, \mR)$.  It is only when $B \in \mR \setminus \mathbb Q$ that these irreps are absolute for the univesal cover  $\widetilde{SL}(2, \mR)$.   } Such unitary representations admit a well-defined associated Hermitian inner-product and the Hamiltonian is a Hermitian operator.   Up to a constant shift, their energies are real and are given by the $\SL2$ Casimir in \eqref{eq:irreps-labeling-SL(2,R)},  $E_{\lambda} = -(\lambda-1/2)^2$. 

When making $\tilde B \in \mathbb C \setminus \mR$ the Hamiltonian is no longer Hermitian and the representations of $\SL2$ do not admit a well defined Hermitian inner-product. However, the partition function defined by the path-integral \eqref{eq:particle-in-magnetic-field-Lagr} is convergent.  As we explain in Section~\ref{sec:search-for-the-gg}, if we analytically continue the Plancherel measure and Casimir to imaginary $\tilde B \to i \infty$, the thermal partition function in this limit reproduces that of the Schwarzian theory \eqref{eq:Schw-partition-function}. Thus, the theory makes sense in Euclidean signature where the correlation function of different  observables is convergent, but a more  careful treatment is needed in Lorentzian signature.\footnote{A more detailed discussion about the properties of the Schwarzian and of JT gravity in Lorentzian signature is forthcoming in \cite{workInProgress2019-2}. }

\subsubsection*{An $\SL2$  chemical potential} 
\label{sec:SL(2,R)-background}

 While the classical computation performed in Section~\ref{sec:recovering-schw} suggests the equivalence between imposing a non-trivial $PSL(2, \mR)$ twist for the Schwarzian field and the gauge theory \eqref{eq:total-action} with a non-trivial holonomy around its boundary this equivalence does not persist quantum mechanically.  Instead, in the presence of a non-trivial holonomy, the gauge theory is  equivalent to the non-relativistic particle in the magnetic field \eqref{eq:particle-in-magnetic-field-Lagr} with $\tilde B \rightarrow i \infty$ and in the presence of an $\SL2$ chemical potential. Note that in the derivation performed above, in order to prove the equivalence between the Schwarzian and the action  \eqref{eq:particle-in-magnetic-field-Lagr} with $\tilde B \rightarrow i \infty$, we have assumed that the field $F(u)$ is periodic: specifically, if one assumes a $PSL(2, \mR)$ twist around the thermal circle for the field $F(u)$, one can no longer use the equality in \eqref{eq:saddle-point-integrals}. Specifically,  \eqref{eq:saddle-point-integrals} assumes that when adding a total derivative to the action, the integral of that derivative around the thermal circle vanishes -- this is no longer true in the presence of a non-trivial twist for the Schwarzian field.

 In order to study \eqref{eq:particle-in-magnetic-field-Lagr} with $\tilde B \rightarrow i \infty$ in the presence of an $\SL2$ chemical potential, we start by considering the case of $\tilde B \in \mR$ and then we analytically continue to an imaginary magnetic field $\tilde B \in i \mR$.  The partition function is given by
\be
\label{eq:partition-function-w-holonomy}
Z_{iB}(\tilde g,\, \beta) & \sim \int ds \rho_B(s) e^{-\frac{\beta}{2C}s^2 }\sum_{m=-\infty}^\infty \langle \frac{1}2 + is,\, m | \tilde g |  \frac{1}2 + is,\, m\rangle + \text{discrete series contributions} \nn \\ &= \int_0^\infty  ds \rho_B(s) \chi_s(\tilde g) e^{-\frac{\beta}{2C}s^2 } + \text{disrete series contributions}\,, 
\ee
where $\chi_s(\tilde g) = \text{Tr}_s(\tilde g)$ is the $\SL2$ character of the principal series representation labelled by $\l = 1/2+i s$ (see Appendix~\ref{app:harmonic-analysis-SL(2,R)} for the explicit character $\chi_s(\tilde g)$). To recover the partition function when $\tilde B  = -{iB\over 2\pi}\rightarrow i \infty$ we again perform the analytic continuation used to obtain \eqref{eq:Schw-partition-function}. Once again the discrete series states have a contribution of $\cO(B e^{-\beta B^2/C})$ and can be neglected. Thus, up to a proportionality factor
\be
Z_{iB}(\tilde g, \,\beta) & \propto  \int_0^\infty  ds \rho(s) \chi_s(\tilde g) e^{-\frac{\beta}{2C}s^2 }\,.
\ee
This formula generalizes \eqref{eq:Schw-partition-function} for any $\tilde g $ and matches up to an overall proportionality factor, with the result obtained in the gauge theory in Section~\ref{sec:search-for-the-gg} (see \eqref{eq:parition-func-w-holonomy}). 

\section{Comparison between  compact  and non-compact groups}
\label{app:comparison-compact-vs-non-compact}

For convenience, we review the schematic comparison between various formulae commonly used for compact gauge groups (which we will denote by $G$) with finite dimensional unitary irreducible representations  and the analogous formulae that need to be used in the non-compact case (which we denote by $\cG$) with infinite-dimensional unitary irreducible representations: 
\\
\begin{table}[ht]
\begin{center}
\begin{tabular}{|c|c|}
\hline
$\delta(g) = \sum_R \dim R \,\chi_R(g) $&    $\delta(g) = \int dR \rho(R) \,\chi_R(g)$ \\\hline
$\int \frac{dg}{\vol G} U_{R, m}^n(g) U_{R', n'}^{m'}(g^{-1}) = \frac{\delta_{RR'} \delta_{mm'} \delta_{nn'} }{\dim R}$ & $\int \frac{dg}{\vol \cG} U_{R, m}^n(g) U_{R', n'}^{m'}(g^{-1}) = \frac{\delta(R,R') \delta_{mm'} \delta_{nn'} }{\rho(R)}$\\\hline
$\int \frac{dg}{\vol G} \chi_R(g) \chi_{R'}(g^{-1}) = \delta_{RR'}$ & $\int \frac{dg}{\vol \cG}  \chi_R(g) \chi_{R'}(g^{-1}) = \frac{\Xi\,\delta(R,R') }{\rho(R)}$\\\hline
$\int \frac{dg}{\vol G} \chi_{R}( g  h_1 {g}^{-1}  h_2) = \frac{\chi_{R}( h_1)\chi_{R}(  h_2)}{\dim R}  $&  $
\int \frac{dg}{\vol \cG} \chi_{R}( g  h_1 {g}^{-1}  h_2) = \frac{\chi_{R}( h_1)\chi_{R}(  h_2)}{\Xi} 
$\\ \hline 
 $\int \frac{dg}{\vol G} \chi_{R_1}( g  h_1) \chi_{R_2}( g^{-1}  h_1) = \frac{\delta_{R_1, R_2} \chi_{R_1}(h_1 h_2)}{\dim R_1}  $&   $\int \frac{dg}{\vol \cG} \chi_{R_1}( g  h_1) \chi_{R_2}( g^{-1}  h_2) = \frac{\delta(R_1, R_2) \chi_{R_1}(h_1 h_2)}{\rho(R)}  $\\ \hline
\end{tabular}
\end{center}
\end{table}
\\
where $\chi_R(g)$ are the characters of the group $G$ or $\cG$, $U_{R, m}^n(g)$ are the associated matrix elements and $\Xi$ is a divergent factor, which can be evaluated by considering the limit $\lim_{g\to \mathbf 1} \chi_R(g) = \Xi$. In the case of $\SL2$ and $\cG$ the limit needs to be taken from the direction of hyperbolic elements and for the group $\cG_B$ we have shown that $\Xi$ is independent of the representation $R$. We consider an in-depth discussion of the above formulae and their consequences in 2D  gauge theories with the non-compact gauge group $\cG_B$ below.

\section{Harmonic analysis on $\SL2$ and $\mathcal G_B$}
\label{app:SL(2,R)-rep-th}

\label{app:harmonic-analysis-SL(2,R)}
We next describe how to work with the characters of $\SL2$ and its $\mR$ extension, $\cG_B$ (and consequently the group $\cG \equiv \cG_B$ when taking the limit $B\to \infty$). In order to get there we first need to discuss the meaning of the Fourier transform on the group manifold of $\SL2$ or $\cG_B$.  Given a finite function $x(\tilde g)$ with $ \tilde g \in \SL2$,\footnote{Here finite means that it is infinitely differentiable if the group manifold is connected and is constant in a sufficiently small domain if the group manifold is disconnected. } for every unitary representation $U_R$ of the continuous and discrete series we can associate an operator 
\be 
\label{eq:fourier-transform-definition}
U_R(x) = \int x(\tilde g) U_R(\tilde g) d \tilde g \,.
\ee  
The operator $U_R(x)$ is called the Fourier transform of $x(\tilde g)$. Just like in Fourier analysis on $\mathbb R$, our goal will be to find the inversion formula for (\ref{eq:fourier-transform-definition}) and express $x(\tilde g)$ in terms of its Fourier transform. To start, we can express the Delta-function $\delta(\tilde g)$ on the group manifold, in terms of its Fourier components
\be 
\label{eq:expansion-delta-function}
\delta(\tilde g) = \int \rho(R)  \text{tr}(U_R(\tilde g)) d R\,,
\ee
where as we will see later in the subsection that $\rho(R)$ is the Plancherel measure on the group and $\chi_R(\tilde g) \equiv \tr(U_R(\tilde g))$ will define the character of the representation $R$. The integral over $R$ is schematic here (see later section for explicit definitions) and represents the integral over the principal and discrete series of the group. The Delta-function is defined such that, 
\be
\label{eq:delta-function-def}
\int x(\tilde g \tilde g_0) \delta(\tilde g) d \tilde g = x(\tilde g_0)\,.
\ee
Multiplying  (\ref{eq:expansion-delta-function}) by $x(\tilde g \tilde g_0)$ and integrating over the group manifold we find that  
\be
\label{eq:SL(2,R)-inversion-formula}
x(\tilde g_0) =  \int \rho(R) \text{tr}(U_R(x) U_R(\tilde g_0^{-1})) d R \,.
\ee
We will review the calculation of the matrix elements $U_{R, \, n}^{m}(\tilde g)$, characters $\chi_R(\tilde g)$ and of the Plancherel measure $\rho(R)$ in the next subsections.
 
\subsection{Evaluation of the matrix elements and characters}

As explained in \cite{Kitaev:2017hnr}, one can parameterize $\widetilde{SL}(2, \R)$ using the coordinates $(\xi, \phi, \eta)$, where we can  restrict $\phi+ \eta \in [0, 4\pi)$.  The $ \widetilde{SL}(2, \R)$ element $\tilde g$ takes the form $\tilde g = e^{\phi P_0} e^{\xi P_1} e^{-\eta P_0}$, where the generators $P_i$ are given by \eqref{eq:sl(2,R)-algebra}. 
In this parameterization, the metric is 
 \es{Metric}{
  ds^2 = d\xi^2 - d \phi^2 - d \eta^2 + 2 \cosh \xi d \phi d \eta 
 } 
and the Haar measure is 
 \es{Haar}{
  d\mu = \sinh \xi \, d\xi\, d\phi\, d\eta \,.
 } 
 For the full group $\cG_B$, we normalize the measure by,
 \ie
 d\tilde g\equiv d\mu d\theta
 \fe

As shown in \cite{Kitaev:2017hnr}, the matrix elements in the representation with quantum numbers $\lambda$ and $\mu$ are given by
\be
\label{eq:SL(2,R)-matrix-el}
U^m_{\lambda, n}(\tilde g) = 
e^{i (n \phi -m \eta)}(1-u)^{\lambda } u^{\frac{n-m}{2}}  
   \sqrt{\frac{\Gamma (n-\lambda +1) \Gamma (n+\lambda )}{\Gamma (m-\lambda
   +1) \Gamma (m+\lambda )}} \, \mathbf F(\lambda -m,n+\lambda
   ,-m+n+1;u)\,,
\ee
where, $\mathbf F(a, b, c, z) = \Gamma(c)^{-1} {}_2F_1(a, b, c; z)$,  $u = \tanh^2 (\xi / 2)$ and $m, n \in \mu + \mZ$.  We can similarly parametrize elements  $\cG_B$ by $g=(\theta, \tilde g)$ where $x$ is an element of $\mR$. The matrix element for the representation ($\l$,  $\mu  = - \frac{Bk}{2\pi}+ q$, $k$) in $\cG_B$ is thus given by, 
\be 
\label{eq:cGB-matrix-el}
U^m_{(\lambda, \,\mu  = - \frac{Bk}{2\pi}+ q,\, k),\, n}(g) = e^{ik\theta} U^m_{\lambda, n}(\tilde g)\,.
\ee 
Once again, this expression depends on $\mu$ only in that $m,\, n,\, k \in \mu + \Z$.  
The diagonal elements are thus given by
\be
\label{eq:SL(2,R)-matrix-el-diag}
U^m_{(\lambda, \mu, k), m}(g)= 
(1-u)^{\lambda } e^{i m (\phi -\eta)} e^{i k \theta} \, _2F_1(\lambda -m,\lambda + m
   ;1;u) \,.
\ee

The characters of the various representations are obtained by summing \eqref{eq:SL(2,R)-matrix-el-diag} over $m$.  Because the characters are class functions, they must be functions of the eigenvalues $x,\, x^{-1}$ of the $\SL2$ matrix $\tilde g$, when $\tilde g$ is expressed in the two-dimensional representation. $x$ can be obtained from the angles $\phi$, $\eta$ and $\xi$ for any representation to be\footnote{(We wrote two distinct formulas depending on whether $u$ is greater or smaller than $\sin^2 \frac{\phi - \eta}{2}$ in order to make explicit the choice of branch cut we use for the square root.)} 
 \es{Gotx}{
  x = \begin{cases}
   \displaystyle{\frac{\cos \frac{\phi - \eta}{2} \pm  \sqrt{u - \sin^2 \frac{\phi - \eta}{2} }}{\sqrt{1 - u}} }\,, & \text{if } u \geq \sin^2 \frac{\phi - \eta}{2}   \,, \\[10pt]
     \displaystyle{\frac{\cos \frac{\phi - \eta}{2} \pm i \sin \frac{\phi - \eta}{2}  \sqrt{1  - \frac{u}{\sin^2 \frac{\phi - \eta}{2} }}}{\sqrt{1 - u}}} \,, & \text{if } u < \sin^2 \frac{\phi - \eta}{2}   \,,
  \end{cases}
 }
where one of the solutions represents $x$ and the other $x^{-1}$.    Note that for hyperbolic elements, $x  \in \R$, which happens whenever $u > \sin^2 \frac{\phi - \eta}{2}$.   Simple examples of hyperbolic elements have $\phi = \eta = 0$, and in this case $ x = e^{\pm \xi/2}$. For elliptic elements, we have $\abs{x} = 1$ (with $x \notin \R$), which happens whenever $ u < \sin^2 \frac{\phi - \eta}{2}$.  Simple examples of elliptic elements have $u = \eta =  0$, and in this case $x = e^{\pm i \phi /2}$.  Lastly, for parabolic elements, we have $x = \pm 1$, and in this case $ u = \sin^2 \frac{\phi - \eta}{2}$. For convenience, from now on we choose $x$ such that $|x|>1$ and $|x^{-1}| < 1$ for hyperbolic elements. For elliptic elements, we choose $x$ to be associated with the negative sign in the 2nd equation of \eqref{Gotx}.

\subsubsection*{Continuous series}

To obtain the characters for the continuous series, we should set $\lambda = \frac 12 + i s$ and sum over all values of $m = \mu + p$ with $p \in \Z$.  The sum is given by
 \es{chisFirst}{
  \chi_{s, \mu, k}(g) 
   = (1-u)^{\frac 12 + i s } e^{ik\theta}  \sum_{p \in \Z} \, e^{i (\mu+p)(\phi-\eta)} {}_2F_1(\frac 12 + i s -\mu - p , \frac 12 + i s + \mu + p 
   ;1;u)\,,
 }
where we consider $\phi - \eta \in [2\pi(n-1), 2\pi n)$, with $n \in \mZ$. This sum can be evaluated using the generating formula for the ${}_2F_1$ hypergeometric function. Evaluating the sum defined in \eqref{Gotx} yields, in terms of the eigenvalue $x$ associated to $\tilde g$ group element, the $\mR$ element $\theta$ and the branch number, $n$, for the angle $\phi - \eta$, 
 \es{chisFinal}{
  \chi_{s, \mu, k}(g) =  \begin{cases}
  e^{i k \theta} e^{2\pi i \mu n}\left(\frac{|x|^{1 - 2 \lambda} + |x|^{-1 + 2 \lambda}}{|x - x^{-1}|}  \right)\,,\qquad \text{ for } \tilde g \text{ hyperbolic,}\\
  0 \,, \hspace{5.35cm}   \text{ for } \tilde g \text{ elliptic,}
  \end{cases}
 }
where $\lambda = \frac 12 + i s$ and, we remind the reader about the restriction that $\mu = \frac{-Bk}{2\pi} + \mZ$.

\subsubsection*{ Discrete series}

For the positive discrete series, we have $\mu = \lambda$ and the sum over $m$ goes over values equal to $\lambda + p$ with $p \in \mathbb Z^+$: 
 \es{chiPositive}{
  \chi_{ \lambda, k}^+(g) = e^{ik \theta} \sum_{p=0}^\infty U^{\lambda + p}_{\lambda, \lambda + p}(g) 
   = (1-u)^{\lambda } e^{ik \theta} e^{i \lambda (\phi -\eta)}   \sum_{p=0}^\infty e^{i p (\phi -\eta)} P_{p}^{(0, 2 \lambda - 1)}(1 - 2u) \,,
 }
where $P_n^{(\alpha, \beta)}(x)$ are the Jacobi polynomials.  We can once again evaluate the sum using the generating formula for the Jacobi polynomial to find that in terms of the eigenvalue $x$, the character is given by
 \es{chip}{
  \chi_{k, \lambda}^+(g) = \frac{e^{ik \theta}x^{1 - 2 \lambda}}{x - x^{-1}} 
 }
for both hyperbolic and elliptic elements.  This expression is identical to the first term in \eqref{chisFinal}. For the negative discrete series, we have $\mu = -\lambda$ and so we should take $m = - \lambda - p$, with $p \in \mZ^+$, and sum over $p$:
 \es{chiNegative}{
  \chi_{k, \lambda}^-(g) =  (1-u)^{\lambda } e^{ik \theta}e^{-i \lambda (\phi -\eta)}  \sum_{p=0}^\infty e^{-i p (\phi -\eta)} P_{p}^{(0, 2 \lambda - 1)}(1 - 2u) \,.
 }
Comparing \eqref{chiNegative} with \eqref{chiPositive}, we conclude that
 \es{chiNegativeFinal}{
  \chi_{\lambda,k }^-(g) = e^{i k \theta}\left( \chi_{ \lambda}^+(\tilde g) \right)^* 
   =  e^{i k \theta} \left( \frac{x^{1 - 2 \lambda}}{x - x^{-1}} \right)^* \,.
 }
This expression is identical to the second term in \eqref{chisFinal}.

Before we end this subsection, we summarize a few identities satisfied by the characters above. We have
\ie
\overline{\chi_{R}(g)}=\chi_{R}(g^{-1})
\fe
which follows from the unitarity of the representations. We also have
\ie
\chi_{s, \mu,k}(g^{-1})=\chi_{s, -\mu,-k}(g),\quad
\chi_{k, \lambda}^+(g^{-1})=\chi_{-k, \lambda}^-(g) \ \ .
\fe

\subsection{The Plancherel inversion formula}
\label{app:plancherel-inversion-form}

The normalization of the matrix elements $U_R$ given by \eqref{eq:SL(2,R)-matrix-el} - \eqref{eq:SL(2,R)-matrix-el-diag} can be computed following \cite{Kitaev:2017hnr}. For the continuous series one finds that, 
\be 
\label{eq:normalization-U-prinicipal-series}
\langle U_{(\frac{1}2+is,\mu, k), n}^m&|U_{(\frac{1}2+is', \mu', k'), n'}^{m'}\rangle = \int dg\, U_{( \frac{1}2 + is, \mu, k), n}^m(g) U_{(\frac{1}2 + is',\mu', k'), m'}^{n'}(g^{-1}) \nn \\ &=\
 4\pi^2 B 
\frac{\cosh(2\pi s) + \cos(B k)}{ s \sinh(2\pi s)} \delta(s-s') \delta(\mu-\mu') \delta_{kk'} \delta_{nn'} \delta_{mm'}\,,\nn \\ 
&\text{with }s, \,s'>0, \qquad \frac{-1}2 \leq \mu \leq \frac{1}2\,,\nn\\& \hspace{1.07cm} k,\,k' \in -\frac{2\pi(\mu+\mZ)}{B}\, ,\qquad \, m,\, n,\,m',\,n' \in \mu '+ \mZ\,. 
\ee
 Similarly, for the positive/negative discrete series one finds that, 
\be
\label{eq:normalization-U-discrete-series}
\langle U_{(\l, k), n)}^{m}&|U_{(\l', k'), n'}^{m'}\rangle = \frac{ 8\pi^2 B }{2\l-1} \delta(\l-\l')\delta_{kk'} \delta_{mm'}\delta_{nn'}\nn \\& \text{ with } \l,\, \l' > \frac{1}2, \qquad k,\, k'\in - \frac{2\pi(\pm \l +\mZ)}{B}\,, \qquad m,\, n,\, m',\, n' \in \pm (\l + \mZ^+)\,. 
\ee
Given the orthogonality of the matrix elements one can then write the $\delta$-function in \eqref{eq:expansion-delta-function} as, 
\be 
\label{eq:delta-function-on-GB}
\delta(g)  &= {1\over 2\pi}\int_{-\infty}^{\infty} {dk\, ds \over (2\pi)^2}\frac{ s \sinh(2\pi s)}{\cosh(2\pi s)+ \cos(B k)} \chi_{(s, \mu = -\frac{Bk}{2\pi}, k )}(g) +\nn\\ &+\int_{\frac{1}2}^{\infty} {d\l \over (2\pi)^2 B} \, \left(\l-\frac{1}2\right) 
\sum_{q=-\infty}^{\infty}\left(\chi_{\left(\l, k = -\frac{2\pi (\l + q)}{B} \right))}^+( g) + \chi_{\left(\l, k = -\frac{2\pi (-\l + q)}{B} \right)}^-( g) \right)\,,
\ee
For the purpose of evaluating the partition function of the gauge theory in Section~\ref{sec:SL(2,R)-Yang-Mills} it is more convenient to write all the terms in \eqref{eq:delta-function-on-GB} under a single $k$-integral. To do this one can perform a contour deformation \cite{Matsushita} to find that $\delta(g)$ can also be expressed as
\be
\label{eq:alternative-delta-g}
\delta(g) = -i\sum_{p \in \mZ}\int_{-\infty}^\infty dk \,\int_{-\infty}^\infty ds\, \left(\frac{Bk}{2\pi}+ p+ i s\right) \tanh(\pi s)  \,U_{\left(\frac{Bk}{2\pi}+ p + i s + \frac{1}2, \frac{Bk}{2\pi} + q,  k \right) \frac{Bk}{2\pi}+ p}^{\frac{Bk}{2\pi}+ p}(\tilde g) \,,
 \ee
with $q \in \mZ$. Using $\delta(g)$ from \eqref{eq:delta-function-on-GB}, the Plancherel inversion formula for $\SL2$ can be generalized to functions acting on the group $\cG_B$, 
\be 
\label{eq:plancherel-inversion-formula}
x(\mathbf{1}) &= {1\over 2\pi }\int_{-\infty}^{\infty} {dk\, ds\,\over (2\pi)^2} \frac{  s \sinh(2\pi s)}{\cosh(2\pi s)+ \cos(2\pi k)} \chi_{(s, \mu = -\frac{Bk}{2\pi}, k )}(x) +\nn\\ &+\int_{\frac{1}2}^{\infty} \frac{d\l}{(2\pi)^2 B}\, \left(\l-\frac{1}2\right) 
\sum_{q=-\infty}^{\infty}\left(\chi_{\left(\l, k = -\frac{2\pi (\l + q)}{B} \right))}^+(x) + \chi_{\left(\l, k = -\frac{2\pi (-\l + q)}{B} \right)}^-(x) \right)\,,
\ee
with 
\be 
\chi_R(x) \equiv \int d\tilde g \int_0^B d \theta x(g) \chi_R(g^{-1})\,.
\ee
In practice, in order to keep track of divergences evaluating the characters on a trivial we introduce the divergent factor $\Xi$, for which $\chi_{(s, \mu = -\frac{Bk}{2\pi}, k )}(x) = \Xi$. One can check this $s$-independent divergence by taking the limit 
\be
\lim_{\tilde g \rightarrow e}  \chi_{(s, \mu)}(\tilde g)=  \lim_{x \rightarrow 1,\, \theta \rightarrow 0} e^{i k \theta} \frac{x^{2is}+x^{-2is}}{x-x^{-1}} = \lim_{x \rightarrow 1} \frac{1}{|x-x^{-1}|}= \Xi\,.
\ee
Similarly, for $n \in \mZ$, 
\be
\lim_{\tilde g \rightarrow e^{2\pi i n\ell_0}}  \chi_{(s, \mu)}(\tilde g)= e^{2\pi i \mu n} \lim_{x \rightarrow \pm 1} \frac{1}{|x-x^{-1}|} = e^{2\pi i \mu n } \Xi \,.
\ee

Another operation that proves necessary for the computations performed in Section~\ref{sec:SL(2,R)-Yang-Mills} is performing the group integral
\be
\frac{1}{\vol \cG_B}\int dg \chi_{s, k = i}( g  h_1 {g}^{-1}  h_2) = \frac{1}{\Xi} \chi_{s, k= i}( h_1)\chi_{s, k= i}(  h_2) \,,
\ee
for principal series representation $s$ and for group elements $h_1$ and $h_2$. The normalization of this formula is set by taking the limit $h_1 \rightarrow e$ and $h_2 \rightarrow e$ and using the normalization for the matrix elements $U_R$,  \eqref{eq:normalization-U-prinicipal-series} and \eqref{eq:normalization-U-discrete-series}.

\subsection{An example: Isolating the principal series representation}

The goal of this appendix is to use the techniques presented in the previous subsections to show that we can isolate the contribution of principal series representations in the partition function. Specifically, we want to show that the regularization procedure suggested in Section~\ref{sec:search-for-the-gg} by adding higher powers of the quadratic Casimir leads to suppression of the discrete series.  Using the rewriting of $\delta(g)$ as in \eqref{eq:alternative-delta-g} we find that the partition function with an overall $\cG_B$ holonomy $g$ is given by, 
\be 
Z(g, \, e \beta) \sim &{-i} \sum_{p \in \mZ} \,\int_{-\infty}^\infty dk  \int_{-\infty}^\infty ds\, \left(-\frac{Bk}{2\pi} + p+is\right) \tanh\left(\pi s \right)  U_{-\frac{Bk}{2\pi}+ p+ i s + \frac{1}2, -\frac{Bk}{2\pi}+ p}^{-\frac{Bk}{2\pi}+ p}(\tilde g)  \nn\\ &\times e^{i k \theta}e^{\frac{e \beta}2\left[\left(p + is\right)^2 -\,\, \cdots\right]} \,,
\ee
where $g = (\tilde g,\, \theta)$ and $\cdots$  captures the contribution of higher powers of the quadratic Casimir. Setting the boundary condition $\phi^\mR= k_0 = -i$, we find that the partition function becomes
\be 
\label{eq:parition-function-transition-step}
Z_{k_0}(\tilde g, e\beta) \sim &{-i} \sum_{p \in \mZ} \,\int_{-\infty}^\infty ds\, \left(p+is\right) \tanh\left(\pi s - \frac{B}2\right)  U_{ p+ i s + \frac{1}2, \frac{Bi}{2\pi}+ p}^{\frac{Bi}{2\pi}+ p}(\tilde g) \nn\\ &\times e^{\frac{e \beta}2\left[\left(p + is\right)^2 -\,\, \cdots\right]} \,,
\ee
where, in order to obtain \eqref{eq:parition-function-transition-step}, we have also performed the contour re-parametrization $s \rightarrow s - \frac{B}{2\pi}$.  The form of higher order terms captured by $\cdots$ is given by higher powers of the quadratic Casimir: thus, for instance the first correction given by the square of the quadratic Casimir  is given by $\sim (p+is)^4/B$.  For each term in the sum, we can now deform the contour as $s \rightarrow s - i p$. Such a deformation only picks up poles located at $s_* = \frac{1}{2\pi} B - \frac{(2n+1)i}2$ with $n \in \mZ$ and $2n +1< p$.\footnote{The only poles in \eqref{eq:parition-function-transition-step} come from the measure factor $\tanh(\pi s - B/2)$.} The residue of each such pole gives rise to the contribution of the discrete series representations to the partition function. However, by choosing the negative sign for the fourth order and higher order terms in the potential the resulting contribution is suppressed as  $\cO(B e^{-\frac{e \beta B^2}2})$. This is the reason why the partition function is finite and is  solely given by the contribution of principal unitary series representations. \be 
\label{eq:partition-function-with-holonomy-p<B}
Z_{k_0}(\tilde g,\, e\beta) \sim &  \sum_{p \in \mZ}\int_{-\infty}^\infty ds\, s \tanh\left(\pi s - \frac{B}2\right) U_{  i s + \frac{1}2, \frac{Bi}{2\pi}+ p}^{\frac{Bi}{2\pi}+ p}(\tilde g)   e^{-\frac{e \beta s^2}2} +\nn \\ &+ \cO(B e^{-\frac{ e \beta B^2}2})\,.
\ee
Note that the integral is even in $s$ and that $\tanh\left(\pi s - \frac{B}2\right)  = (\sinh(2\pi s)-\sinh(B))/(\cosh(2\pi s) + \cosh(B))$. Thus, when considering the $B \rightarrow \infty$ limit the Plancherel measure becomes $ds s \sinh(2\pi s)/e^{-B}$. Thus, summing up all matrix coefficients in \eqref{eq:partition-function-with-holonomy-p<B} we recover the fact that the partition function only depends on characters, and we recover the result in Section~\ref{sec:partition-function}.

\section{Clebsch-Gordan coefficients, fusion coefficients and 6-j symbols}
\label{app:fusion-coeff}

The purpose of this section is to derive the fusion coefficients and the 6-j symbols needed in the main text.  To do so, we find it convenient to represent the states in the unitary representation $(\mu, \lambda)$ of $\SL2$ as functions $f(\phi)$ on the unit circle obeying the twisted periodicity condition
 \es{Twisted}{
  f(\phi+2\pi) &= e^{2\pi i \mu}   f(\phi)\,,
 }
with the rule that under a diffeomorphisms $V \in \widetilde{\text{Diff}}_+(S^1)$ of the unit circle, these functions transform as
 \es{DiffeoTransf}{
  (V   f)(\phi) &= \left(\partial_\phi V^{-1}(\phi)\right)^\l    f(V^{-1}\left(\phi\right))\,.
 }
Such a transformation property can be thought of arising from a ``$\mu$-twisted $\l$-form,'' namely an object formally written as $ f(\phi) (d\phi)^\l$.  We denote the space of such forms as $\cF_\lambda^\mu$.   In infinitesimal form, a diffeomorphism is described by a vector field $v(\phi) = v^\phi (\phi) \partial_\phi$, which acts on $f$ via the infinitesimal from of \eqref{DiffeoTransf}:
 \es{Infinitesimal}{
  v   f = -v^\phi \partial_\phi   f - \l (\partial_\phi v^\phi)  f\,.
 }

To see why the space $\cF_\lambda^\mu$ is isomorphic with the representation $(\mu, \lambda)$ of $\SL2$, note that \eqref{Infinitesimal} implies that the vector fields $L_n^\phi = -i e^{i n \phi}$ with $n=-1, 0, 1$ obey the commutation relations
 \es{CommutRel}{
  [L_{\pm 1}, L_0 ] = \pm L_{\pm 1} \,, \qquad
   [L_1, L_{-1}] = 2 L_0
 } 
so the transformations \eqref{Infinitesimal} corresponding to them generate an $\SL2$ subalgebra of $\widetilde{\text{Diff}}_+(S^1)$.  By comparison with \eqref{eq:sl(2,R)-algebra}, we can identify $\ell_0 = L_0$, $\ell_+ = L_1$, $\ell_- = L_{-1}$ when acting on $\cF_\lambda^\mu$.  From \eqref{Infinitesimal}, we can also determine the action of the quadratic Casimir
 \es{QuadCasimir}{
  \hat C_2 f = \left(-L_0^2 + \frac{L_1 L_{-1} + L_{-1} L_1}{2} \right) f 
   = \lambda(1 - \lambda) f \,.
 }
This fact, together with $e^{-2 \pi i L_0} f(\phi) = e^{2\pi \partial_\phi} f(\phi) = f(\phi + 2\pi) = e^{2\pi i \mu}   f(\phi) $ implies that $\cF^\mu_\lambda$ should be identified with the representation $(\lambda, \mu)$ (or with the isomorphic representation $(1-\lambda, \mu)$) of $\SL2$.

Let us now identify the function corresponding to the basis element $|m \rangle$ in the $(\mu, \lambda)$ representation.  This basis element has the property that $L_0 |m \rangle = -m |m \rangle$, which becomes $i \partial_\phi f = -m f$, so it should be proportional to $f_{\lambda, m} = e^{i m \phi}$.  (Recall that $m \in \mu + \Z$ for the irrep $(\mu, \lambda)$, so $f_{\lambda, m}$ obeys the twisted periodicity \eqref{Twisted}.)  In other words 
 \es{mRep}{
  |m \rangle  \qquad \text{corresponds to} \qquad
   c_{\lambda, m} f_{\lambda, m}(\phi) \equiv \langle \phi | m \rangle
 }
for some constant $c_{\lambda, m}$.  To determine $c_{\lambda, m}$, note that from \eqref{Infinitesimal}, we obtain 
 \es{LnBasis}{
  L_n  f_{\l, m} = -(m+n \l)  f_{\l, m+n}\.
 }
By comparison with the action \eqref{eq:irreps-states} of the raising and lowering operators on the states $|m \rangle$, we conclude that $c_{m, \lambda}$ obeys the recursion relation
 \es{mtof}{
  c_{\lambda, m+1} = c_{\lambda, m} \frac{( \lambda + m) }{\sqrt{(\lambda + m)(1 - \lambda + m)}}
 }
with the solution\footnote{The recursion formula only fixes $c_{\lambda,m}$ (similarly for $c_{\lambda,m}^-$ in \eqref{cnd}) up to an $m$ independent constant that could depend on $\lambda$. Here we have chosen a particular normalization for convenience. The physical observables we compute are however independent of such normalizations.}
 \es{cSol}{
  c_{\lambda, m} = \frac{\Gamma(\lambda + m)}{\sqrt{\Gamma(\lambda + m) \Gamma( 1- \lambda + m)}} \,.
 } 
Note that this expression holds both for the continuous series which we will denote as $c_{\lambda, m}$ and for the positive discrete series $c^+_{\lambda, m}$.
For negative discrete series we have instead
\ie
c^-_{\lambda, m-1} = c^-_{\lambda, m} \frac{( m-\lambda ) }{\sqrt{(m-\lambda )(m-1+ \lambda  )}}\,,
\fe
which leads to
\ie
c^-_{\lambda, m} =& (-1)^{m-\mu}\frac{\sqrt{\Gamma(1-m-\lambda ) \Gamma( \lambda -m)}}{\Gamma(1-\lambda-m)}
\label{cnd}
\fe
for $m=-\lambda,-\lambda-1,-\lambda-2,\dots$.

From these expressions and $\langle m | n \rangle = \delta_{mn}$, we can infer the inner product on the space $\cF_\lambda^\mu$.  Indeed, any two functions $f$ and $g$ obeying \eqref{Twisted} can be expanded in Fourier series as
 \es{FourierExpansion}{
  f (\phi) &= \sum_{m} a_m e^{i m \phi}  \qquad
   \Longleftrightarrow \qquad a_m = \frac{1}{2 \pi} \int d \phi\, e^{-i m \phi} f(\phi) \,, \\
  g(\phi) &= \sum_{m} b_m e^{im\phi}  \qquad \Longleftrightarrow \qquad b_m = \frac{1}{2 \pi} \int d \phi\, e^{-i m \phi} g(\phi) \,.
 }
Then we can write
 \es{f12Inner}{
  \langle f | g \rangle 
   = \sum_{m, n} \frac{a_m^* b_n}{c_{\lambda, m}^* c_{\lambda, n}} \langle m | n \rangle 
    =  \sum_{m} \frac{a_m^* b_m}{ \abs{c_{\lambda, m}}^2} \,.
 }
Writing $a_m$ and $b_m$ in terms of $f_1$ and $f_2$ using the Fourier series inversion formula, we obtain
 \es{f12Inner2}{
   \langle f | g \rangle
    =  \int d\phi_1\, d\phi_2\, f(\phi_1)^* g(\phi_2) G(\phi_1 - \phi_2) 
 } 
where $G(\phi)$ given by
 \es{GDef}{
  G(\phi) = \frac{1}{4 \pi^2} \sum_m \frac{e^{i m \phi}}{\abs{c_{\lambda, m}}^2} \,.
 }
For the continuous series, $\abs{c_{\lambda, m}}^2 = 1$, and the sum is over $m \in \mu + \Z$.  We obtain
  \es{ContinuousInner}{
  \text{continuous series:}  \qquad G(\phi) = \frac{1}{4 \pi^2} e^{i \mu (\phi_1 - \phi_2) } D \left(\frac{\phi_1 - \phi_2}{2 \pi} \right) \,,
 }
where $D(x) = \sum_{k \in \Z} \delta(x - k)$ is a Dirac comb with unit period.  For the positive discrete series, $m \in \lambda + \Z_+$ and $\mu = \lambda > 0$.  We find that \eqref{GDef} evaluates to 
 \es{DiscreteSeriesInner}{
  \text{positive discrete series:}  \qquad G(\phi) 
   = \frac{e^{i \lambda \phi} {}_2 F_1 (1, 1, 2 \lambda, e^{i \phi})}{4 \pi^2 \Gamma(2 \lambda)} 
  \,.
 }

To obtain the fusion coefficients, we need to consider tensor products of representations.   As a warm-up, let us consider the tensor product
 \es{TensorProd}{
  {\cal C}_{\frac 12 + is, \mu} \otimes {\cal C}_{\frac 12 + i s, -\mu}
 }
and identify the state corresponding to the identity representation.  This state is
 \es{identityState}{
  \sum_{m \in \mu + \Z} (-1)^m |m \rangle | -m \rangle \,,
 }
and it can be obtained as the unique state invariant under $L_n^{(1)} + L_n^{(2)}$, where the $L_n^{(i)}$ (with $n = -1, 0, 1$ and $i = 1, 2$) are the $\SL2$ generator acting on the $i$th factor of the tensor product.  

The state \eqref{identityState} can also be found in a more indirect way by first constructing the two-variable function $Y(e^{i \phi_1}, e^{i \phi_2})$  representing it.  This function obeys the conditions 
 \es{Conditions}{
   \sum_{i=1}^2 \partial_{\phi_i} Y(e^{i \phi_1}, e^{i \phi_2}) = 0 \,, \qquad
  \sum_{i=1}^2 \left(i e^{\pm i \phi_i}\partial_{\phi_i}  \mp \lambda  e^{\pm i \phi_i}   \right) Y (e^{i \phi_1}, e^{i \phi_2}) = 0 \,, 
 }
(with $\lambda = \frac 12 + i s$) representing the invariance under the $\SL2$ generators, as well as the periodicity conditions \eqref{Twisted} in $\phi_1$ and $\phi_2$ individually.  When $0< \phi_1 - \phi_2  < 2 \pi$, the solution of the equations \eqref{Conditions} is
 \es{YSol}{
  Y(e^{i \phi_1}, e^{i \phi_2}) = C \sin \left(\frac{\phi_1 - \phi_2}2\right)^{-2 \l}
 }
for some constant $C$.  Away from this interval, the expression \eqref{YSol} should be extended using the periodicity condition \eqref{Twisted}.  The state corresponding to this function is generally of the form $\sum_{m_1 \in \mu + \Z} \sum_{m_2 \in -\mu + \Z} C_{m_1, m_2} |m_1 \rangle | m_2 \rangle$, with coefficients $C_{m_1, m_2}$ obtained by taking the inner product with the basis elements:
 \es{Cm1m2}{
  C_{m_1, m_2} = \frac{1}{ 4 \pi^2} \int d\phi_1 \int d\phi_2 \, c_{\lambda, m_1}^* c_{\lambda, m_2}^*  e^{-i m_1 \phi_1} e^{-i m_2 \phi_2} Y(e^{i \phi_1}, e^{i \phi_2})
 }
Because $Y$ depends only on $\phi_1 - \phi_2$, the only non-zero $C_{m_1, m_2}$ are those with $m_1 = - m_2$.
Using
 \es{Integral}{
  \int_0^{2 \pi}  d\phi \, e^{-i m \phi} \left( \sin \frac{\phi}2\right)^{-2 \l} = 
{-2 e^{-i m \pi}\sin (m \pi)\Gamma( 1- 2\lambda ) \Gamma( \lambda - m)  \over \Gamma(1- \lambda-m) } \,,
 }
and $\lambda = \frac 12 + i s$, the expression \eqref{Cm1m2} with $m_1 = -m_2 = m$ evaluates to 
 \es{CmFinal}{
  C_{m, -m} = e^{- i \pi m }  C{\sin (\pi\mu ) \over   2s \sin (\pi (\mu-\lambda)) \sinh(2 \pi s) \Gamma(2 i s)} \sqrt{ \frac{\cos (2 \pi \mu) + \cosh(2 \pi s)}{2}} \,.
 }
We see that up to an $m$-independent constant, $C_{m, -m} \propto (-1)^m$, so \eqref{CmFinal} agrees with \eqref{identityState}.

\subsection{Clebsch-Gordan coefficients: $\mC_{\lambda_1 = \frac{1}2+i s_1}^{\mu_1} \otimes \mathcal D_{\lambda_2}^\pm \to {\cal C}_{\lambda = \frac 12 + i s}^{\mu}$}
\label{sec:CG-coefficients}

In \cite{Kitaev:2017hnr} a general recipe was outlined for obtaining the ``Clebsch-Gordan'' coefficients for $\SL2$.\footnote{Alternatively, see \cite{Repka1} and \cite{Repka2} for a more mathematical approach.} and, in particular, Ref.~\cite{Kitaev:2017hnr} constructed the decomposition of the tensor products ${\cal D}^+_{\l_1} \otimes {\cal D}^+_{\l_2}$ and ${\cal D}^+_{\l_1} \otimes {\cal D}^-_{\l_2}$.  Here we follow the same recipe to determine the Clebsch-Gordan coefficients and fusion coefficients between two continuous series representations and a positive/negative discrete series representation:
\es{CD}{
	\mC_{\l_1=  \frac{1}2+is_1}^{ \mu_1} \otimes \mathcal D_{\lambda_2}^\pm  
	\to  {\cal C}_{\lambda = \frac 12 + i s}^{\mu}\,, 
}
with $\mu = \mu_1 \pm \lambda$.  The state $|s, m\rangle$ that is part of ${\cal C}_{\lambda = \frac 12 + i s}^{\mu}$ in the tensor product \eqref{CD} must take the form
\es{StateInTensor}{
	|s, m \rangle = \sum_{m_2 = \pm (\lambda + \mZ^+) } C^{s_1,\, \l_2^\pm,\, s}_{m-m_2, m_2, m} |m - m_2 \rangle |m_2 \rangle 
}
where $C^{s_1,\, \l_2^\pm,\, s}_{m-m_2, m_2, m}$ is the Clebsch-Gordan coefficient  and the range of $m_2$ depends on whether it comes from the positive or negative discrete series.

As in the previous section, we determine $C^{s_1,\, \l_2^\pm,\, s}_{m-m_2, m_2, m} $ in a rather indirect way by first constructing the functions $Y_{s, m}(e^{i \phi_1}, e^{i \phi_2})$ that represent the state \eqref{StateInTensor}.  This function can be found using the conditions that 
\es{ConditionsYsm}{
	L_0 Y_{s, m} &= - m Y_{s, m} \,, \\
	\left( -L_0^2 + \frac{L_1 L_{-1} + L_{-1} L_1}{2} \right) Y_{s, m} &= \lambda( 1 - \lambda) Y_{s, m} \,,
}
where $L_n = L_n^{(1)} + L_n^{(2)}$ and $\lambda = \frac 12 + i s$.   Let us first solve these equations for $0< \phi_1 < 2 \pi$ and $0 < \phi_1 - \phi_2 < 2 \pi$. (The expression for $Y$ can then be continued away from this range using the appropriate periodicity condition \eqref{Twisted} in both $\phi_1$ and $\phi_2$.)

The first equation in \eqref{ConditionsYsm} implies that $Y_{s, m}$ equals $e^{i m \phi_1}$ times a function of $\phi_1 - \phi_2$.  The second condition gives a second order differential equation for this function of $\phi_1 - \phi_2$ with two linearly independent solutions
\es{YsmM}{
	Y^-_{s, m}( e^{i \phi_1}, e^{i \phi_2}) &= B^-_{s, m} e^{i m \phi_1}  e^{i \lambda_2 (\phi_1 - \phi_2)} \left( 1 - e^{i (\phi_1 - \phi_2)} \right)^{\lambda - \lambda_1 - \lambda_2} \\
	&{}\times 
	{}_2 F_1( \lambda - \lambda_1 + \lambda_2, \lambda + m, 1 + m - \lambda_1 + \lambda_2, e^{i (\phi_1 - \phi_2)}) \,.
}
and
\es{YsmP}{
	Y^+_{s, m}( e^{i \phi_1}, e^{i \phi_2}) &=  B^+_{s, m} e^{i m \phi_2}  e^{i (\lambda_2-m) (\phi_2 - \phi_1)} \left( 1 - e^{i (\phi_2 - \phi_1)} \right)^{\lambda - \lambda_1 - \lambda_2} \\
	&{}\times 
	{}_2 F_1( \lambda -\lambda_1 + \lambda_2, \lambda - m, 1 - m - \lambda_1 + \lambda_2, e^{i (\phi_2 - \phi_1)}) \,.
}
for some constant $B^\pm_{s, m}$. Both of the solutions are linearly dependent under $s\to -s$, thus from now on, we will restrict to $s> 0$. As suggested by the notation, this specific basis of solutions correspond   precisely to the generating functions for Clebsch-Gordon coefficients for the tensor product $\mC_{\lambda_1 = \frac{1}2+i s_1}^{\mu_1} \otimes \mathcal D_{\lambda_2}^\pm$. This is fixed by requiring that $Y^+_{s,m}(z,w)w^{-\lambda_2}$ and $Y^-_{s,m}(z,1/w)w^{-\lambda_2}$ to be holomorphic inside the unit disk $|w|<1$, as suggested by the one-side bounded sum in $m_2$ in \eqref{StateInTensor}, with $m_2 = \pm(\l_2 + \mZ^+)$ \cite{Kitaev:2017hnr}.

The dependence of $B^-_{s, m}$ on $m$ is fixed by requiring $Y^-_{s, m}$ to transform appropriately under the action of the raising and lowering operators.  Explicit computation shows that
\es{LpYsm}{
	L_1 Y^-_{s, m} = - \frac{(\lambda + m)(1 - \lambda + m)}{1 + m - \lambda_1 + \lambda_2} \frac{B^-_{s, m}}{B^-_{s, m+1}} Y^-_{s, m + 1} \,.
}
Comparing with the desired relation $L_1 Y^-_{s, m} = - \sqrt{(\lambda + m)(1 - \lambda + m)} Y^-_{s, m+1}$, we obtain the recursion formula
\es{RecursionBRatio}{
	B^-_{s, m+1} = \frac{ \sqrt{(\lambda + m)(1 - \lambda + m)} }{1 + m - \lambda_1 + \lambda_2} B^-_{s, m} \,.
}
Up to an overall constant which we denote by $B^-_s$, this recursion is solved by
\es{GotBsm}{
	B^-_{s, m} = \frac{\sqrt{\Gamma(\lambda + m) \Gamma( 1- \lambda + m)}}{\Gamma(1 - \lambda_1 + \lambda_2 + m)} B^-_s \,.
}
Similarly we can determine $ B^+_{s,m}$ by recursion relations
\es{LpYsm2}{
	L_1 Y^+_{s, m} = -(\lambda_1-\lambda_2+m) \frac{B^+_{s, m}}{B^+_{s, m+1}}  Y^+_{s, m + 1} \,.
}
to be
\es{GotBsm2}{
	B^+_{s, m} = {\Gamma(\lambda_1 - \lambda_2 + m)\over \sqrt{\Gamma(\lambda + m) \Gamma( 1- \lambda + m)}} B^+_s \,.
}

\subsubsection*{Normalization}

We would like to compute the normalization  constant ${\cal N}(s)$ for the inner product of states \eqref{StateInTensor},
\ie
\langle s, m | s', m' \rangle ={\cal N}(s) \delta (s-s')\delta_{mm'}
\label{NormExpectation}
\fe
For this purpose, it is sufficient to consider $m = m' $, and take the inner product of the functions representing the LHS of \eqref{NormExpectation}.  Using \eqref{StateInTensor}, we can write this inner product as
\es{InnerProdComplete}{
	\langle s, m | s', m \rangle
	= \sum_{m_2}  (C^{s_1,\, \l_2^\pm,\, s}_{m-m_2, m_2, m} )^* C^{s_1,\, \l_2^\pm,\, s'}_{m-m_2, m_2, m} 
}
(The answer should be independent of $m$.)  The expected delta functions in \eqref{NormExpectation} arise from the large $m_2$ terms in the sum.  Thus, let us compute 
\es{CDef}{
	C^{s_1,\, \l_2^\pm,\, s}_{m-m_2, m_2, m} 
	= \langle m-m_2 | \langle m_2,^\pm | s , m \rangle
	=  \frac{1}{c^\pm_{\lambda_2, m_2} e^{i m_2 \phi_2}}  \frac{1}{2 \pi}  \int d \phi_1 \, c_{\lambda_1, m - m_2}^* e^{-i (m - m_2)  \phi_1} Y^\pm_{s, m}(e^{i \phi_1}, e^{i \phi_2})  
}
at large $m_2$.  

We first start by considering $\l_2$ in the negative discrete series. After plugging in the expression for $Y$ and writing $\phi_1 = \phi_2 + \phi$, we obtain
\es{CAgain}{
	C^{s_1,\, \l_2^-,\, s}_{m-m_2, m_2, m} 
	&=B_s \frac{c_{\lambda_1, m - m_2}^*}{2 \pi c^-_{\lambda_2, m_2}} \frac{\sqrt{\Gamma(\lambda + m) \Gamma( 1- \lambda + m)}}{\Gamma(1 - \lambda_1 + \lambda_2 + m)} 
	\int_0^{2 \pi}  d \phi \,  
	e^{i m_2 \phi} e^{i \lambda_2 \phi} \left( 1 - e^{i \phi} \right)^{\lambda - \lambda_1 - \lambda_2} \\
	&{}\times 
	{}_2 F_1( \lambda - \lambda_1 + \lambda_2, \lambda + m, 1 + m - \lambda_1 + \lambda_2, e^{i \phi})  \,.
}
The large $m_2$ behavior of the $\phi$ integral comes from the regions where the integrand is singular or non-analytic (because the $\phi$ integral extracts a Fourier coefficient, and in general, Fourier coefficients with large momenta come from singularities in position space).  In this case, the singularities of the integrand are at $e^{i \phi} = 1$, where the integrand is approximately
\es{ApproximateIntegrand}{
	e^{i (m_2+\lambda_2) \phi} \left[  (1  - e^{i \phi} )^{\lambda - \lambda_1 - \lambda_2}  \frac{\Gamma(1 - 2 \lambda) \Gamma( 1+ m - \lambda_1 + \lambda_2)}{\Gamma( 1+ m - \lambda) \Gamma(1 - \lambda - \lambda_1 + \lambda_2)}  + \  \text{$\left( \lambda \leftrightarrow 1 - \lambda \right)$}  \right]\,.
}
The integral $\int_0^{2 \pi} d\phi \,  e^{-i k \phi} (1  - e^{i \phi} )^\alpha$ has the same large $k$ asymptotics as the integral 
\ie
\int_{0}^\infty d\phi\,  e^{-i k \phi} (-i)^\alpha \abs{\phi}^\alpha+ \int_{-\infty}^0 d\phi\,  e^{-i k \phi} i^\alpha \abs{\phi}^\alpha\,.
\fe 
Using the formula $\int_0^\infty d\phi \,\phi^\alpha e^{-ik \phi-\epsilon \phi}={\Gamma(1+\alpha)\over (\epsilon+i k)^{1+\alpha}}$, the integral in \eqref{CAgain} gives, approximately at large $m_2$, 
\es{m2Asymp}{
	&
	-2|m_2|^{\lambda_1 + \lambda_2 - \lambda - 1}   \sin {\pi (\lambda-\lambda_1-\lambda_2) }
	\Gamma( \lambda - \lambda_1 - \lambda_2 + 1)
	\\
	&{}\times \frac{\Gamma(1 - 2 \lambda) \Gamma( 1+ m - \lambda_1 + \lambda_2)}{\Gamma( 1+ m - \lambda) \Gamma(1 - \lambda - \lambda_1 + \lambda_2)} + \  \text{$\left( \lambda \leftrightarrow 1 - \lambda \right)$} \,.
}
The prefactor in \eqref{CAgain} gives 
\es{Pref}{
	\frac{B^-_s c_{\lambda_1, m-m_2}^*e^{-i\pi(m_2+\l_2)}}{2 \pi}    \frac{\sqrt{\Gamma(\lambda + m) \Gamma( 1- \lambda + m)}}{\Gamma(1 - \lambda_1 + \lambda_2 + m)}
	|m_2|^{ -\lambda_2 + \frac 12} 
}
In total, we have 
\es{Total}{
\lim_{m_2 \rightarrow -\infty }C^{s_1,\, \l_2^-,\, s}_{m-m_2, m_2, m} 	&=  -\frac{ c_{\lambda_1, m-m_2}^*}{ \pi}  B^-_s e^{-i\pi(m_2+\l_2)} \sin {\pi (\lambda-\lambda_1-\lambda_2) }
	\bigg(
	\frac{\Gamma( \lambda - \lambda_1 - \lambda_2 + 1) }{ \Gamma(1 - \lambda - \lambda_1 + \lambda_2)}   \\
	&{}\times 
	  \Gamma(1 - 2 \lambda) \frac{\sqrt{\Gamma(\lambda + m) }}{ \sqrt{ \Gamma( 1- \lambda + m)}}
	|m_2|^{\lambda_1 - \lambda -  \frac 12} 
	+ \  \text{$\left( \lambda \leftrightarrow 1 - \lambda \right)$}\bigg)
}

Thus, the large $m_2$ asymptotics of the product $(C^{s_1,\, \l_2^-,\, s}_{m-m_2, m_2, m})^* C^{s_1,\, \l_2^-,\, s'}_{m-m_2, m_2, m} $ are, 
\es{ProdAsymp}{
	&   {\abs{B_s}^2} \bigg[ |m_2|^{i (s - s') - 1}   
	\abs{\frac{\Gamma(- 2 is) }{ \Gamma( is_1 -is + \lambda_2 ) \Gamma(- is - is_1 + \lambda_2)}  }^2 
	+\begin{pmatrix}
		s \to -s \\ s'\to -s'
	\end{pmatrix}\bigg]
	\,,
}
where we kept $s \neq s'$ only in the power of $m_2$, anticipating that the sum over $m_2$ gives a term proportional to $\delta(s - s')$. To see why the sum $\sum_{m_2} (m_2)^{-1 + i \alpha}$ gives a delta function, note that we can regularize the sum by taking $\epsilon > 0$, thus writing $\sum m_2^{-1 + i \alpha - \epsilon} = \zeta(1 - i \alpha - \epsilon)$.  Close to $\alpha = 0$, this becomes $\frac{i}{\alpha + i \epsilon} \to P \frac{i}{\alpha} + \pi \delta(\alpha)$ as $\epsilon \to 0$.  The $P \frac{i}{\alpha}$ cancels from the final answer.  We finally find 
\es{GotN}{
	{\cal N}_{s_1,\, \l_2^-,\, s} =  &   2\abs{B^-_s}^2  
	\abs{\frac{\Gamma(- 2 is) }{ \Gamma(- is \pm  is_1 + \lambda_2)}  }^2 
	\,.
}

Similarly, we compute $\cN^+$ by focusing on the large $m_2$ limit of $(  C^{s_1,\, \l_2^+,\, s}_{m-m_2, m_2, m})^* C^{s_1,\, \l_2^+,\, s}_{m-m_2, m_2, m}$ with
\es{CAgain2}{
	C^{s_1,\, \l_2^+,\, s}_{m-m_2, m_2, m}
	&=B^+_s \frac{c_{\lambda_1, m - m_2}^*}{2 \pi c^+_{\lambda_2, m_2}} \frac{\Gamma(   \lambda_1 - \lambda_2 + m)}{\sqrt{\Gamma(\lambda + m) \Gamma( 1- \lambda + m)}} 
	\int_0^{2 \pi}  d \phi \,  
	e^{i (\lambda_2-m_2)\phi} \left( 1 - e^{i \phi} \right)^{\lambda - \lambda_1 - \lambda_2} \\
	&{}\times 
	{}_2 F_1( \lambda - \lambda_1 +\lambda_2, \lambda - m, 1 -m - \lambda_1 + \lambda_2, e^{i \phi})  \,.
}
 We find after similar manipulations that when fixing $\l_2$ to be in the positive discrete series,
\es{GotNP}{
	{\cal N}_{s_1,\, \l_2^+,\, s} =  &   2\abs{B^+_s}^2  
	\abs{\frac{\Gamma(- 2 is) }{ \Gamma(- is \pm  is_1 + \lambda_2)}  	 {\sin (\pi (\mu_1+\lambda_2+\lambda))\over \sin (\pi (\mu_1+ \lambda_1))} }^2 
	\,.
} 

\subsubsection*{Clebsch-Gordan coefficients in the $\mu_1 \to i \infty$ limit}

In order to compute the expectation of Wilson lines value once fixing the the value of $\phi^\mR=-i$ along the boundary we are interested in analytically continuing the product of Clebsch-Gordan coefficients for imaginary values of $\mu_1$. Specifically, we would like to compute
\ie
I^{s_1,\, \l_2^\pm,\, s}_{m-m_2, m_2, m} \equiv ({\cal N}_{s_1,\, \l_2^\pm,\, s})^{-1} C^{s_1,\, \l_2^\pm,\, s}_{m-m_2, m_2, m} (C^{s_1,\, \l_2^\pm,\, s}_{m-m_2, m_2, m})^*
\fe
in the limit $\mu_1 \rightarrow i \infty$, with $m-m_2 = \mu_1 + \mZ$. Note that in the above expression we will first take conjugate, and then take the limit $\mu_1\to i \infty$.

We start with \eqref{CAgain} and \eqref{CAgain2}
and use
\ie
\lim_{x\to \infty}{}_2 F_1(a,b+x,c+x,z) = (1-z)^{-a}\,,
\fe
which holds away from $z=1$.
In this limit the Clebsch-Gordan coefficients become
\ie
C^{s_1,\, \l_2^-,\, s}_{m-m_2, m_2, m}
&\sim B^-_s \frac{c_{\lambda_1, m - m_2}^*}{2 \pi c^-_{\lambda_2, m_2}} \frac{\sqrt{\Gamma(\lambda + m) \Gamma( 1- \lambda + m)}}{\Gamma(1 - \lambda_1 + \lambda_2 + m)} 
\int_0^{2 \pi}  d \phi \,  
e^{i m_2 \phi} e^{i \lambda_2 \phi} \left( 1 - e^{i \phi} \right)^{- 2\lambda_2} \,,\\
\\
(	C^{s_1,\, \l_2^-,\, s}_{m-m_2, m_2, m})^*
&\sim (B^-_s)^* \frac{c_{\lambda_1, m - m_2}}{2 \pi (c^-_{\lambda_2, m_2})^*} \frac{\sqrt{\Gamma(\lambda + m) \Gamma( 1- \lambda + m)}}{\Gamma( \lambda_1 + \lambda_2 + m)} 
\int_0^{2 \pi}  d \phi \,  
e^{-i m_2 \phi} e^{-i \lambda_2 \phi} \left( 1 - e^{-i \phi} \right)^{  - 2\lambda_2} \,,\\
\fe

Now using
\ie
\int_0^{2 \pi}  d \phi \,  
e^{i a \phi}   \left( 1 - e^{i \phi} \right)^{ b}={i(1-e^{2 \pi i a})\Gamma(a)\Gamma(b+1)\over \Gamma(1+a+b)}
=
{2\pi e^{ \pi i a} \Gamma(b+1)\over\Gamma(1-a) \Gamma(1+a+b)}\,,
\fe
valid by analytic continuation in $b$,
and
\ie
\lim_{z\to \infty, z\notin \mR_-} \Gamma(z)\sim e^{-z} z^z \sqrt{2\pi \over z}(1+\cO(1/z))\,,
\fe
we have that in the limit $\mu_1 \rightarrow i \infty$, and consequently in the limit $m \rightarrow i \infty$, 
\ie
C^{s_1,\, \l_2^-,\, s}_{m-m_2, m_2, m}
&\sim {B^-_s\over   c^-_{\lambda_2, m_2}}  
m^{-\lambda_2} e^{\pi i (m_2+\lambda_2)}
{  \Gamma(1-2\lambda_2)\over \Gamma(1-\lambda_2\pm m_2)}\,,
\\
(	C^{s_1,\, \l_2^-,\, s}_{m-m_2, m_2, m})^*
&\sim{(B^-_s)^*\over  (c^-_{\lambda_2, m_2})^*}  
m^{-\lambda_2} e^{\pi i (m_2+\lambda_2)}
{\Gamma(1-2\lambda_2)\over \Gamma(1-\lambda_2\pm m_2)}\,.
\fe

Putting this together, we obtain 
\ie
\label{eq:I-for-l-2+}
I^{s_1,\, \l_2^-,\, s}_{m-m_2, m_2, m}
 &
\sim 
=  \mu_1 ^{-2\lambda_2} {\Gamma(1-m_2-\lambda_2)\over\Gamma( \lambda_2-m_2) }  {  \Gamma(1-2\lambda_2)^2\over \Gamma(1-\lambda_2\pm m_2)^2} \cI
\\
 &
 =  \mu_1 ^{-2\lambda_2} (-1)^{m_2+\lambda_2}  {  \Gamma(1-2\lambda_2) \over \Gamma( 2\lambda_2 ) \Gamma(1-\lambda_2\pm m_2)} \cI\,,
\fe
with
\ie
\cI=& {1\over 2}\abs{\frac{ \Gamma(- is \pm  is_1 + \lambda_2)}  {\Gamma(- 2 is) } }^2 
=
{s \sinh (2\pi s)\over \pi}\Gamma(\pm is \pm  is_1 + \lambda_2)\,.
\fe

Similarly
we have in this limit
\ie
C^{s_1,\, \l_2^+,\, s}_{m-m_2, m_2, m}
&\sim B^+_s \frac{c_{\lambda_1, m - m_2}^*}{2 \pi c_{\lambda_2, m_2}} 
\frac{\Gamma(   \lambda_1 - \lambda_2 + m)}{\sqrt{\Gamma(\lambda + m) \Gamma( 1- \lambda + m)}} 
\int_0^{2 \pi}  d \phi \,  
e^{i (\lambda_2-m_2) \phi} \left( 1 - e^{i \phi} \right)^{- 2\lambda_2}\,,
\fe
which, together with the conjugate relation, yields in the limit $\mu_1 \rightarrow i \infty$ and $m - m_2 \rightarrow i \infty $,
\ie
C^{s_1,\, \l_2^+,\, s}_{m-m_2, m_2, m}
&\sim {B^+_s\over   c_{\lambda_2, m_2}}  
m^{-\lambda_2}e^{\pi i (m_2-\lambda_2)}
{  \Gamma(1-2\lambda_2)\over \Gamma(1-\lambda_2\pm m_2)}
\\
( C^{s_1,\, \l_2^+,\, s}_{m-m_2, m_2, m})^*
&\sim{(B^+_s)^*\over  c_{\lambda_2, m_2}^*}  
m^{-\lambda_2}e^{\pi i (m_2-\lambda_2)}
{\Gamma(1-2\lambda_2)\over \Gamma(1-\lambda_2\pm m_2)}\,,
\fe
and
\ie
\label{eq:I-for-l-2-}
I^{s_1,\, \l_2^+,\, s}_{m-m_2, m_2, m}\sim& \mu_1^{-2\lambda_2} {\Gamma(m_2+1-\lambda_2)\over\Gamma(m_2+\lambda_2) }  {  \Gamma(1-2\lambda_2)^2\over \Gamma(1-\lambda_2\pm m_2)^2} \cI
\\
=&
 \mu_1 ^{-2\lambda_2} (-1)^{m_2-\lambda_2}  {  \Gamma(1-2\lambda_2) \over \Gamma( 2\lambda_2 ) \Gamma(1-\lambda_2\pm m_2)} \cI\,,
\fe
which is identical to $I^{s_1,\, \l_2^-,\, s}_{m+m_2, -m_2, m}$.

\subsection{Fusion coefficient as $\mu \rightarrow i \infty$}
We are interested in generalizing the simple Clebsch-Gordan decomposition of the product of matrix element for some group element $g$ (given by $U_{R_1,\, n}^m(g) U_{R_1,\, n'}^{m'}(g) $) for compact groups, to the case of $\SL2$. To do this we start by inserting two complete set of states to re-express the product of two $\SL2$ matrix elements
\be 
U_{\left(\l_1 = \frac{1}2+is_1,\,\mu_1\right),\,  n_1}^{m_1}(g) &U_{\lambda_2^\pm,\,  n_2}^{m_2}(g) = \langle (\l_1, \mu_1), m_1; \l_2^\pm, m_2|g| (\l_1, \mu_1), n_1; \l_2^\pm, n_2 \rangle = \nn\\ &=  \int \frac{ds}{{\cal N}_{s_1,\, \l_2^\pm,\, s} } \frac{ds'}{{\cal N}_{s_1,\, \l_2^\pm,\, s'}}   \langle (\l_1, \mu_1), m_1; \l_2^\pm, m_2| (\l, \mu_1\pm \l_2), m_1+m_2 \rangle   \nn\\& \times \langle (\l_1, \mu_1), n_1; \l_2^\pm, n_2| (\l', \mu_1\pm\l_2), n_1+n_2 \rangle^*  \langle \l, m_1+m_2 |   g   | \l', n_1+n_2 \rangle \nn \\ &+ \text{ discrete series contributions .}
\ee
Thus, the product of two matrix elements is given by 
\be 
\label{eq:product-of-U}
U_{\left(\l_1 = \frac{1}2+is_1,\,\mu_1\right),\, n_1}^{m_1}&(g) U_{\lambda_2,\, n_2}^{m_2}(g)  =  \int \frac{ds}{{\cal N}_{s_1,\, \l_2^+,\, s}} C_{m_1, m_2, m_1+m_2}^{s_1, \l_2^\pm, s} (C_{n_1, n_2, n_1+n_2}^{s_1, \l_2^\pm, s})^*\,U^{m_1+m_2}_{(\l = \frac{1}2+is, \mu + \l_2),\,n_1+n_2 }(g) \nn \\ &+ \text{ discrete series contributions . }
\ee
In the limit $\mu_1 \rightarrow i \infty$ we are interested in computing the fusion between the regular character $\chi_{(s_1, \mu_1)}(g) $ and the truncated character $\bar \chi_{\l_2^\pm}(g) $ defined in \eqref{eq:regularized-character}. Thus, the product of characters is given by
\be 
\label{eq:product-of-characters}
\chi_{(s_1, \mu_1)}(g) \bar \chi_{\l_2^\pm}(g) &= \int ds \left(\sum_{k=0}^\Xi  I^{s_1,\, \l_2,\, s}_{\mu_1 + \tilde k, \,\pm (\l_2 + k),\, \mu_1 + \tilde k \pm (\l_2 + k)}\right) \chi_{(s, \mu_1+\l_2)}(g)\,\nn \\ &+ \text{ discrete series contributions }\,,
\ee
where we identify $m_1 = \mu_1 +\tilde k$ and $m_2 = \pm(\l_2+ k)$ with $\tilde k \in \mZ$ and $k \in \mZ^+$. We note that the sum over  $k$ yields a result that is independent of $\tilde k$, therefore leading to the separation of the sums in \eqref{eq:product-of-U}. Alternatively, the results above can be recasted  as the group integral of three matrix elements given by
\be 
\int dg\, &U_{\left(is_1,\,\mu_1\right),\, n_1}^{m_1}(g) U_{\lambda_2,\, n_2}^{m_2}(g)  U^{n_1+n_2}_{(s, \mu_1 + \l_2),\,m_1+m_2}(hg^{-1}) \nn\\  &= \frac{ C_{m_1, m_2, m_1+m_2}^{s_1, \l_2^\pm, s} (C_{n_1, n_2, n_1+n_2}^{s_1, \l_2^\pm, s})^*U^{n+n' }_{(\l = \frac{1}2+is, \mu + \l_2),\,m+m'}(h) }{\rho(s, \mu+\l_2){\cal N}_{s_1,\, \l_2^+,\, s}}\,,
\ee
where $\rho(s, \mu+l_2)$ is the $\SL2$ Plancherel measure in \eqref{eq:plancherel-inversion-formula}, and where we note that the product $\rho(s, \mu+\l_2){\cal N}_{s_1,\, \l_2^+,\, s}$ is symmetric under the exchange of $s_1$ and $s$. 
  Consequently, the product of two regular continuous series  characters and a regularized discrete series character is given by
\be 
\int dg\, &\chi_{\left(s_1,\,\mu_1\right)}(g) \bar \chi_{\lambda_2^\pm}(g)  \chi_{(s, \mu_1 + \l_2)}(h g^{-1}) = \frac{ \chi_{(s, \mu_1 + \l_2)}(h)}{\rho(s, \mu_1+\l_2)}\sum_{m_1 - m_2}  I_{m_1, m_2, m_1+m_2}^{s_1, \l_2^\pm, s}\,.
\ee
Using Eq.~\eqref{eq:I-for-l-2+} and \eqref{eq:I-for-l-2-} we thus find that by taking the $\mu_1 \rightarrow i \infty$ limit and truncating the sum over $m_1 - m_2$, 
\be
\label{eq:fusion-coeff}
\lim_{\mu_1 \rightarrow i \infty }\int dg\,\chi_{\left(s_1,\,\mu_1\right)}(g) \bar `\chi_{\lambda_2^\pm}(g)  \chi_{(\l = \frac{1}2+is, \mu_1 + \l_2)}(h g^{-1})   &= \frac{N_{\l_2^\pm} N^{s}{}_{s_1, \l_2^\pm}}{\rho(s, \mu_1+\l_2)}  \chi_{(s, \mu_1 + \l_2)}(h)\,,
\ee
where we define the fusion coefficient $N^{s_1, \l_1}{}_{s}$ in the $\mu_1 \to i \infty$ limit, 
\be
N^{s}{}_{s_1, \l_2^\pm} \equiv  \frac{|\Gamma(\l_2 + is_1 - i s)\Gamma(\l_2+i s_1 + is)|^2}{\Gamma(2\l_2) }\,,
\ee 
up to a $\l_2^\pm$ dependent normalization constant,  
\be 
\label{eq:funsion-coeff}
N_{\l_2^\pm}=   \sum_{ k=0}^{\Xi} \mu_1^{-2\l_2}(-1)^{ k}{  \Gamma(1-2\lambda_2) \over  \Gamma(1+k)\Gamma(1-k-2\l_2)} 
= \frac{(-1)^{\Xi} \mu_1^{-2\l_2}  \Gamma(-2\l_2)}{\Xi! \Gamma(-\Xi-2\l_2)}\,,
\ee
As we take the cut-off, $\Xi \to \infty$, the normalization constant becomes  
 \be
 \label{eq:normalization-factor-W-loop} 
 N_{\l_2^\pm}= \frac{\mu_1^{-2\l_2} \Xi^{2\l_2}}{\Gamma(1+2\l_2)}
 \ee
 Using the fusion coefficient, together with the normalization factor, we compute the expectation value of the Wilson lines in Section~\ref{sec:wilson-loop-and-bi-local-op}.

\subsection{6-j symbols}
\label{app:6-j symbols}
To obtain the OTO-correlator in Section~\ref{sec:OTO4pt} we need to consider the integral of six characters in \eqref{eq:intersecting-Wilson-loop-exp-value},
\be
&\int dh_1 dh_2 dh_3 dh_4\, \chi_{s_1}(h_1 h_2^{-1}) \chi_{s_2}(h_2 h_3^{-1}s) \chi_{s_3}(h_3 h_4^{-1}) \chi_{s_4}(g h_4 h_1^{-1}) \bar \chi_{\l_1^\pm}(h_1 h_3^{-1})\bar \chi_{\l_2^\pm}(h_2 h_4^{-1}) = \nn\\
&= \int  dh_1 dh_2 dh_3 dh_4\, \sum_{m_i, n_i, q_i, \tilde m_i} U_{s_1, n_1}^{m_1}(h_1) U_{s_1, m_1}^{n_1}(h_2^{-1}) U_{s_2, n_2}^{m_2}(h_2)  U_{s_2, m_2}^{n_2}(h_3^{-1}) U_{s_3, n_3}^{m_3}(h_3)  \nn \\ &\times   U_{s_3, m_3}^{n_3}(h_4^{-1}) U_{s_4, n_4}^{m_4}(g) U_{s_4, q_4}^{n_4}(h_4) U_{s_4, m_4}^{q_4}(h_1^{-1}) U_{\l_1^\pm, \tilde n_1}^{\tilde m_1}(h_1) U_{\l_1^\pm, \tilde m_1}^{\tilde n_1}(h_3^{-1}) U_{\l_2^\pm, \tilde n_2}^{\tilde m_2}(h_2) U_{\lambda_2^\pm, \tilde m_2}^{\tilde n_2 }(h_4)\,, \ee
where, for the case of interest in Section~\ref{sec:OTO4pt}, $s_1$, $s_2$, $s_3$, and $s_4$ label continuous series representations, and $\l_1^\pm$ and $\l_2^\pm$ label representations in the positive/negative discrete series.  As in the case of computing the time-ordered correlators of the Wilson lines we first consider the result when $\mu_1 \in \mR$ and only afterwards analytically continue the final result to $\mu_1 \to i \infty$. 

The sums over $\tilde m_1$ and $\tilde n_1$, as well as that over $\tilde m_2$ and $\tilde n_2$ are truncated according to the regularization prescription for the characters associated to the Wilson lines.  Evaluating the integrals we find  
\be
\label{eq:OTO-sum-over-CG-coeff}
& \sum_{m_i, n_i, \tilde m_i, \tilde n_i, q_4}  U_{s_4, q_4}^{m_4}(g) \frac{C_{m_1, \,\tilde m_1, \, m_4}^{s_1, \l_1^\pm, s_4}
(C_{n_1, \,\tilde n_1,\, q_4 }^{s_1, \l_1^\pm, s_4})^*}{\rho(s_4,\, \mu_4){\cal N}_{s_1,\, \l_1^\pm,\, s_4}}
\frac{C_{m_1,\,\tilde m_2, \, m_1}^{s_2, \lambda_2^\pm, s_1 }
(C_{n_2,\,\tilde n_1, \, n_1}^{s_2, \lambda_2^\pm, s_1})^*}{\rho(s_1,\, \mu_1){\cal N}_{s_2,\, \l_2^\pm,\, s_1}}
\nn\\ & \hspace{2.7cm} \times 
\frac{C_{m_3,\,\tilde m_1, \, m_2 }^{s_3, \lambda_1^\pm, s_2}
(C_{n_3, \,\tilde n_1,\, n_2}^{s_3, \lambda_1^\pm, s_2})^*}{\rho(s_2,\, \mu_2){\cal N}_{s_3,\, \l_1^\pm,\, s_2}}
  \frac{C_{n_4, \,\tilde n_2, \, m_3}^{s_4,  \lambda_2^\pm, s_3}
(C_{q_4, \,\tilde m_2, \, n_3}^{s_4,  \lambda_2^\pm, s_3})^*}{\rho(s_3,\, \mu_3){\cal N}_{s_4,\, \l_2^\pm,\, s_3}}
  \,,
\ee
Performing the sums over the $n_1$, $n_2$, $n_3$, $\tilde n_1$ and $\tilde n_2$ states we obtain the 6-j symbol associated to the six representations $s_1,\, s_2,\,s_3,\,s_4,\,\l_1^\pm$, and $\l_2^\pm$. Furthermore, the sum also imposes the constraint $m_4 = q_4$. The remaining sum over four Clebsch-Gordan coefficient yields the square root for the factor present in \eqref{eq:fusion-coeff}. Specifically, we obtain that \eqref{eq:OTO-sum-over-CG-coeff} equals 
\be 
\label{eq:going-to-6-j symbols}
N_{\l_1^\pm} N_{\l_2^\pm}\chi_{s_4}(g)   \sqrt{N^{s_4}{}_{\l_1^\pm, s_1}  N^{s_3}{}_{\l_1^\pm, s_2}  N^{s_3}{}_{\l_2^\pm, s_1}  N^{s_4}{}_{\l_2^\pm, s_2} } R_{s_3 s_4}\! \left[\; {}^{s_2}_{s_1} \;{}^{\l_2}_{\l_1}\right] \,.
\ee
The 6-j symbol for $\SL2$ is given by \cite{groenevelt2006wilson} 
\be
\label{eq:6-j symbol}
R_{s_3 s_4}\! \left[\; {}^{s_2}_{s_1} \;{}^{\l_2}_{\l_1}\right] \, & \, = \,\,\mathbb{W}(s_3, s_4 ; \l_1 + i s_2,\l_1 - i s_2, \l_2 - i s_1,\l_2 + i s_1) \\[3.5mm]
 &\times\,\sqrt{\Gamma(\l_2 \pm i s_1 \pm is_3)\Gamma(\l_1 \pm i s_2 \pm is_3)\Gamma(\l_1 \pm is_1\pm is_4)\Gamma(\l_2 \pm i s_2 \pm i s_4)}\nonumber\,,
\ee 
where the Wilson function $\mathbb{W}(s_a, s_b ; \l_1 + i s_2,\l_1 - i s_2, \l_2 - i s_1,\l_2 + i s_1) $  is given by \cite{groenevelt2003wilson}
\be 
\label{eq:wilson-function}
\mathbb{W}(\a, \b, a,b,c,d) \equiv \frac{\Gamma(d-a){}_4F_3\left[\; {}^{a+i\b}_{a+b} \;{}^{a-i \b}_{a+c} \;{}^{a-i \b}_{a+c} \;{}^{\tilde a + i \a}_{1+a-d}\;{}^{\tilde a - i \alpha}; \,1\right]}{\Gamma(a+b)\Gamma(a+c)\Gamma(d\pm i \b)\Gamma(\tilde d\pm i \alpha )} + (a \leftrightarrow d)\,, 
\ee
with $\tilde a = (a+b+c-d)/2$ and $\tilde d = (b+c+d-a)/2$. The normalization for the 6-j symbol in \eqref{eq:6-j symbol} is obtained by imposing the orthogonality relation \eqref{eq:6-j symbol-ortho-main-text } using the orthogonality properties of the Wilson function \cite{groenevelt2003wilson,groenevelt2006wilson}. Such an orthogonality condition on the 6-j symbol follows from its definition in terms of a sum of Clebsch-Gordan coefficients as that shown in \eqref{eq:OTO-sum-over-CG-coeff}. 

Firstly we note that the result is the same when considering $\l_1$ or $\l_2$ in the positive or negative discrete series. Furthermore, since the result is explicitly independent of $\mu_1,\, \mu_2, \,\mu_3$ and $\mu_4$ one can easily perform the analytic continuation to  $\mu_1,\, \mu_2, \,\mu_3, \mu_4 \to i \infty$ as required by our boundary conditions on the field $\phi^\mR$. Putting this together with the analytic continuation of the fusion coefficients presented in the previous sub-section we find the final results from Section~\ref{sec:wilson-loop-and-bi-local-op}.

\section{Wilson lines as probe particles in JT gravity}
\label{sec:wilson-loop-and-external-matter}

As mentioned in Section~\ref{sec:wilson-loop-and-bi-local-op}, the insertion of a Wilson loop in 3D Chern-Simons theory with gauge algebra $\mathfrak{so}(2,2)$ (or an isomorphic algebra) can be interpreted as the effective action of a massive probe in AdS$_3$ (or other spaces with an isomorphic symmetry algebra) \cite{witten1989topology, carlip1989exact, vaz1994wilson, de1990spin, skagerstam1990topological, Ammon:2013hba}. 
In this Appendix we extend this interpretation to 2D. 
Specifically, we outline the proof of the equivalence, as stated in section~\ref{sec:gravitational-interp-wilson-line}, between the boundary-anchored Wilson line observables ${\cal W}_\lambda({\cal C}_{\tau_1\!\tau_2})$ in the $\cG=\cG_B$ BF theory formulation, and the boundary-to-boundary propagator of a massive particle in the metric formulation of JT gravity. The latter is given by the functional integral over all paths $x(s)$ diffeomorphic to the curve ${\cal C}_{\tau_1\!\tau_2}$ weighted with the standard point particle action (here $\dot{x}^\mu = \frac{d x^\mu}{ds}$)
\be
S[x, g_{\mu\nu}] =  m \int_{{\cal C}_{\tau_1\!\tau_2}}\! ds\, \sqrt{g_{\mu\nu} \dot{x}^\mu \dot{x}^\nu}\,.
\label{2particle}
\ee
 Concretely, we would like to demonstrate that 
\be
\label{equality}
\widehat{\cal W}_{\lambda,k=0}({\cal C}_{\tau_1\!\tau_2})= \tr_{\lambda, k=0} \Bigl({\cal P} \exp \int_{{{\cal C}_{\tau_1\tau_2}}}\! A \Bigr)   \cong \!\!\int\limits_{{\rm paths}\, \sim \, {\cal C}_{\tau_1\!\tau_2}}\!\!\!\!\!\! [dx] \, e^{-S[x, g_{\mu\nu}]},
\ee
where the mass of the particle is determined by the $\cG_B$ representation $(\lambda,k=- 2\pi \lambda/B)$ as $m^2 = \lambda(\lambda-1) = -C_2(\lambda)$.\footnote{For notational simplicity, we take all Wilson lines to be in the positive discrete series representations in this section. We also emphasize that the Wilson line in the representation $(\lambda,k)$ is a defect operator (external probe), thus $k$ is not constrained to be $k_0$.} In the equation above, we have taken the limit $B \rightarrow \infty$ thus set $k=0$. Consequently  the Wilson line $\widehat{\cal W}_{\lambda,k=0}({\cal C}_{\tau_1\!\tau_2})$ only couples to the $\mathfrak {sl}(2,\mR)$-components of the $\cG_B$ gauge field. In the rest of this Appendix, we will implicitly assume that $A$ take values in $\mathfrak{sl}(2,\mR)$. For notation convenience, we will refer to these Wilson lines as $\widehat{\cal W}_{\lambda}({\cal C}_{\tau_1\!\tau_2})$ from now on.\footnote{Equivalently, one can think of the boundary-anchored Wilson lines $\widehat{\cal W}_{\lambda}({\cal C}_{\tau_1\!\tau_2})$ as $PSL(2,\mR)$ Wilson lines in the discrete series representation $\lambda$ (projective for $\lambda\notin \mZ$) of $PSL(2,\mR)$.}

 The congruence symbol $\cong$ in \eqref{equality} indicates that we want to prove an operator equivalence inside the functional integral of JT gravity.  Indeed, the right-hand side of \eqref{equality} depends only on the diffeomorphism class of the path  ${\cal C}_{\tau_1\!\tau_2}$, whereas the Wilson line operator $\widehat{\cal W}_{\lambda}({\cal C}_{\tau_1\!\tau_2})$ on the left-hand side follows some given path. So in writing \eqref{equality},  we implicitly assume that  $\widehat{\cal W}_{\lambda}({\cal C}_{\tau_1\!\tau_2})$  is evaluated inside the functional integral of a diffeomorphism invariant $BF$ gauge theory.

To start proving \eqref{equality},  following \cite{Witten:1988hf, Beasley:2009mb}, we rewrite the Wilson line $\widehat{\cal W}_{\lambda}({\cal C}_{\tau_1\!\tau_2})$ around a given space-time contour ${\cal C}_{\tau_1\!\tau_2}$, parametrized by an auxiliary variable $s$,  as a functional integral over paths $g(s) \in PSL(2,\mR)$ via \footnote{Note that coadjoint orbits of a connected semisimple Lie group are identical with those of the  universal cover groups, as evident from the definition \eqref{orbit} for the $PSL(2,\mR)$ case and its coverings.}
\be
\label{wloop}
\tr_\lambda \left({\cal P} \exp \oint_{{\cC_{\tau_1\tau_2}} }\!\!\!\! A \right) = \int_{\cC_{\tau_1\tau_2}} \! [dg]_{\pmb \alpha} \, e^{-S_{\pmb  \alpha}[g,A]}\ee
where $S_{\pmb  \alpha}[g,A]$ denotes the (first order) coadjoint orbit action of the representation $\lambda$, coupled to a background $\mathfrak{sl}(2,\mR)$ gauge field~$A_s(s) \equiv  A_\mu(x(s)) \dot x^\mu(s)$
\begin{eqnarray}
\label{coadjoint}
S_{\pmb \alpha}[g,A] \, = \, \int_{\cC_{\tau_1\tau_2}}\! ds \,\tr \left({\pmb \alpha}\, g^{-1} D_{\! A} g\right) \, = \,  
\int_{\cC_{\tau_1\tau_2}} \! ds \,\left(\tr({\pmb \alpha}\, g^{-1} \partial_s g)  - \tr(A_s g {\pmb \alpha} g^{-1}) \right) . 
\end{eqnarray}
Here ${\pmb \alpha} = \alpha_i P^i  \in \mathfrak{sl}(2, \mR)$ denotes some {\it fixed} Lie algebra element with specified length squared equal to the second Casimir  
\be
\label{plength}
\tr ({\pmb \alpha}^2) = -C_2(\lambda) =- \lambda (\lambda-1)
 \ee
 The classical phase space in \eqref{coadjoint} is over the (co)adjoint orbit of the Lie algebra element~${\pmb \alpha}$
 \ie
 \cO_{\pmb \alpha}\equiv \{g{\pmb \alpha}g^{-1}|g\in PSL(2,\mR)\}\ \,
 \label{orbit}
 \fe
 Consequently the path integral is over maps from $\cC_{\tau_1\tau_2} \to \cO_{\pmb \alpha}$ which can be equivalently described by their lift $g: \cC_{\tau_1\tau_2} \to PSL(2,\mR)$ up to an identification due to local right group action by the stabilizer of $\pmb \alpha$ on $g$. This is the meaning of path integral measure $[dg]_{\pmb \alpha}$ in \eqref{wloop}.

Let us briefly recall why equation \eqref{wloop} holds.  Expanding $g$ around a base-point, with $g = e^{x^a(s)P_a} g(s_0) $, we find from \eqref{coadjoint} that the canonical momenta associated to $x^a(s)$ are give by 
\ie
\pi_{x^i} = \tr (P^i g {\pmb \alpha} g^{-1}),
\fe
which are in fact the generators of the $PSL(2,\mR)$ symmetry which acts by left multiplication on $g$, as $g \rightarrow U g$. 
  The Casimir associated to $\mathfrak{sl}(2, \mR)$ component of $\cG_B$  is given by $\hat C_2^{\mathfrak{sl}(2, \mR)} = - \eta^{ij} \pi_{x^i} \pi_{x^j}= - \tr( \pmb \alpha^2) $. The Hilbert space of the theory is spanned by functions on the group $\cG_B$ which are invariant under   right group actions that stabilize ${\pmb \alpha}$. The Hilbert space of the quantum mechanics model on $\cO_{\pmb \alpha}$ thus forms an irreducible (projective) $PSL(2, \mR)$ representation~$\lambda$. Since the functional integral  around a closed path $g(s) \in PSL(2,\mR)$ amounts to taking the trace over the Hilbert space, we arrive at the identity \eqref{wloop}.\footnote{This is because we are considering a boundary condition with $A_\tau=0$. Consequently, the boundary-anchored Wilson line has the same expectation value as a Wilson loop that touches the boundary.}

  Since the identity \eqref{wloop} holds for any choice of Lie algebra element $\pmb \alpha$ with length squared given by \eqref{plength}, we are free to include in  the definition of ${\cal W}_\lambda({{\cal C}_{\tau_1\!\tau_2}})$ a functional integral over all Lie algebra elements of the form
\be
{{\pmb \alpha}}(s) = \alpha_a(s) P^a = \alpha_1(s) P^1 +\alpha_2(s) P^2
\ee
subject to the constraint \eqref{plength}. This leads to the identity (up to an overall factor that does not depend on $A$)
\be
\hat {\cal W}_\lambda({\cal C}_{\tau_1\!\tau_2}) \sim \int  \! [d\alpha_{1,2} dg d\Theta]\, e^{-S_{\pmb \alpha}[g,\Theta,A]}
\label{Padjlm}
\ee
with 
\be
S_{\pmb \alpha}[g,\Theta,A] = \oint_{{\cal C}_{\tau_1\!\tau_2}}\!\!\! \! ds \, \left(\tr \left({{\pmb \alpha}}\, g^{-1} D_{\! A} g\right) +i \Theta (\eta^{ab}\alpha_a \alpha_b - m^2)\right) \,.
\label{Sadjlm}
\ee
Here $m^2 = \lambda(\lambda-1)$ and  $\Theta$ denotes a Lagrange multiplier that enforces the constraint \eqref{plength}. This already looks closely analogous to the world line action of a point particle of mass $m$.

So far we have considered a general background gauge field  $A$ in the bulk. In the context in which we make $A$ dynamical and perform the path integral in the BF-theory in the presence of a defect \eqref{eq:total-action}, the path integral (after integrating out the adjoint scalar $\phi$)  localizes to configurations of flat $A$, away from the defect. Similarly, on the JT gravity side (in the metric formulation), integrating out the dilaton $\Phi$ forces the ambient metric on the disk to be that of AdS$_2$. 
Thus for the purpose of proving \eqref{equality}, we can take $A$ to be flat on the BF theory side, and the metric to be AdS$_2$ on the JT gravity side.

The action \eqref{Sadjlm} is invariant under gauge transformations $U(s)$ for which $g\rightarrow U(s) g$, together with the corresponding gauge transformation of $A$ which leaves the connection flat. Note however the gauge transformation mixes the components of $A$ associated to the frames and spin connection.  We can always (partially) gauge fix by setting  $g = \mathbf 1$ by choosing $U(s) = g^{-1}(s)$ along the curve $\mC_{\tau_1 \tau_2}$ and smoothly extending this gauge transformation onto the entire disk.\footnote{There's no obstruction for such extensions since $\cG_B$ is simply connected.} After such a gauge fixing, the action \eqref{Sadjlm} simply becomes,
\be
\label{eq:worldline-action_2}
S_1[x, k, \l, g^{\mu \nu}]
&\equiv  \int_{\cC_{\tau_1\tau_2}} ds (k_\mu \dot x^\mu + i \Theta(g^{\mu \nu}k_\mu k_\nu - m^2))
\\
 &=\int_{\cC_{\tau_1\tau_2}}\, ds \, \left(  \eta_{ab}{\alpha}^a \tilde e^b_\mu \, \dot x^\mu + i \Theta (  \eta^{ab}\alpha_a \alpha_b - m^2)\right) \nn 
 \,,
\ee
where $g_{\mu \nu} =  \eta_{ab}    e_\mu^a   e_\nu^b $ is the AdS$_2$ metric associated to the background flat connection $A$ and $k_\mu \equiv \alpha_a e^a_\mu$. The action \eqref{eq:worldline-action_2} agrees with the first order action for a particle moving on the world-line ${\cal C}_{\tau_1\tau_2}$. To finish the proof, we need to show that the path integral over flat $A$ in the BF theory reproduces the integral over paths diffeomorphic to $\cC_{\tau_1\tau_2}$ for the particle in the JT gravity.

 As mentioned in Section~\ref{sec:rewriting-JT-gravity}, space-time diffeomorphisms can be identified with field dependent gauge transformations in the BF theory when the gauge field is flat
 \be
\label{eq:gauge-transf-same-as-diffeo}
\delta^{\text{diff}}_{\xi} = \delta^{\text{gauge}}_\epsilon \,,
\ee
where the  the vector field $\xi^\mu(x)$ generating the diffeomorphism transformation and the infinitesimal gauge transformation parameter $\epsilon^a(x)$ (vanish on the boundary) are related by 
\be
\label{infid}
\epsilon^a(x) = e^a_\mu(x) \xi^\mu(x)\,, \qquad 
\epsilon^0(x) = \omega_\mu(x) \xi^\mu(x)\,.
\ee
Since flat connections $A$ are generated by gauge transformations, the equivalence \eqref{eq:gauge-transf-same-as-diffeo} acting on $\tilde e^b_\mu$ implies that, 
\ie
&\int_{\cC_{\tau_1\tau_2}}\, ds \, \left(  \eta_{ab}{\alpha}^a (\tilde e^b_\mu)_\epsilon \, \dot x^\mu + i \Theta (  \eta^{ab}\alpha_a \alpha_b - m^2)\right) \nn 
=
\int_{\cC^\xi_{\tau_1\tau_2}}\, ds \, \left(  \eta_{ab}{\alpha}^a  \tilde e^b_\mu \, \dot x^\mu + i \Theta (  \eta^{ab}\alpha_a \alpha_b - m^2)\right) \nn 
\fe
where  $ (\tilde e^b_\mu)_\epsilon$ denotes the finite gauge transformation of $\tilde e^b_\mu$ generated by $\epsilon$, and $\cC^\xi_{\tau_1\tau_2}$ denotes a path diffeomorphic  to $\cC_{\tau_1\tau_2}$ generated by displacement vector field $\xi$.  Consequently integrating over flat connections $A$ of the BF theory in the presence of the Wilson line insertion is equivalent to integrating over all paths diffeomorphic to the curve $\mC_{\tau_1 \tau_2}$, which  precisely gives the first order form of \eqref{2particle} that describes a particle propagating between boundary points in ${\rm AdS}_2$.\footnote{Note that in the world-line action \eqref{eq:worldline-action_2}, the fields $(x^\nu,k_\mu(x))$ take values in the co-tangent bundle $T^*\Sigma$. The path integration measure is the natural one induced by the symplectic structure of $T^*\Sigma$.}

Alternatively, to get the second order formulation for the world-line action we can directly  perform the Gaussian integration over $\alpha_a$ in \eqref{Sadjlm} and then integrate out the Lagrange multipler $\Theta$. The world-line path integral \eqref{Padjlm} becomes (up to an $A$ independent factor), 
\be
\label{eq:path-integral-over-g}
\hat{\cal W}_\lambda({\cal C}_{\tau_1\!\tau_2}) \sim \int  \! [ dg ]\, e^{-S_{\rm 2}[g, A]}\,,
\ee
where the action $S_2[g, A]$ is specified by
\be
\label{eq:action-2-coadjoint}
S_{\rm 2}[g,A] = m  \int_{{\cal C}_{\tau_1\!\tau_2}}\!\!\! \! ds \, \sqrt{  \eta_{ab} (g^{-1} D_{\! A} g)^a(g^{-1} D_{\!A} g)^b}\,.
\ee
Due to the integration over $g(s)$, this is a gauge invariant observable as expected.  Note that while \eqref{eq:action-2-coadjoint} is exact on-shell in order for the path-integral \eqref{eq:path-integral-over-g} to agree with \eqref{wloop} one has to appropriately modify the measure $[d g]$ in \eqref{eq:path-integral-over-g}.  

Once again performing the gauge transformation with  $U(s) = g^{-1}(s)$ along the curve $\mC_{\tau_1 \tau_2}$ to gauge fix $g(s)=1$ and smoothly extending the gauge transformation onto the entire disk, the action \eqref{eq:action-2-coadjoint} simply becomes,
\be
\label{eq:worldline-action}
S_{\rm 2}[g,A] =m  \int_{{\cal C}_{\tau_1\tau_2}}\,ds \, \sqrt{  \eta_{ab}    e^a_\a  e^b_\b \dot x^\a \dot x^\b}
=m  \int_{{\cal C}_{\tau_1\tau_2}}\,ds \, \sqrt{ g_{\alpha\beta} \dot x^\a \dot x^\b}\,,
\ee
which agrees with the 2nd order action \eqref{2particle} for a particle moving on the world-line ${\cal C}_{\tau_1\tau_2}$. Following the same reasoning as before, the gauge transformation can be mapped to a diffeomorphism, and integrating over flat connections in the BF theory path integral with the Wilson line insertion, is once again equivalent to integrating over all paths diffeomorphic to the curve $\mC_{\tau_1 \tau_2}$. Using this, we finally arrive at the desired equality between the Wilson line observable and the worldline representation of the boundary-to-boundary propagator given by \eqref{equality}.\footnote{As usual in AdS/CFT, the worldline observable (boundary-to-boundary propagator) requires appropriate regularization and renormalization due to the infinite proper length near the boundary of AdS$_2$. Here in the gauge theory description, we also require a proper renormalization of the boundary-anchored Wilson line to remove the divergence due to the infinite dimensional representation carried by the Wilson line (see \eqref{eq:regularized-character}). It would be interesting to understand the precise relation between the two renormalization schemes.}

\bibliographystyle{ssg}
\bibliography{Biblio}

\end{document}